\documentclass[11pt,fleqn]{article}

\usepackage{bm}
\usepackage{amsmath}
\usepackage{graphicx}
\usepackage{cite}
\usepackage[margin=1.25in]{geometry}
\usepackage{booktabs}
\usepackage{caption}
\usepackage{subcaption}
\usepackage{float}

\usepackage{color}

\newcommand{\tensr}[1]{\bm{\mathsf{#1}}}
\newcommand{\ms}{\scriptscriptstyle}
\newcommand{\um}{\scalebox{0.75}[1.0]{\( - \)}}

\newcommand{\K}{\kappa}

\newcommand{\PP}{\tensr{P}}
\newcommand{\F}{\tensr{F}}

\newcommand{\subscript}[1]{_{\ms #1}}

\usepackage{color}
\definecolor{orange}{rgb}{1,0.5,0}
\definecolor{darkorchid}{rgb}{0.6,0.196,0.8}
\definecolor{olivedrab}{rgb}{0.42,0.56,0.14}

\makeatletter
\newcommand{\mathleft}{\@fleqntrue\@mathmargin0pt}
\newcommand{\mathcenter}{\@fleqnfalse}
\makeatother

\newcommand{\Keq}[1]{\kappa\subscript{#1}^{eq}}

\newcommand{\wphi}[1]{\omega\subscript{#1}^{\phi}}

\definecolor{olivedrab}{rgb}{0.42,0.56,0.14}
\definecolor{oxfordblue}{rgb}{0.0, 0.13, 0.28}
\definecolor{deepsaffron}{rgb}{1.0, 0.6, 0.2}

\title{\vspace{-2.0cm} Thermocapillary Convection in Superimposed Layers of Self-Rewetting Fluids: Analytical and Lattice Boltzmann Computational Study}
\author{{Bashir Elbousefi, William Schupbach, Kannan N. Premnath, Samuel W.J. Welch}\\Department of Mechanical Engineering\\ University of Colorado Denver\\ 1200 Larimer Street, Denver, CO 80204, U.S.A.}

\begin{document}

\maketitle

\begin{abstract}
Self-rewetting fluids (SRFs), such as aqueous solutions of long-chain alcohols, exhibit anomalous quadratic dependence of surface tension on temperature having a minimum and with a positive gradient. When compared to the normal fluids (NFs) that have negative gradient of surface tension on temperature, the SRFs can be associated with significantly modified interfacial dynamics, which have recently been exploited to enhance flow and thermal transport in various applications, including those involving microgravity and microscale transport systems. In this work, first, we develop a new analytical solution of thermocapillary convection in superimposed two SRF layers confined within a microchannel that is sinusoidally heated on one side and maintained at a uniform temperature on the other side under the creeping flow regime and at small Marangoni and capillary numbers. Then, a robust central moment lattice Boltzmann method using a phase-field model involving the Allen-Cahn equation for interface tracking, two-fluid motion, and the energy transport for numerical simulations of SRFs with nonlinear surface tension variations is constructed. The analytical and computational techniques are generally shown to be in good quantitative agreement with one another in the simulation of superimposed SRFs in a microchannel. Moreover, the effect of the various characteristic parameters on the magnitude and the distribution thermocapillary-driven motion is studied. The thermocapillary flow patterns in SRFs are shown to be strikingly different when compared to the NFs: For otherwise the same conditions, the SRFs result in eight periodic counterrotating thermocapillary convection rolls, while the NFs exhibit only four such vortices. Moreover, the direction of the circulating fluid motion in such vortical structures for the SRFs is found to be towards the hotter zones on the interfaces, which is opposite to that in NFs. These features are found to be sustained even as the interfaces deforms in simulations. By tuning the sensitivity coefficients of the surface tension on temperature, it is shown that not only the magnitude of the thermocapillary velocity can be significantly manipulated, but also the overall flow patterns as well. It is also demonstrated that the thermocapillary convection can be enhanced if the SRF layer adjacent to the nonuniformly heated wall is made relatively thinner or has higher thermal conductivity ratio or has smaller viscosity when compared to that of the other fluid layer.
\end{abstract}

\section{Introduction}
Surface tension forces arising at the interface between fluids play prominent role in many multiphase and thermal transport processes~\cite{de2004capillarity}. Their variations can be caused by changes in the local interfacial temperature or with the addition of surface active materials (i.e., surfactants). The surface tension gradients result in the so-called Marangoni stresses~\cite{scriven1960marangoni}, which, via the viscous effects of the fluids, induce their convective motions in the vicinity of the interfaces~\cite{probstein2005physicochemical}. If they are set up due to local temperature variations, they are referred to as the thermocapillary convection. Since the seminal study by Young \emph{et al.}~\cite{young1959motion}, who demonstrated the ability of a bubble to migrate towards hot regions in the absence of gravity due to Marangoni stresses, thermocapillary effects have been exploited in controlling the motion of dispersed phases (bubbles or drops) in fluids, especially in microgravity applications~\cite{subramanian2002motion} (see e.g.,~\cite{welch1998transient,ma2011direct} for related numerical investigations). On the other hand, in micro-electro-mechanical-systems, as the scales of the devices are reduced, the interfacial forces dominate, and the thermocapillary convection can be utilized to manipulate the motion of fluid streams and thermal transport phenomena in microchannels (see e.g.,~\cite{darhuber2005principles,karbalaei2016thermocapillarity}). In particular, an analysis of convective currents set up via thermally induced surface tension gradients in superimposed fluids which are bounded by differentially heated walls represents study for a prototypical configuration in this regard~\cite{pendse2010analytical} (see e.g.,~\cite{gambaryan2015modulation} for a review of thermocapillary convection in layers of fluid films).

\begin{figure}[H]
  \begin{subfigure}{0.48\textwidth}
    \centering
    \includegraphics[width=\textwidth]{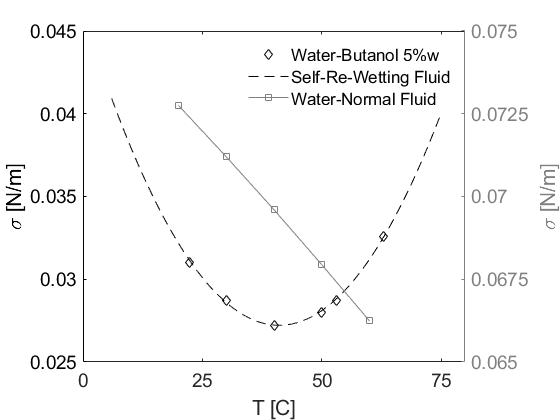}
    \caption{}
\end{subfigure}
\begin{subfigure}{0.48\textwidth}
    \centering
    \includegraphics[width=\textwidth]{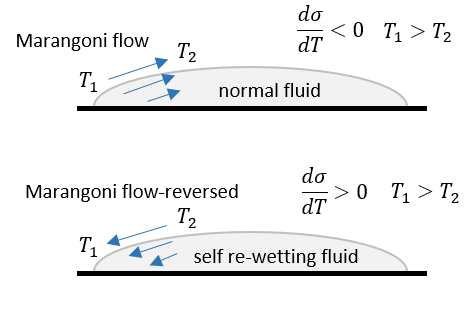}
    \caption{}
\end{subfigure}
\caption{$(a)$ Surface tension variation with respect to temperature for a normal fluid (NF) and for a self-rewetting fluid (SRF). The parabolic functional dependence of the surface tension for the example SRF (butanol aqueous solution) with temperature is based on a curve fit generated from the data given in~\cite{savino2009surface}. $(b)$ Differences in the thermocapillary motions in the vicinity of interfaces for NFs and SRFs.}
\label{fig1}
\end{figure}

It is generally known that common fluids, such as water, air, and various oils have the property of surface tension that decreases somewhat linearly with increasing temperatures. On the other hand, certain fluids, such as aqueous solutions of long-chain (i.e., ``fatty") alcohols, some liquid metallic alloys, and nematic liquid crystals exhibit anomalous nonlinear parabolic dependence of surface tension on temperature with a range involving its positive gradient. In particular, Vochten and Petre~\cite{vochten1973study} performed measurements in non-azeotropic, high-carbon alcohol solutions (such as n-butanol), and demonstrated that they have a property that beyond a certain threshold temperature, the surface tension will increase with further increase in temperature; the surface tension becomes a minimum at this threshold temperature, whose value increases for alcohols with longer carbon chains; and for a particular alcohol, the minimum surface tension decreases monotonically with its concentration. These findings were corroborated by follow on experimental studies reported in~\cite{petre1984experimental,limbourg1986thermocapillary,villers1988temperature}. In essence, for a range of temperature, such fluids have a positive gradient of surface tension with temperature and they have been named as ``self-rewetting" fluids (SRFs) by Abe~\emph{et al.}~\cite{abe2004microgravity} due to a significantly altered thermocapillary convection promoting a desired wetting effect when compared to the common or normal fluids (NFs). In particular, the Marangoni stresses induce the motion of fluids in the vicinity of the interfaces towards higher temperatures in SRFs (where the fluids are associated with higher surface tension), which is opposite to that observed in NFs. See the schematic in Fig.~\ref{fig1}, which illustrates the differences in the behavior of both these types of fluids.

As such, the self-rewetting fluids have the ability to generate vigorous inflow of liquids near high temperature regions, e.g., towards nucleating sites during boiling thereby preventing the onset of dry patches at such hot spots. These and other peculiar features arising from the thermocapillarity associated with SRFs involving dispersed phases have provided strong impetus for their investigations as novel classes of fluids to enhance transport in various thermal management applications during the last two decades. They have been proposed as working fluids for various technological applications in both terrestrial and microgravity environments~\cite{abe2004microgravity,abe2007terrestrial}. The use of SRFs has been shown to improve heat transfer efficiency in heat pipes~\cite{savino2009surface,savino2013some,hu2014heat,wu2017study,cecere2018experimental,zhu2020thermal}, flow boiling~\cite{sitar2015heat} and evaporation~\cite{sefiane2020heat} in microchannels, pool boiling processes~\cite{hu2015heat,hu2018marangoni,hu2019marangoni,kim2022pool}, and two-phase heat transfer devices using self-rewetting gold nanofluids~\cite{zaaroura2021thermal}. Moreover, the peculiar characteristics of the migration of bubbles in SRFs have been experimentally studied in~\cite{shanahan2014recalcitrant,mamalis2017bubble}.

As noted in a recent review involving the use of SRFs~\cite{hu2018review}, only limited analytical and numerical studies involving the SRFs, which can provide fundamental insights into the details of the transport phenomena, have been performed. Analytical investigations into the behavior of thin films of SRFs were presented in~\cite{oron1994nonlinear,batson2017thermocapillary,yu2018thermocapillary} and a similarity solution of the motion of SRFs in an unbounded domain was discussed in~\cite{slavtchev1998thermocapillary}. Theoretical analysis of the migration of a bubble in a SRF was presented in~\cite{tripathi2015non} and that of an elongated slug in~\cite{duffy2018unsteady}. More recently, numerical studies on the migration of a bubble in SRFs were performed in~\cite{balla2019non,majidi2020single,mitchell2021computational}.

Among the various computational methods, the lattice Boltzmann (LB) method, a technique inspired from kinetic theory~\cite{benzi1992lattice,yu2003viscous,lallemand2021lattice}, has shown promising capabilities for simulating multiphase flows (see e.g.,~\cite{he1999lattice,he2002thermodynamic,lee2005stable,premnath2007three,hajabdollahi2021central}). Its popularity stem from its ability to readily represent the relevant physics at the mesoscopic scales using kinetic models, its natural parallelization features, and its efficient and ease of implementation using the collide-and-stream steps. The LB methods have also been applied to simulate thermocapillary flow problems (see e.g.,~\cite{liu2012modeling,majidi2020single,mitchell2021computational}). More recently, using robust collision models~\cite{geier2006cascaded}, the LB method has been extended to simulate multiphase flows at high density ratios and including Marangoni stresses~\cite{hajabdollahi2021central}, which will serve as a basis for further extension for its application to an interesting configuration involving thermocapillary flows in SRFs as discussed below.

One of the important applications of exploiting thermocapillarity is in manipulating the motion of continuous streams of fluids confined within microchannels. In this regard, in the case of two superimposed normal fluids (NFs), Pendse and Esmaeeli~\cite{pendse2010analytical} presented a theoretical analysis for thermocapillary convection driven by periodic heating from the bounding walls, representing, for example, micropatterned walls. Under the assumption of creeping flow limit and in the absence of gravity, and taking the surface tension to decrease linearly with temperature, they developed an analytical solution for the thermocapillary-driven flow field demonstrating the existence of a pair of periodic convection cells in each fluid.

In this work, we generalize the above mentioned configuration reported in Ref.~\cite{pendse2010analytical} and develop a new analytical solution for thermocapillary convection in two superimposed layers of self-rewetting fluids (SRFs) confined within a microchannel and subjected to periodic heating on the lateral walls. Such an investigation yields a new pathway to enhance mixing and transport by tuning thermocapillary effects in SRFs when compared to NFs in differentially heated microchannels. We consider a nonlinear (parabolic) dependence of the surface tension on temperature representing the general class of SRFs, and derive analytical solutions for the thermocapillary convection currents under the assumptions of small capillary and Marangoni numbers and in the creeping flow limit, which are representative of situations in microchannels. The flow field will be represented in terms of the streamfunctions for each of the SRFs, which are parameterized by the thickness ratio of the fluids and the ratios of the thermal conductivities as well as that of viscosities, and the coefficients of the functional dependence of surface tension on temperature. As a second objective, we will also present a numerical simulation approach for such a configuration based on a robust central moment LB scheme using a phase field model based on the conservative Allen-Cahn equation. It involves computing the evolution of three distribution functions, one each for the flow field, temperature field and the capturing of the interfaces via a order parameter, and with an attendant surface tension equation of state for SRFs. It will involve an improved and simpler implementation strategy when compared to our recent work in this regard~\cite{hajabdollahi2021central}. As a third objective, we will compare the predictions based on our analytical solution against the results from our LB computational approach, thereby demonstrating qualitative as well as quantitative consistency between the two approaches, and thus establishing the validity of our analysis. Finally, we aim to present a study of the effect of the various characteristic parameters (such as ratios of the fluid thicknesses and the ratios of the fluid properties) on the vortical convection patterns in SRFs, in terms of both the number of convection cells and their sense of direction of motion and comparing and contrasting them with that of NFs, and on the magnitude of the thermal convection velocities. These contributions not only serve in elucidating the physics of a thermally induced capillary phenomenon in SRFs in a fundamental configuration, but the analytical solutions developed herein could also serve as a benchmark for any new computational techniques for simulating thermocapillary flows in SRFs in future. Moreover, the numerical algorithm based on the LB method presented in this work, while applied here in what follows for SRF layers in a microchannel, can also be readily extended for other situations including those involving tracking the motion of any dispersed phase in SRFs.

This paper is organized as follows. In the next section (Section~\ref{Sec.1}), we will discuss the problem setup of the thermocapillary flow in superimposed layers of SRFs in a microchannel and the attendant governing equations for incompressible two-fluid motion, energy transport and the interfacial equation of state. A derivation of the new analytical solution of thermocapillary motion in superimposed layers of SRFs is presented in Section~\ref{Sec.2}. The computational model equations for the LB schemes for multiphase flows using a phase field model are given in~\ref{Sec.3}. The discretized central moment LB algorithms for simulating multiphase flows of SRFs are summarized in Sec.~\ref{Sec.LBschemes}, with Secs.~\ref{Sec.4.1},~\ref{Sec.4.2} and~\ref{Sec.4.3} discussing the LB schemes for fluid motion, the interfacial dynamics, and the energy transport, respectively. Section~\ref{Sec.5} presents a numerical validation of the computational approach against an existing benchmark problem for thermocapillary flow in a NF. The results and discussion of the effect of various characteristic parameters on the thermocapillary convection in superimposed layers of SRFs in a microchannel are presented in Section~\ref{Results and Discussion}, where our new analytical solutions are also compared against the results obtained using our central moment LB schemes in this regard; moreover, the utility of the computational method in simulating such flows with interfacial deformations at higher capillary numbers is also demonstrated. The main findings and contributions of this work are summarized in Sec.~\ref{Sec.9}. Additional supporting details are given in the appendices.

\section{Problem setup, Governing equations, and Interface equation of state} \label{Sec.1}
\subsection{Problem setup} \label{Subsec.1.1}
\begin{figure}[H]
\centering
\includegraphics[trim = 0 0 0 0,clip, width = 90mm]{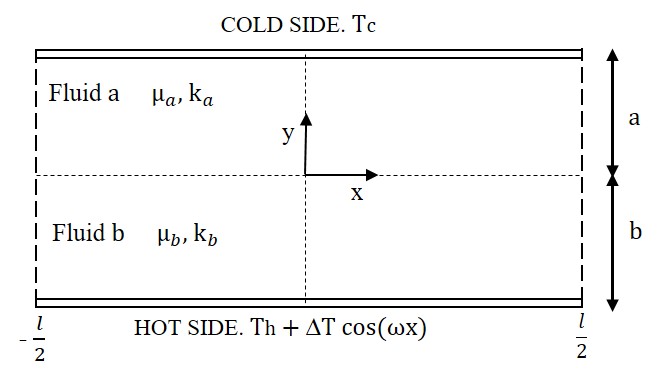}
\caption{\label{fig:Model_setup} Schematic of the geometric setup for two superimposed self-rewetting fluid (SRF) layers within a horizontal microchannel with a periodic heating at the bottom wall.}
\label{Model_setup}
\end{figure}

A schematic of the geometric configuration of two superimposed SRF layers confined within a microchannel is shown in Fig~\ref{Model_setup}. The channel is of horizontal length $l$ and whose walls are separated by a lateral distance ($a+b$). A sinusoidal temperature variation is imposed on the hot bottom wall side, while cold bottom wall side is maintained at a uniform temperature. The channel is filled with two immiscible SRFs, fluid `a' on the top side and fluid `b' on the bottom side with thicknesses $a$ and $b$, respectively; the viscosities and thermal conductivities of the top fluid are denoted by $\mu_a$ and $k_a$, respectively, while those for the bottom fluid are represented by $\mu_b$ and $k_b$, respectively. The origin of the coordinate system axes is located on their interface at the midsection in the horizontal direction as shown.
The upper wall is set with a constant reference cold temperature $T_c$, while the lower wall is prescribed with a spatially varying hot temperature based on a sinusoidal profile involving a reference temperature $T_h$ and a peak amplitude $\Delta T$ for the variation. Thus, the corresponding boundary conditions at these two sides can be written as
\mathleft
\begin{equation}
\qquad T^a(x,a)=T_c, \label{two}
\end{equation}
and
\mathleft
\begin{equation}
\qquad T^b(x,-b)=T_h + \Delta T \cos(\omega x),   \label{one}
\end{equation}
where $\omega = 2\pi / l$ is the wavenumber based on the channel length $l$ and $\Delta T>0$, and assume $T^b(x,-b)\ge T^a(x,a)$ for any $x$. Here, and in what follows, we use a superscript notation with `a' or `b' to label any quantity associated with a top or bottom fluid, respectively. The imposed spatial variations in the temperature results in a heat diffusion into the bulk regions of the fluids with sets up a nonuniform distribution of the temperature along the interface. The surface tension $\sigma = \sigma(T)$ at the interface between the SRFs also then varies locally, which, via the viscous actions in the bulk fluids, induce a thermocapillary convection. The resulting flow field is then subject to the no-slip condition for the velocity components at the bounding walls.
\subsection{Bulk fluid motion and energy transport} \label{Subsec.1.2}
The thermocapillary convection in the SRFs obey the equations of mass and momentum (i.e., Navier–Stokes equations (NSE)) along with the equation of the energy transport. Each of the two SRFs obey such conservation equations. They can be respectively written as follows:
\mathleft
\begin{subequations}
\begin{equation}
\qquad \bm{\nabla} \cdot {\bm{u}}=0,
\end{equation}
\begin{equation}
\qquad \rho \left( \frac{\partial \bm{u}}{\partial t} + \bm{\nabla} \cdot {(\bm{uu})} \right) = - \bm{\nabla} p + \bm{\nabla} \cdot \left[ \mu (\bm{\nabla} \bm{u} + \bm{\nabla} \bm{u}^{\dagger})\right],
\end{equation}
\begin{equation}    \label{energy eqn}
\qquad  \frac{\partial {T}}{\partial t} + \bm{u} \cdot \bm{\nabla}T   =  \bm{\nabla} \cdot \left(\alpha \bm{\nabla}T \right),
\end{equation}
\end{subequations}
where $\rho$, $\mu$ and $\alpha$ are the fluid density, dynamic viscosity, and thermal diffusivity of the fluid, respectively, with $\alpha=k/(\rho c_p)$ based on the thermal conductivity $k$ and specific heat $c_p$. In the above, $\bm{u}$, $p$, and $T$ denote the velocity, pressure, and temperature fields of the fluids, respectively, and the superscript symbol $\dagger$ represents taking transpose of the dyadic velocity gradient $\bm{\nabla} \bm{u}$.
\subsection{Interface equation of state for surface tension} \label{Subsec.1.3}
At the interface, we need to impose an equation for the surface tension relating it to the variations in the local temperature. For the SRF, we consider the following nonlinear (parabolic) dependence of surface tension on temperature:
\mathleft
\begin{equation}\label{ST_SRF}
\qquad \sigma (T) = \sigma_{0} + \sigma_T (T-T_{ref})+ \sigma_{TT} (T-T_{ref})^2,
\end{equation}
where $\sigma_{0}$ denotes the value of the surface tension at a reference temperature $T_{ref}$, $\sigma_{T}=\frac{d\sigma}{dT}\big\vert_{T_{ref}}$ and $\sigma_{TT}=\frac{1}{2}\frac{d^2\sigma}{dT^2}\big\vert_{T_{ref}}$ are the coefficients of the surface equation of state, expressing the sensitivity of the surface tension temperature. It should noted for a SRF, $\sigma_{TT}\neq 0$, while for a NF, $\sigma_{TT} = 0$ where only $\sigma_T$ is non-zero. In general, $\sigma_0$, $T_{ref}$, $\sigma_T$, and $\sigma_{TT}$ are properties, which are unique to a chosen SRF. In addition, at the interface, a relation between the Marangoni stress due to the nonuniform tangential surface tension gradient and the viscous fluid stress, along with the interfacial continuity conditions for the flow and thermal fields need to be imposed. These will be accounted for in what follows.

When the above governing equations are nondimensionalized using a reference velocity scale $U$ and a length scale $b$ corresponding to the thickness of the bottom SRF layer, we obtain the following dimensionless groups: Reynolds number $\mbox{Re}$, Marangoni number $\mbox{Ma}$, and the capillary number $\mbox{Ca}$, which can be defined as
\mathleft
\begin{equation}
\qquad \mbox{Re} = \frac{U b} {\nu_b},    \quad  \quad     \mbox{Ma} = \frac{U b} {\alpha_b} = \mbox{Re}\mbox{Pr},     \quad \textrm{and} \quad     \mbox{Ca}= \frac{U \mu_b} {\sigma_{0}}.                    \label{ReMaCa}
\end{equation}
respectively. Here, $\nu=\mu/\rho$ is the kinematic viscosity, and $\mbox{Pr}$ is the Prandtl number ($\mbox{Pr}=\nu/\alpha$). In addition, the thermocapillary convection in SRFs is governed by the following ratios of the bulk material properties
\mathleft
\begin{equation}
\qquad \tilde{\rho} =  \frac{\rho_a}{\rho_b},   \quad \quad   \tilde{\mu} =  \frac{\mu_a}{\mu_b},  \quad \quad  \tilde{k} = \frac{ k_a}{k_b}, \quad \textrm{and} \quad  \tilde{c}_p =  \frac{c_{p_a}}{c_{p_b} },  \label{three}
\end{equation}
and the dimensionless parameters for the interface equation of state for the surface tension
\begin{equation}
\qquad M_1 =\sigma_T\left(\frac{\Delta T}{\sigma_0}\right),\quad M_2 =\sigma_{TT}\left(\frac{{\Delta T}^2}{\sigma_0}\right). \label{three-prime}
\end{equation}
As in the above, taking reference values for the properties using those for the bottom fluid, it may be noted that by balancing the scale for the viscous shear stress $\mu_bU/b$ with that of the Marangoni stress due to the surface tension gradient $|d\sigma/dT|(\Delta T/l)$, we can estimate the scale for the reference velocity $U$ of thermocapillary convection used in the above via $U\sim |d\sigma/dT|(\Delta T/\mu_b)(b/l)$, where $\Delta T$ is set to be equal to the maximum amplitude in the spatial variation in the imposed temperature at the bottom wall $T_0$. Here, $|d\sigma/dT|$ can be obtained from Eq.~(\ref{ST_SRF}) in terms of the surface tension-temperature relation properties of the SRF.

\section{Analytical Solution for Thermocapillary Convection in Superimposed SRF Layers in a Microchannel } \label{Sec.2}
We will now derive a new analytical solution for thermocapillary convection in superimposed SRF layers in the Stokes flow regime relevant to microchannels. In this regard, we consider the fluids to be incompressible, immiscible, and Newtonian, and assume that the Reynolds and Marangoni numbers are much less than one (i.e., $\mbox{Re}\ll 1$ and $\mbox{Ma}\ll 1$), the convective transport of momentum and energy can be neglected. Also, the capillary number is also taken to be much less than one (i.e., $\mbox{Ca}\ll 1$), so that we can consider the interface to be nearly flat, and the established thermocapillary convection patterns are steady. All these typical assumptions are consistent with an earlier theoretical analysis involving the NFs reported previously~\cite{pendse2010analytical}. Based on these considerations, all the conservation equations given above simplify considerably. The mass conservation read as
\mathleft
\begin{equation} \label{mass}
\qquad  \bm{\nabla}\cdot {\bm{u}}= 0,
\end{equation}
while the momentum equation now reduces to
\mathleft
\begin{equation} \label{momentum}
\qquad -\bm{\nabla} {p} + \mu \bm{\nabla}^{2} {\bm{u}}= 0,
\end{equation}
and the balance of thermal energy equation is given as
\mathleft
\begin{equation} \label{energy}
\qquad \bm{\nabla}^2{T}= 0,
\mathleft
\end{equation}
where $\bm{\nabla}^{2}= \frac{\partial^2}{\partial x^2} + \frac{\partial^2}{\partial y^2}$. These bulk transport equations need to be solved in conjunction with the interface continuity conditions for the flow and temperature fields and the Marangoni stress condition at the interface between the SRFs (see below for details).

\subsection{Temperature field}
The thermal energy equation Eq.~(\ref{energy}), which satisfies the wall boundary conditions given in the previous section, can be solved readily and is independent of the nature of the fluid; the specific details involved in the solution procedure are given in Appendix~\ref{Appendix A} of this paper. The solution for the temperature field is summarized here as follows:
In the upper fluid $a$,
\mathleft
\begin{equation}
\qquad T^{a}(x,y) = \frac{(T_c-T_h)y+T_c \tilde{k} b+T_h a}{(a+b\tilde{k})} + \Delta T f(\tilde a,\tilde b,\tilde k)\sinh(\tilde a-\omega y)\cos(\omega x),       \label{eq:Ta}
\end{equation}
and in the lower fluid $b$,
\mathleft
\begin{equation}
\begin{split}
\qquad &  T^{b}(x,y) = \frac{\tilde{k}(T_c-T_h)y+T_c \tilde{k} b+T_h a}{(a+b\tilde{k})}\\
\qquad  & \qquad\qquad + \Delta T f(\tilde a,\tilde b,\tilde k)   \left[\sinh(\tilde a) \cosh(\omega y) -   \tilde{k} \sinh(\omega y)\cosh(\tilde a)\right]\cos(\omega x),       \label{eq:Tb}\\
\end{split}
\end{equation}
where $\tilde{k}=k_a/k_b$, $\tilde{a}=a\omega$, and $\tilde{b}=b\omega$ are the dimensionless parameters, and the expression for the function $f(\tilde a,\tilde b,\tilde k)$ is given in Eq.~(\ref{f}) in Appendix~\ref{Appendix A}.

\subsection{Flow field: Stream function}
Next, for obtaining the flow field driven by thermocapillary effects in SRFs, for convenience, we introduce the stream function $\psi$ defined based on the components of the velocity field $\bm{u}=(u,v)$ as
\begin{equation}\label{eq:streamfunctiondefinition}
\qquad {u} = -\frac{\partial \psi}{\partial y}, \quad \text{and} \quad {v} = -\frac{\partial \psi}{\partial x},
\end{equation}
so that the continuity equation Eq.~(\ref{mass}) is satisfied automatically, and the momentum equation (Eq.~(\ref{momentum})) can be entirely rewritten in terms of a single scalar variable $\psi$. For the latter purpose, taking the of `curl' Eq.~(\ref{momentum}) and using $\bm{\nabla}\times\bm{\nabla}p=0$ and using the invoking the above definition of the velocity field in terms of $\psi$ (Eq.~(\ref{eq:streamfunctiondefinition})) we finally obtain the following biharmonic equation for the stream function~\cite{langlois1964slow}:
\mathleft
\begin{equation} \label{psi}
\qquad \bm{\nabla}^4 \psi = \bm{\nabla}^2(\bm{\nabla}^2 \psi)= 0.
\end{equation}
Since Eq.~(\ref{psi}) is linear, we can apply the method of separation of variables by assuming the solution of $\psi$ to be product of two solutions $X(x)$ and $Y(x)$ in the two respective coordinate directions as
\mathleft
\begin{equation}
\qquad \psi(x,y) = X(x)Y(y). \nonumber
\end{equation}
Since the thermocapillary flow is established by the tangential stress at the interface, we can establish the form of the solution $X(x)$ by considering the Marangoni interfacial condition reflecting a balance between the viscous shear stress and the surface tension gradient given by
\mathleft
\begin{equation}\label{eq:Marangonistresscondition}
\qquad \left( \tau_{xy}^b - \tau_{xy}^a \right) \Biggr\rvert_{y=0} = \frac{d \sigma}{d T} \frac{\partial T}{\partial x} \Biggr\rvert_{y=0},
\end{equation}
where $\tau_{xy}=\mu\left(\frac{\partial u}{\partial y}+\frac{\partial v}{\partial x}\right)$ is the viscous shear stress and $d \sigma/dT$ for SRFs follows from Eq.~(\ref{ST_SRF}) as
\mathleft
\begin{equation}
\qquad \frac{d \sigma}{d T} = \sigma_T + 2\sigma_{TT}\left( T-T_{ref} \right) = \left( \sigma_T - 2\sigma_{TT}T_{ref}\right) + 2\sigma_{TT}T. \nonumber
\end{equation}
Now, from Eq.~(\ref{eq:Ta}), it follows that $\left.\frac{\partial T}{\partial x} \right|_{y=0} \sim \sin(\omega x)$ and from the last equation together with using Eq.~(\ref{eq:Ta}) for $T(x,y=0)$, we have $d\sigma/dT \sim \cos(\omega x)$. Using these two estimates for the horizontal spatial variations in Eq.~(\ref{eq:Marangonistresscondition}), it can be readily inferred that $\tau_{xy} \sim \alpha_1\sin(\omega x) + \alpha_2\sin(\omega x) \cos(\omega x)$, where $\alpha_1$ and $\alpha_2$ are some lumped constants; this suggests that the stream function to be split into a linear combination of two distinct product solutions with known spatial distribution in the $x$ direction as in
\mathleft
\begin{equation}\label{eq:formofstreamfunction}
\qquad \psi(x,y)=f(y)\sin(\omega x)+g(y) \sin(\omega x) \cos(\omega x),
\end{equation}
where $f(y)$ and $g(y)$ are the two unknown functions varying along the $y$ direction, which will to be determined in what follows.
Here, it should be noted that the first term in the last equation (Eq.~(\ref{eq:formofstreamfunction})), $f(y) \sin(\omega  x)$ arises from the linear part of the surface tension equation of state (which recovers the special case of the NFs given in~\cite{pendse2010analytical}), while the second term $g(y)\cos(\omega  x)\sin(\omega  x)$ emerges from including the quadratic term for $\sigma(T)$ to encompass the more general SRFs.

Substituting Eq.~(\ref{eq:formofstreamfunction}) in Eq.~(\ref{psi}) and simplifying results in the following two 4th order differential equations for the unknown functions $f(y)$ and $g(y)$:
\mathleft
\begin{subequations}
\begin{equation}
\qquad f''''-2\omega^2f''+\omega^4f=0,   \label{ff}
\end{equation}
\begin{equation}
\qquad g''''-8\omega^2g''+16\omega^4g=0.  \label{g}
\end{equation}
\end{subequations}
Equation~(\ref{ff}) has solutions of the form $f(y)=e^{my}$, where $m$ is a constant to be determined from the characteristic equation $(m^2-\omega^2)^2=0$, giving $m=\pm \omega$. The four solutions of $f(y)$ are $e^{\omega y}$, $ye^{\omega y}$, $e^{-\omega y}$, and $ye^{-\omega y}$  because it has double roots. Similarly, for Eq.~(\ref{g}), the solutions are of the form $g(y)=e^{ny}$, with the characteristic equation $(n^2-4\omega^2)^2=0$, yielding the four possible solutions of $g(y)$ as $e^{2\omega y}$, $ye^{2\omega y}$, $e^{-2\omega y}$, and $ye^{-2\omega y}$. Because the vertical direction is finite, it is convenient to employ hyperbolic functions in lieu of the exponential functions. As a result, the general form of the stream function $\psi(x,y)$ for the upper fluid can be written as
\mathleft
\begin{equation}
\begin{split}
\qquad &  \psi^a=U_{t}[(C_1^a+C_2^ay)\cosh(\omega y)+(C_3^a+C_4^ay)\sinh(\omega y)]\sin(\omega x)+ \\
& \qquad  \frac{1}{2} U_{tt}[(D_1^a+D_2^ay)\cosh(2\omega y)+(D_3^a+D_4^ay)\sinh(2\omega y)]\sin(2\omega x), \label{psia}
\end{split}
\end{equation}
and for the lower fluid, it reads as
\begin{equation}
\begin{split}
\qquad&  \psi^b=U_{t}[(C_1^b+C_2^by)\cosh(\omega y)+(C_3^b+C_4^by)\sinh(\omega y)]\sin(\omega x)+ \\
&  \qquad  \frac{1}{2} U_{tt}[(D_1^b+D_2^by)\cosh(2\omega y)+(D_3^b+D_4^by)\sinh(2\omega y)]sin(2\omega x). \label{psib}
\end{split}
\end{equation}
Here, $C_j^\gamma$ and $U_{t}$ (for the first term in each of the last two equations), and $D_j^\gamma$ and $U_{tt}$ (for the corresponding second term), where $\gamma = a, b$ and $j=1,2,3,4$, are the constants which will be determined through the specification of the boundary conditions next.

The constants $C_j^\gamma$ and $D_j^\gamma$, where $\gamma = a,b$ and $j=1,2,3,4$ can be evaluated by using the following boundary conditions: \newline
\noindent
i) No-slip, no-through boundary condition at the lower wall:
\mathleft
\begin{equation}
\qquad u^b(x,-b) =  v^b (x,-b) = 0.\nonumber
\end{equation}
ii) No-slip, no-through boundary condition at the upper wall:
\mathleft
\begin{equation}
\qquad u^a(x,a) = v^a (x,a) = 0.\nonumber
\end{equation}
iii) Continuity of the tangential component of the velocity at the interface:
\mathleft
\begin{equation}
\qquad u^a(x,0) = u^b (x,0) = U_{t} \sin(\omega x) + \frac{1}{2}U_{tt} \sin(2\omega x). \nonumber
\end{equation}
iv) No through flow boundary condition at the interface:
\mathleft
\begin{equation}
\qquad v^a(x,0) = v^b (x,0) = 0.\nonumber
\end{equation}
As a result, we obtain the following expressions:
\begin{equation}
\begin{split}
\qquad &C_1^b = C_1^a = 0 ,\\
\qquad &C_2^b = \frac{\sinh^2(\tilde b)}{\sinh^2(\tilde b)-\tilde b^2} ,       \quad   \quad  \quad C_2^a = \frac{\sinh^2(\tilde a)}{\sinh^2(\tilde a)-\tilde a^2},\\
\qquad &C_3^b = \frac{-b\tilde b}{\sinh^2(\tilde b)-\tilde b^2} ,         \quad    \quad \quad         C_3^a = \frac{-a\tilde a}{\sinh^2(\tilde a)-\tilde a^2},\\
\qquad &C_4^b = \frac{\sinh(2\tilde b)-2\tilde b}{2(\sinh^2(\tilde b)-\tilde b^2)} , \quad \quad C_4^a = -\frac{\sinh(2\tilde a)-2\tilde a}{2(\sinh^2(\tilde a)-\tilde a^2)}. \nonumber \\
\end{split}
\end{equation}
and
\mathleft
\begin{equation}
\begin{split}
\qquad &D_1^b = D_1^a = 0 ,\\
\qquad &D_2^b = \frac{\sinh^2(2\tilde b)}{\sinh^2(2\tilde b)-4\tilde b^2} ,       \quad   \quad  \quad D_2^a = \frac{\sinh^2(2\tilde a)}{\sinh^2(2\tilde a)-4\tilde a^2},\\
\qquad &D_3^b = \frac{-2b\tilde b}{\sinh^2(2\tilde b)-4\tilde b^2} ,         \quad    \quad \quad        D_3^a = \frac{-2a\tilde a}{\sinh^2(2\tilde a)-4\tilde a^2},\\
\qquad &D_4^b = \frac{\sinh(4\tilde b)-4\tilde b}{2(\sinh^2(2\tilde b)-4\tilde b^2)} , \quad \quad D_4^a = -\frac{\sinh(4\tilde a)-4\tilde a}{2(\sinh^2(2\tilde a)-4\tilde a^2)}.    \nonumber \\
\end{split}
\end{equation}
where $\tilde a = a \omega$ and $\tilde b = b \omega$.

Lastly, by applying the following fifth boundary condition corresponding to the Marangoni stress balance condition at the interface, which is applied to both parts of the solution for $\psi(x,y)$ given above simultaneously, the proportionality constants $U_{t}$ and $U_{tt}$ can be obtained in terms of the other constants and dimensionless parameters given above:\newline\noindent
v) Balance of net viscous shear stress and Marangoni stress:
\mathleft
\begin{equation}
\qquad \mu_b \frac{\partial u^b}{\partial y}\Biggr\rvert_{y=0} -\mu_a \frac{\partial u^a}{\partial y}\Biggr\rvert_{y=0} = \Big\{\sigma_T + 2\sigma_{TT}[T(x,y=0)-T_{ref}] \Big\} \frac{\partial T}{\partial x}\Biggr\rvert_{y=0}. \nonumber
\end{equation}
Then, the expression for $U_{t}$  reads as
\mathleft
\begin{equation}\label{eq:Umaxexpression}
\qquad U_{t}=-\left(\frac{\Delta T}{\mu_b} \right)g(\tilde a,\tilde b,\tilde k)h(\tilde a,\tilde b,\tilde \mu)\left[\sigma_{T}+2\sigma_{TT}\left(\frac{T_c\tilde kb+T_ha}{(a+b \tilde k)}-T_{ref}\right) \right],
\end{equation}
where
\mathleft
\begin{equation}
\qquad g(\tilde a,\tilde b,\tilde k)=\sinh(\tilde a)f(\tilde a,\tilde b,\tilde k).\nonumber
\end{equation}
Here, the function $f(\tilde a,\tilde b,\tilde k)$ is given in Eq.~(\ref{f}), and $h(\tilde a,\tilde b,\tilde \mu)$ in Eq.~(\ref{eq:Umaxexpression}) reads as
\mathleft
\begin{equation}
\qquad h(\tilde a,\tilde b,\tilde \mu)=\frac{\left(\sinh^2(\tilde a)-\tilde a^2 \right)\left(\sinh^2(\tilde b)-\tilde b^2 \right)}{\tilde \mu\left(\sinh^2(\tilde b)-\tilde b^2 \right)\left(\sinh(2\tilde a)-2\tilde a \right)+ \left(\sinh^2(\tilde a)-\tilde a^2 \right)\left(\sinh(2\tilde b)-2\tilde b \right)}.\nonumber
\end{equation}
Moreover, the functional relationship for $U_{tt}$ is given by
\mathleft
\begin{equation}
\qquad U_{tt}=-\left(\frac{\sigma_{TT}\Delta T^2}{\mu_b}\right)g^2(\tilde a,\tilde b,\tilde k)\tilde h_1(\tilde a,\tilde b,\tilde \mu),
\end{equation}
where
\mathleft
\begin{equation}
\qquad \tilde h_1(\tilde a,\tilde b,\tilde \mu)=\frac{\left(\sinh^2(2\tilde a)-4\tilde a^2 \right)\left(\sinh^2(2\tilde b)-4\tilde b^2 \right)}{\tilde \mu\left(\sinh^2(2\tilde b)-4\tilde b^2 \right)\left(\sinh(4\tilde a)-4\tilde a \right)+ \left(\sinh^2(2\tilde a)-4\tilde a^2 \right)\left(\sinh(4\tilde b)-4\tilde b \right)}. \nonumber
\end{equation}
That is, $\tilde h_1(\tilde a,\tilde b,\tilde \mu)=\tilde h(2\tilde a,2\tilde b,\tilde \mu)$. Finally, substituting for the constants in Eqs.~(\ref{psia}) and (\ref{psib}), we can arrive at the following analytical solution for the stream function in the upper and lower fluids:
\begin{equation}
\begin{split}
\psi^a=&\frac{U_{t}/\omega}{\sinh^2(\tilde a)-\tilde a^2} \times \\
& \qquad \qquad \bigg\{ \sinh^2(\tilde a)(\omega y)\cosh(\omega y) -\frac{1}{2}\left[2\tilde a^2+\left(\sinh(2\tilde a)-2\tilde a\right)(\omega y)\right]\sinh(\omega y)\bigg\}\sinh(\omega y)\nonumber\\
&+\frac{1}{2}\frac{U_{tt}/\omega}{\sinh^2(2\tilde a)-4\tilde a^2} \times \\
& \qquad \qquad \bigg\{ \sinh^2(2 \tilde a)(\omega y)\cosh(2\omega y) -\frac{1}{2}\left[4\tilde a^2+\left(\sinh(4\tilde a)-4\tilde a\right)(\omega y)\right]\sinh(2\omega y)\bigg\}\sinh(2\omega y), \nonumber
\end{split}
\end{equation}
and
\begin{equation}
\begin{split}
\psi^b=&\frac{U_{t}/\omega}{\sinh^2(\tilde b)-\tilde b^2} \times \\
& \qquad \qquad \bigg\{ \sinh^2(\tilde b)(\omega y)\cosh(\omega y) -\frac{1}{2}\left[2\tilde b^2-\left(\sinh(2\tilde b)-2\tilde b\right)(\omega y)\right]\sinh(\omega y)\bigg\}\sinh(\omega y). \nonumber\\
&+\frac{1}{2}\frac{U_{tt}/\omega}{\sinh^2(2\tilde b)-4\tilde b^2} \times \\
& \qquad \qquad \bigg\{ \sinh^2(2 \tilde b)(\omega y)\cosh(2\omega y) -\frac{1}{2}\left[4\tilde b^2-\left(\sinh(4\tilde b)-4\tilde b\right)(\omega y)\right]\sinh(2\omega y)\bigg\}\sinh(2\omega y). \nonumber
\end{split}
\end{equation}
In addition, the analytical solutions for thermocapillary-driven velocity field components in SRFs $u^\gamma(x,y)$ and $v^\gamma(x,y)$ (for $\gamma=a,b$) can be recovered from the stream function via Eq.~(\ref{eq:streamfunctiondefinition}), i.e., using $u^\gamma = -\partial \psi^\gamma/\partial y$ and $v^\gamma = -\partial \psi^\gamma/\partial x$, which yields the following for the upper fluid
\begin{eqnarray}
\qquad {u}^{a}(x,y) \! \! \! &=& \! \! \! U_{t}  \{ \left[C_2^a + \omega(C_3^a+C_4^a y)\right] \cosh(\omega y)+ (C_4^a+\omega C_2^a y) \sinh(\omega y)  \} \sin(\omega x)\nonumber \\[2mm]
& &+ \frac{1}{2} U_{tt}  \{ \left[D_2^a + 2\omega( D_3^a+ D_4^a y)\right] \cosh(2\omega y)+ (D_4^a+2\omega D_2^a y) \sinh(2\omega y)  \} \sin(2\omega x),\nonumber\label{ua_SRF}
\end{eqnarray}
\mathleft
\begin{equation}
\begin{split}
\qquad & {v}^{a}(x,y)=- \omega U_{t}  \left[C_2^a y \cosh(\omega y) + (C_3^a+C_4^a y) \sinh(\omega y)  \right] \cos(\omega x)- \\
& \qquad\qquad\qquad  \omega U_{tt}  \left[D_2^a y \cosh(2\omega y) + (D_3^a+D_4^a y) \sinh(2\omega y)  \right] \cos(2\omega x),      \label{va_SRF}
\end{split}
\end{equation}
and for the lower fluid as
\begin{eqnarray}
\qquad {u}^{b}(x,y) \! \! \! &=& \! \! \! U_{t}  \{[C_2^b + \omega(C_3^b+C_4^b y)] \cosh(\omega y)+ (C_4^b+\omega C_2^b y) \sinh(\omega y)  \} \sin(\omega x) \nonumber \\[2mm]
& &+ \frac{1}{2} U_{tt}  \{ [D_2^b + 2\omega( D_3^b+ D_4^b y)] \cosh(2\omega y)+ (D_4^b+2\omega D_2^b y) \sinh(2\omega y)  \} \sin(2\omega x), \nonumber\label{ub_SRF}
\end{eqnarray}
\mathleft
\begin{equation}
\begin{split}
\qquad & {v}^{b}(x,y)=- \omega U_{t} [C_2^b y \cosh(\omega y) + (C_3^b+C_4^b y) \sinh(\omega y)] \cos(\omega x)- \\
& \qquad\qquad\qquad  \omega U_{tt}  [D_2^b y \cosh(2\omega y) + (D_3^b+D_4^b y) \sinh(2\omega y)] \cos(2\omega x).       \label{vb_SRF}
\end{split}
\end{equation}
From Eqs.~(\ref{va_SRF}) and (\ref{vb_SRF}), it can be inferred that the parameters $U_{t}$ and $U_{tt}$ represent measures of the scales for the thermocapillary velocity contributions arising from the linear and quadratic part of the surface tension variation with the temperature $\sigma(T)$ for the SRFs. When the coefficient $\sigma_{TT}$ for the quadratic contribution in $\sigma(T)$ becomes zero (see Eq.~(\ref{ST_SRF})), the above results reduce to that presented in~\cite{pendse2010analytical} applicable for the NFs.

\section{Computational Modeling for LBM: Interface capturing and motion of binary fluids driven by thermocapillary effects} \label{Sec.3}
We will now discuss a modeling formulation suitable for the development of a numerical approach based on the LBM for simulation of thermocapillary convection in SRFs presented in the next section. The phase-field lattice Boltzmann approach based on the conservative Allen-Cahn equation (ACE)~\cite{chiu2011conservative} is considered in this study to capture interfacial dynamics while maintaining the segregation of two immiscible fluids, which is an improvement over an earlier model~\cite{sun2007sharp} based on a counter term approach \cite{folch1999phase}. The binary fluids are distinguished by an order parameter or the phase field variable $\phi$. The fluid $A$ is identified by $\phi = \phi_{A}$, while fluid $B$ by $\phi = \phi_{B}$. The interface-tracking equation based on the conservative ACE in terms of the phase field variable is given as
\mathleft
\begin{equation}\label{eqn1}
\qquad \frac{\partial \phi}{\partial t} + \bm{\nabla} \cdot (\phi \bm{u}) = \bm{\nabla} \cdot [M_\phi(\bm{\nabla} \phi - \theta \bm{n})],
\end{equation}
where $\bm{u}$ is the fluid velocity, $M_{\phi}$ is the mobility, and $\bm{n}$ is the unit normal vector, which can be calculated using the order parameter $\phi$ as  $\bm{n} = \bm{\nabla}{\phi}/|\bm{\nabla}{\phi}|$. Here, the parameter $\theta$ can be expressed as
$\theta = -4\left(\phi - \phi_{A} \right) \left(\phi - \phi_{B} \right)/[W \left(\phi_{A} - \phi_{B} \right)]$, where $W$ is the width of the interface. Essentially, the term $M_\phi\theta \bm{n}$ in Eq.~(\ref{eqn1}) serves as the interface sharpening term counteracting the diffusive flux $-M_\phi\bm{\nabla} \phi$ following the advection of $\phi$ by the fluid velocity. At equilibrium, the conservative ACE reduces the order parameter to a hyperbolic tangent profile across the diffuse interface, which is given by $\phi\left( \zeta \right)= \frac {1}{2}\left(\phi_{A} + \phi_{B} \right)+ \frac {1}{2}\left(\phi_{A} - \phi_{B} \right)\tanh\left(2\zeta/W\right)$, where $\zeta$ is a spatial coordinate along the normal with the origin at the interface.

Now, for ease of implementations, the interfacial surface tension effects can be incorporated within a diffuse interface via a distributed or smoothed volumetric force term in a single-field formulation representing the motion of binary fluids. Then, the corresponding single-field incompressible Navier-Stokes equations for binary fluids can be written as
\mathleft
\begin{equation}\label{eqn4}
\qquad \bm{\nabla} \cdot {\bm{u}}=0,
\end{equation}
\mathleft
\begin{equation}\label{eqn5}
\qquad \rho \left( \frac{\partial \bm{u}}{\partial t} + \bm{\nabla} \cdot {(\bm{uu})} \right) = - \bm{\nabla} p + \bm{\nabla} \cdot \left[ \mu (\bm{\nabla} \bm{u} + \bm{\nabla} \bm{u}^{\dagger})\right] + \bm{F}_s + \bm{F}_{ext},
\end{equation}
where $\bm{F}_{s}$ is the surface tension force, and $\bm{F}_{ext}$ is any external body force. Here, surface tension force effectively exerts itself in both the normal and tangential directions to the interface as surface tension varies with temperature. To accommodate this, a geometric technique known as the continuous surface force approach~\cite{brackbill1992continuum} can be used, which can be expressed by the following equation involving the Dirac delta function $\delta_s$:
\mathleft
\begin{equation}\label{eqn6}
\qquad \bm{F}_{s} = \left( \sigma \kappa \bm{n} + \bm{\nabla}_{s} \sigma \right)\delta_{s},
\end{equation}
where $\bm{n}$ and $\bm{\kappa}$ are the unit vector normal and the interface curvature, respectively; they can be obtained from the order parameter via $\bm{n} = \bm{\nabla}{\phi}/|\bm{\nabla}{\phi}|$ and ${\kappa} = \bm{\nabla} \cdot \bm{n}$. In the right side of Eq.~(\ref{eqn6}), the first term is the normal or capillary force acting on the interface, and the second term involving the surface gradient operator $\bm{\nabla}_{s}$ is the tangential or Marangoni force induced by surface tension gradients. Because the surface tension only acts on the interface, the delta function $\delta_{s}$ is required to satisfy $\int_{-\infty}^{+\infty} \delta_{s} dy = 1$. One formulation of $\delta_{s}$, which localizes the smoothed surface tension force suitable within the phase-field modeling framework is given by $\delta_{s}  = 1.5 W |\bm{\nabla} \phi|^2$.

The surface gradient $\bm{\nabla}_{s}$ in Eq.~(\ref{eqn6}) is given by $\bm{\nabla}_{s} = \bm{\nabla} - \bm{n}(\bm{n} \cdot \bm{\nabla})$. Therefore, the Cartesian components of the surface tension force in Eq.~(\ref{eqn6}) can then be expressed as
\mathleft
\begin{eqnarray}
\qquad F_{sx} &=& -\sigma(T) |\bm{\nabla} \phi|^2   (\bm{\nabla} \cdot \bm{n}) {n}_x  +|\bm{\nabla} \phi|^2 \left[ ( 1-{n}_x^2 ) \partial_x \sigma(T)  -  {n}_x {n}_y  \partial_y \sigma(T)  \right],\nonumber\\
\qquad F_{sy} &=& -\sigma(T) |\bm{\nabla} \phi|^2   (\bm{\nabla} \cdot \bm{n}) {n}_y  +|\bm{\nabla} \phi|^2 \left[ ( 1-{n}_y^2 ) \partial_y \sigma(T)  -  {n}_x {n}_y  \partial_x \sigma(T)  \right]. \label{eq:surfacetensionforcecomponents}
\end{eqnarray}
Here, the functional dependence of the surface tension on temperature for the SRF is obtained from the nonlinear (parabolic) equation given in Eq.~(\ref{ST_SRF}). In numerical implementations, in this work, the required spatial gradients $\partial_x \sigma(T) $ and $\partial_y \sigma(T)$ in Eq.~(\ref{eq:surfacetensionforcecomponents}) are calculated using an isotropic finite differencing scheme \cite{kumar2004isotropic}. Here, we note that temperature field $T$ is computed by solving the energy transport equation given earlier in Eq.~(\ref{energy eqn}). Finally, the jumps in fluid properties across the interface, such as density and viscosity can be expressed as a continuous function of the phase field variable, which can then be employed Eq.~(\ref{eqn5}). We use the following linear interpolation to account for the interfacial variations of fluid properties in this study (see e.g., \cite{ding2007diffuse}):
\mathleft
\begin{equation}\label{eqn11}
\qquad \rho  = \rho_{B} + \frac {\phi - \phi_{A}}{\phi_{A} - \phi_{B}} \left(\rho_{A} - \rho_{B} \right), \quad
\mu  = \mu_{B} + \frac {\phi - \phi_{A}}{\phi_{A} - \phi_{B}} \left(\mu_{A} - \mu_{B} \right),
\end{equation}
where $\rho_{A}$, $\rho_{B}$ and $\mu_{A}$, $\mu_{B}$ are the densities and the dynamic viscosities in the fluid phases, respectively and denoted by $\phi_{A}$ and $\phi_{A}$. An equation similar to Eq.~(\ref{eqn11}) will also be utilized for distributing the interfacial jump in the thermal conductivity in solving the energy equation. In this study, we use $\phi_{A}=0$ and $\phi_{B}=1$.

\section{Central Moment Lattice Boltzmann Schemes for Interface Tracking, Two-Fluid Motion and Energy Transport\label{Sec.LBschemes}}
In this section, we will present a numerical LB approach based on more robust collision models involving central moments~\cite{geier2006cascaded,premnath2009incorporating,premnath2007three,hajabdollahi2021central} for solving the equations of the phase-field model for tracking the interface (Eq.~(\ref{eqn1})) and the binary fluid motions (Eqs.~(\ref{eqn4})-(\ref{eq:surfacetensionforcecomponents})) given in the previous section, along with transport of energy presented in Eq.~(\ref{energy eqn}) earlier. Solving these three equations requires evolving three separate distribution functions on the standard two-dimensional, square lattice (D2Q9) lattice, which involve performing a \emph{collision step} based on the relaxation of different central moments of the distribution function to their equilibria, which is followed by a lock-step advection of the distribution functions to their adjacent nodes along the characteristic directions in the \emph{streaming step}. Then, the macroscopic variables, viz., the order parameter, the fluid pressure and velocity, as well as the temperature field, are obtained via taking the moments of the respective distribution functions. It should be noted that since the collision step is performed using central moments while the streaming step is performed by means of the distribution functions, this requires the use of appropriate mappings that transform between these quantities pre- and post-collision step. The central moment LB methods are shown to be more robust (e.g., enhanced numerical stability) when compared to the other collision models in the LB framework (see~\cite{hajabdollahi2021central,yahia2021central,yahia2021three} for recent examples). While the recent central moment LB scheme for two-fluid interfacial flows~\cite{hajabdollahi2021central} was constructed using an orthogonal moment basis, in what follows, we will present an improved formulation involving the non-orthogonal moment basis.

\subsection{LB scheme for phase-field based interface capturing} \label{Sec.4.1}
We will now discuss a central moment LB technique to solve the conservative ACE given in Eq.~(\ref{eqn1}) by evolving a distribution function $f_\alpha$, where $\alpha=0,1,2,\ldots,8$ represent the discrete particle directions, on the D2Q9 lattice. Generally, during collision, the set of distribution functions $\mathbf{f}=(f_0,f_1,f_2,\ldots,f_8)^\dagger$ relax to the corresponding equilibrium distribution functions given by $\mathbf{f}^{eq}=(f_0^{eq},f_1^{eq},f_2^{eq},\ldots,f_8^{eq})^\dagger$, which needs to be implement via their central moments in what follows.

In this regard, first, the components of the particle velocities of this lattice can be represented by the following vectors in standard Dirac's bra-ket notation as
\mathleft
\begin{subequations}
\begin{equation}     
\qquad \left| \bm{e}_x \right> = ( 0, 1, 0, -1, 0, 1, -1, -1, 0)^\dagger, \nonumber 
\end{equation}
\begin{equation}    
\qquad \left| \bm{e}_y \right> = ( 0, 0, 1, 0,-1, 1, 1, -1, -1)^\dagger.  \nonumber
\end{equation}
\end{subequations}
We also need the following 9-dimensional vector to define the zeroth moment of $f_\alpha$:
\mathleft
\begin{eqnarray} 
\qquad \left|\mathbf{1}\right> = (1,1,1,1,1,1,1,1,1)^{\dag}. \nonumber
\end{eqnarray}
That is, its inner product with the set of distribution functions $\left<\mathbf{f}|\mathbf{1}\right>$ should yield the order parameter $\phi$ of the phase-field model. The central moment LB will then be constructed based on the following set of nine non-orthogonal basis vectors (which differs from the approach presented in~\cite{hajabdollahi2021central}):
\mathleft
\begin{gather}
\qquad \left| P_0 \right> = \left| \mathbf{1} \right>, \quad
\left| P_1 \right> = \left| \bm{e}_x \right>, \quad
\left| P_2 \right> = \left| \bm{e}_y \right>, \nonumber \\[2mm]
\qquad \left| P_3 \right> = \left| \bm{e}_x^2+\bm{e}_y^2 \right>, \quad
\left| P_4 \right> = \left| \bm{e}_x^2-\bm{e}_y^2 \right>, \quad
\left| P_5 \right> = \left| \bm{e}_x \bm{e}_y \right>,\nonumber \\[2mm]
\qquad \left| P_6 \right> = \left| \bm{e}_x^2 \bm{e}_y \right>,\quad
\left| P_7 \right> = \left| \bm{e}_x \bm{e}_y^2 \right>,\quad
\left| P_8 \right> = \left| \bm{e}_x^2 \bm{e}_y^2 \right>.  \nonumber
\end{gather}
Symbols like $ \left| e_x^2 e_y \right> =  \left| e_x e_x e_y \right>$ signify a vector that results from the element-wise vector multiplication of vectors $\left| e_x \right>$,$\left| e_x \right>$ and $\left| e_y \right>$. They can be grouped together in the form of the following matrix that maps the distribution functions to the \emph{raw} moments in terms of the above moment basis vectors:
\mathleft
\begin{equation}  \label{eqn34}
\qquad \mathbf{P} = \left[
\left<P_0\right|,
\left<P_1\right|,
\left<P_2\right|,
\left<P_3\right|,
\left<P_4\right|,
\left<P_5\right|,
\left<P_6\right|,
\left<P_7\right|,
\left<P_8\right|
\; \right].
\end{equation}
Here, it should be noted that the \emph{central} moments are obtained from the distribution moments by shifting the particle velocity $\bm{e}_\alpha$ by the fluid velocity $\bm{u}$. Given these, we can then formally define the raw moments of the distribution function $f_\alpha$ as well as its equilibrium $f_\alpha^{eq}$ as
\mathleft
\begin{subequations}
\begin{equation}
\qquad \left( \begin{array}{c}\kappa'_{mn}\\[2mm]   \kappa'^{\;eq}_{mn} \end{array} \right)  = \sum_{\alpha = 0}^{8} \left( \begin{array}{c}f_{\alpha} \\[2mm]   f_{\alpha}^{eq} \end{array} \right)  e_{\alpha x}^m   e_{\alpha y}^n,
\end{equation}
and the corresponding central moments as
\mathleft
\begin{equation}
\qquad \left( \begin{array}{c}\kappa_{mn} \\[2mm]   \kappa_{mn}^{eq} \end{array} \right)  = \sum_{\alpha = 0}^{8} \left( \begin{array}{c}f_{\alpha} \\[2mm]   f_{\alpha}^{eq} \end{array} \right) (e_{\alpha x}-u_x)^m  ( e_{\alpha y}-u_y)^n.
\end{equation}
\end{subequations}
Thus, $\kappa'_{mn}$ represents the raw moment of order $(m+n)$, while the corresponding central moment is $\kappa_{mn}$. For convenience, we can group all the possible raw moments and the central moments for the D2Q9 lattice via the following two vectors as
\mathleft
\begin{subequations}
\begin{eqnarray}
\qquad \bm{\kappa^{'}} \! \! \! &=& \! \! \! ( \K_{00}^{'}, \K_{10}^{'},\K_{01}^{'}, \K_{20}^{'}, \K_{02}^{'}, \K_{11}^{'},\K_{21}^{'}, \K_{12}^{'},\K_{22}^{'} ),\label{eqn:4a} \\[3mm]
\qquad \bm{\kappa} \! \! \! &=& \! \! \! ( \K_{00},\K_{10}, \K_{01}, \K_{20}, \K_{02}, \K_{11}, \K_{21}, \K_{12}, \K_{22} ).
\end{eqnarray}
\end{subequations}
It should be noted that one can readily map from the distribution functions to the raw moments via $\bm{\kappa^{'}} = \PP\mathbf{f}$, which can then be transformed into the central moments through $\bm{\kappa} = \F \bm{\kappa^{'}}$, where the $\tensr{F}$ follows readily from binomial expansions of $(e_{\alpha x}-u_x)^m  ( e_{\alpha y}-u_y)^n$ to relate to $e_{\alpha x}^m   e_{\alpha y}^n$ etc. Similarly, the inverse mappings from central moments to raw moments, from which the distribution functions can be recovered via the matrices $\tensr{F}^{-1}$ and $\tensr{P}^{-1}$, respectively. All these mapping relations are explicitly listed in Appendix~\ref{App B}.

As mentioned above, a key aspect of our approach is to perform the collision step such that different central moments shown above relax to their corresponding central moment equilibria. The discrete central moment equilibria $\Keq{mn}$ defined above can be obtained by matching them to the corresponding central moments of the continuous Maxwell distribution function after replacing the density $\rho$ with the order parameter $\phi$; furthermore, the interface sharpening flux terms in the conservative ACE (Eq.~(\ref{eqn1})) need to be accounted for by augmenting the first order central moment equilibrium components with $M_{\phi}\theta n_x$ and $M_{\phi}\theta n_y$~\cite{hajabdollahi2021central}. Thus, we have
\mathleft
\begin{gather}
\qquad \Keq{00} = \phi, \qquad
\Keq{10} = M_{\phi} \theta  n_x,\qquad
\Keq{01} = M_{\phi} \theta  n_y,\nonumber \\[2mm]
\qquad \Keq{20} = c_{s\phi}^2 \phi,\qquad
\Keq{02} = c_{s\phi}^2 \phi,\qquad
\Keq{11} = 0,\nonumber  \\[2mm]
\qquad \Keq{21} = 0,\qquad
\Keq{12} = 0,\qquad
\Keq{22} = c_{s\phi}^4 \phi,
\end{gather}
where $c_{s\phi}^2=1/3$.

Based on the above considerations, inspired from the algorithmic implementation presented in~\cite{geier2015cumulant} (see also~\cite{yahia2021central,yahia2021three}), we can now summarize the central moment LB algorithm for solving the conservative ACE for a time step $\Delta t$ starting from $f_\alpha=f_\alpha(\bm{x},t)$ as follows:
\begin{itemize}
  \item Compute pre-collision raw moments from distribution functions via $\bm{\kappa^{'}} = \PP\mathbf{f}$ (see Eq.~(\ref{eq:tensorP}) in Appendix~\ref{App B} for $\tensr{P}$)
  \item Compute pre-collision central moments from raw moments via $\bm{\kappa} = \F \bm{\kappa^{'}}$ (see Eq.~(\ref{eq:tensorF}) in Appendix~\ref{App B} for $\tensr{F}$)
  \item Perform collision step via relaxation of central moments $\kappa_{mn}$ to their equilibria $\kappa_{mn}^{eq}$: \newline
        \begin{equation}\label{eq:centralmomentrelaxationCACE}
           \qquad \tilde{\kappa}_{mn} = \kappa_{mn} + \wphi{mn}  (\Keq{mn} - \kappa_{mn}),
        \end{equation}
        where $(mn)=(00),(10),(01),(20),(02),(11),(21),(12)$, and $(22)$, and $\wphi{mn}$ is the relaxation parameter for moment of order ($m+n$). Here, the implicit summation convention of repeated indices is not assumed. The relaxation parameters of the first order moments $\omega_{10}^{\phi} = \omega_{01}^{\phi} = \omega^{\phi}$ are related to the mobility coefficient $M_{\phi}$ in Eq.~(\ref{eqn1}) via $M_{\phi}= c_{s\phi}^2 \left( \frac{1}{\omega^{\phi}} - \frac{1}{2}\right)\Delta t$, and the rest of the relaxation parameters are typically set to unity, i.e., $\wphi{mn}=1.0$, where $(m+n) \geq 2$. The results of Eq.~(\ref{eq:centralmomentrelaxationCACE}) are then grouped in $\bm{\tilde{\kappa}}$.
  \item Compute post-collision raw moments from post-collision central moments via $\bm{\tilde{\kappa}^{'}} = \F^{-1} \bm{\tilde{\kappa}}$ (see Eq.~(\ref{eq:tensorFinverse}) in Appendix~\ref{App B} for $\tensr{F}^{-1}$)
  \item Compute post-collision distribution functions from post-collision raw moments via $\mathbf{\tilde{f}} = \PP^{-1}\bm{\tilde{\kappa}^{'}}$ (see Eq.~(\ref{eq:tensorPinverse}) in Appendix~\ref{App B} for $\tensr{P}^{-1}$)
  \item Perform streaming step via $f_{\alpha}(\bm{x}, t+ \Delta t) = \tilde{f}_{\alpha}(\bm{x}-\bm{e}_{\alpha} \Delta t)$, where $\alpha = 0,1,2,...,8$.
  \item Update the order parameter $\phi$ of the phase-field model for interface capturing through \newline
        \begin{equation}
           \qquad \qquad\phi = \sum_{\alpha=0}^{8} f_{\alpha}.
        \end{equation}
\end{itemize}

\subsection{LB scheme for two-fluid motion with capillary and Marangoni forces}\label{Sec.4.2}
Next, we will present a central moment LB scheme to solve the motion of binary fluids with interfacial forces represented in Eqs.~(\ref{eqn4})-(\ref{eq:surfacetensionforcecomponents}) by evolving another distribution function $g_\alpha$, where $\alpha=0,1,2,\ldots,8$. Our approach is based on a discretization of the modified continuous Boltzmann equation and obtaining the discrete central moment equilibria and central moments of the source terms for the body forces via a matching principle with their continuous counterparts as detailed in Ref.~\cite{hajabdollahi2021central}. However, in contrast to Ref.~\cite{hajabdollahi2021central}, where an orthogonal moment basis is employed resulting in the so-called cascaded LB approach, in the following, we consider the simpler, non-orthogonal moment basis vectors as given earlier in Eq.~(\ref{eqn34}).

As in the previous section, we first define the following raw moments and the central moments of the distribution function $g_\alpha$, its equilibrium $g_\alpha^{eq}$, as well as the source term $S_\alpha$, where the latter accounts for the surface tension and body forces, as well as those that arising from the application of a transformation to simulate flows at high density ratios in the incompressible limit (see~\cite{he1999lattice,hajabdollahi2021central}):
\mathleft
\begin{subequations}
\begin{equation}
\qquad \left( \begin{array}{c} {\eta}'_{ mn}\\[1mm]   \eta'^{\;eq}_{ mn}\\[1mm]   {\sigma}'_{ mn} \end{array} \right)  =  \sum_{\alpha = 0}^{8} \left( \begin{array}{c}g_{\alpha} \\[1mm]   g_{\alpha}^{eq}\\[1mm]  S_\alpha \end{array} \right)  e_{\alpha x}^m   e_{\alpha y}^n,
\end{equation}
\mathleft
\begin{equation}
\qquad \left( \begin{array}{c}{\eta}_{ mn}\\[1mm]   \eta^{\;eq}_{ mn}\\[1mm]   {\sigma}_{ mn} \end{array} \right)  = \sum_{\alpha = 0}^{8} \left( \begin{array}{c}g_{\alpha} \\[1mm]   g_{\alpha}^{eq}\\[1mm]  S_\alpha \end{array} \right)  (e_{\alpha x}-u_x)^m  ( e_{\alpha y}-u_y)^n.
\end{equation}
\end{subequations}
For conveniences, we can group the elements of the distribution function, its equilibrium, and the source term for the D2Q9 lattice as the following vectors: $\mathbf{g}=(g_0,g_1,g_2,\ldots,g_8)^\dagger$, $\mathbf{g}^{eq}=(g_0^{eq},g_1^{eq},g_2^{eq},\ldots,g_8^{eq})^\dagger$, and $\mathbf{S}=(S_0,S_1,S_2,\ldots,S_8)^\dagger$. Moreover, we group all the possible raw moments and the central moments defined above for the D2Q9 lattice via the following:
\mathleft
\begin{subequations}
\begin{eqnarray}
\qquad \bm{{\eta}^{'}} \! \! \! &=& \! \! \! ( {\eta}_{00}^{'}, {\eta}_{10}^{'},{\eta}_{01}^{'}, {\eta}_{20}^{'}, {\eta}_{02}^{'}, {\eta}_{11}^{'},{\eta}_{21}^{'}, {\eta}_{12}^{'},{\eta}_{22}^{'} ),\label{eqn:4a} \\[3mm]
\qquad \bm{{\eta}} \! \! \! &=& \! \! \! ( {\eta}_{00},{\eta}_{10}, {\eta}_{01}, {\eta}_{20}, {\eta}_{02}, {\eta}_{11}, {\eta}_{21}, {\eta}_{12}, {\eta}_{22} ),
\end{eqnarray}
\end{subequations}
and similarly for raw moments and the central moments the equilibrium and the source term.

The collision step will be performed such that different central moments shown above relax to their corresponding central moment equilibria, which are augmented by changes in the central moments due to the net forces; the latter is given by sum the surface tension force $\bm{F}_s=(F_{sx},F_{sy})$, which can have contributions from both the capillary and Marangoni forces as represented in Eq.~(\ref{eq:surfacetensionforcecomponents}), and any external force $\bm{F}_{ext}=(F_{ext,x},F_{ext,y})$, i.e., $\bm{F}_{t}=\bm{F}_{s}+\bm{F}_{ext}$ or $(F_{tx},F_{ty})=(F_{sx}+F_{ext,x},F_{sy}+F_{ext,y})$. Moreover, the use of an incompressible transformation as mentioned above leads to a pressure-based formulation, involving the incorporation of a net pressure force $\bm{F}_p$ arising from $\varphi(\rho)=p-\rho c_s^2$, i.e., $\bm{F}_p=-\bm{\nabla}\varphi$, or $(F_{px},F_{py})=(-\partial_x\varphi,-\partial_y\varphi)$ (see~\cite{hajabdollahi2021central} for details). Then, the discrete central moment equilibria $\eta_{mn}$ defined above can be obtained by matching them to the corresponding continuous central moments of the equilibrium that arise from the incompressible transformation, and similarly for the central moments of the source term $\sigma_{mn}$, which then results in the following expressions for the D2Q9 lattice~\cite{hajabdollahi2021central}:
\mathleft
\begin{gather}
\qquad {\eta}_{00}^{eq} = p, \quad {\eta}_{10}^{eq} = -\varphi(\rho) u_x, \quad {\eta}_{01}^{eq} = -\varphi(\rho)u_y, \quad {\eta}_{20}^{eq} =  p c_s^2 + \varphi(\rho)u_x^2,\nonumber \\
\qquad {\eta}_{02}^{eq} =  p c_s^2 + \varphi(\rho)u_y^2, \quad  {\eta}_{11}^{eq} = \varphi(\rho)u_x u_y , \quad {\eta}_{21}^{eq} = -\varphi(\rho)(u_x^2+ c_s^2) u_y,\nonumber \\
\qquad {\eta}_{12}^{eq} = -\varphi(\rho)(u_y^2+ c_s^2) u_x, \quad {\eta}_{22}^{eq} = c_s^6 \rho + \varphi(\rho)(u_x^2+ c_s^2) (u_y^2+ c_s^2).
\end{gather}
and
\begin{gather}
\qquad {\sigma}_{00}= \Gamma_{00}^p, \quad {\sigma}_{10} = c_s^2 F_{tx}-u_x{\Gamma}_{00}^p, \quad {\sigma}_{01} =  c_s^2 F_{ty}-u_y{\Gamma}_{00}^p, \nonumber \\
\qquad {\sigma}_{20} = 2c_s^2 F_{px}u_x+(u_x^2+c_s^2){\Gamma}_{00}^p,\quad {\sigma}_{02} =  2c_s^2 F_{py}u_y+(u_y^2+c_s^2){\Gamma}_{00}^p, \nonumber\\
\qquad {\sigma}_{11} = c_s^2 (F_{px}u_y+F_{py}u_x)+u_x u_y{\Gamma}_{00}^p,\quad {\sigma}_{21} = 0, \quad {\sigma}_{12} = 0, \quad {\sigma}_{22} = 0,
\end{gather}
where $\Gamma_{00}^p=(F_{px}u_x+F_{py}u_y)$.

Using the above developments, we can now summarize the central moment LB algorithm for computing the two-fluid motion with interfacial forces for a time step $\Delta t$ starting from $g_\alpha=g_\alpha(\bm{x},t)$ as follows:
\begin{itemize}
  \item Compute pre-collision raw moments from distribution functions via $\bm{\eta^{'}} = \PP\mathbf{g}$ (see Eq.~(\ref{eq:tensorP}) in Appendix~\ref{App B} for $\tensr{P}$)
  \item Compute pre-collision central moments from raw moments via $\bm{\eta} = \F \bm{\eta^{'}}$ (see Eq.~(\ref{eq:tensorF}) in Appendix~\ref{App B} for $\tensr{F}$)
  \item Perform collision step via relaxation of central moments $\eta_{mn}$ to their equilibria $\eta_{mn}^{eq}$ and augmented with the source terms $\sigma_{mn}$: \newline
      In order to allow for an independent specification of the shear viscosity $\nu$ from the bulk viscosity $\zeta$, the trace of the second order moments ${\eta}_{20} + {\eta}_{02}$ should be evolved independently from the other second order moments. To accomplish this, prior to collision, we combine the diagonal parts of the second order moments as follows (see e.g.,~\cite{geier2015cumulant,yahia2021central,yahia2021three}):
        \mathleft
        \begin{gather}
        \qquad {\eta}_{2s} = {\eta}_{20} + {\eta}_{02}, \qquad {\eta}_{2s}^{eg} = {\eta}_{20s}^{eg} + {\eta}_{02}^{eg}, \qquad {\sigma}_{2s} = {\sigma}_{20s} + {\sigma}_{02},\nonumber \\[2mm]
        \qquad {\eta}_{2d} = {\eta}_{20} -  {\eta}_{02}, \qquad {\eta}_{2d}^{eg} = {\eta}_{20s}^{eg} -  {\eta}_{02}^{eg},  \qquad {\sigma}_{2d} = {\sigma}_{20s} - {\sigma}_{02},\nonumber
        \end{gather}
        and thus ${\eta}_{2s}$ and ${\eta}_{2d}$ will be evolved independently under collision. Then, the post-collision central moments under relaxation and augmentation due to the forces can be computed via
        \mathleft
        \begin{equation} \label{eq:centralmomentrelaxationtwofluidmotion}
        \qquad  \tilde{\eta}_{ mn} =  {\eta}_{mn} + \omega_{mn} \left( {\eta}_{mn}^{eq} -{\eta}_{mn} \right)  + \left(1-\omega_{mn}/2 \right)  {\sigma}_{mn}\Delta t,
        \end{equation}
        where $\omega_{mn}$ is the relaxation time corresponding to the central moment ${\eta}_{ mn}$, and $(mn)= (00), (10), (01), (2s), (2d), (11), (21), (12), \mbox{and}, (22)$. Here, the relaxation parameter $\omega_{2s}$ is related to the bulk viscosity via $\zeta= c_s^2 \left(1/\omega_{2s}- 1/2\right)\Delta t$, while the relaxation parameters  $\omega_{2d}$ and $\omega_{11}$ are related to shear viscosity via $\nu= c_s^2 \left( 1/\omega_{ij} - 1/2\right)\Delta t$ where $(ij)=(2d),(11)$. Typically, $c_s^2=1/3$. In view of Eq.~(\ref{eqn11}) it should be noted that if the bulk fluid properties are different, the relaxation parameters $\omega_{2d}$ and $\omega_{11}$ will then vary locally across the interface. The rest of the relaxation parameters of central moments are generally set to unity, i.e., $\omega_{ij}=1.0$, where $(ij)=(00),(10),(01),(2s),(21),(12),(22)$.\newline
        Also, the combined forms of the post-collision central moments $\tilde{\eta}_{2s}$ and $\tilde{\eta}_{2d}$ are then segregated in their individual components $\tilde{\eta}_{20}$ and $\tilde{\eta}_{02}$ via
        \mathleft
        \begin{equation*}
        \qquad  \tilde{\eta}_{20} = \frac{1}{2} \left(\tilde{\eta}_{2s}+\tilde{\eta}_{2d} \right), \qquad  \tilde{\eta}_{02} = \frac{1}{2} \left(\tilde{\eta}_{2s}-\tilde{\eta}_{2d} \right).
        \end{equation*}
        Finally, the results of Eq.~(\ref{eq:centralmomentrelaxationtwofluidmotion}) by accounting for the above segregation are then grouped in $\bm{\tilde{\eta}}$.
  \item Compute post-collision raw moments from post-collision central moments via $\bm{\tilde{\eta}^{'}} = \F^{-1} \bm{\tilde{\eta}}$ (see Eq.~(\ref{eq:tensorFinverse}) in Appendix~\ref{App B} for $\tensr{F}^{-1}$)
  \item Compute post-collision distribution functions from post-collision raw moments via $\mathbf{\tilde{g}} = \PP^{-1}\bm{\tilde{\eta}^{'}}$ (see Eq.~(\ref{eq:tensorPinverse}) in Appendix~\ref{App B} for $\tensr{P}^{-1}$)
  \item Perform streaming step via $g_{\alpha}(\bm{x}, t+ \Delta t) = \tilde{g}_{\alpha}(\bm{x}-\bm{e}_{\alpha} \Delta t)$, where $\alpha = 0,1,2,...,8$.
  \item Update the pressure field $p$ and the components of the fluid velocity $\bm{u}=(u_x,u_y)$ through \newline
        \begin{equation}
           \qquad \qquad p = \sum_\alpha g_\alpha + \frac{1}{2} \bm{F}_p \cdot {\bm{u}}\Delta t,\quad \rho c_s^2 \bm{u} = \sum_\alpha g_\alpha \bm{e}_\alpha + \frac{1}{2} c_s^2 \bm{F}_{t}\Delta t.
        \end{equation}
\end{itemize}


\subsection{LB scheme for energy equation} \label{Sec.4.3}
Finally, we will now discuss a central moment LB approach for the solution of the energy transport equation (Eq.~(\ref{energy eqn})) by evolving a third distribution function $h_\alpha$, where $\alpha=0,1,2,\ldots,8$, on the D2Q9 lattice. Since Eq.~(\ref{energy eqn}) is an advection-diffusion equation, its construction procedure is quite similar to that of the LB scheme for the conservative ACE presented earlier, albeit without the presence of a term such as the interface sharpening flux term which appears in the latter case.
As before, we first define the following raw moments and central moments, respectively, of the distribution function $h_\alpha$, as well as its equilibrium $h_\alpha^{eq}$:
\mathleft
\begin{subequations}
\begin{equation}
\qquad \left( \begin{array}{c}\chi'_{mn}\\[2mm]   \chi'^{\;eq}_{mn} \end{array} \right)  = \sum_{\alpha = 0}^{8} \left( \begin{array}{c} h_{\alpha} \\[2mm]  h_{\alpha}^{eq} \end{array} \right)  e_{\alpha x}^m   e_{\alpha y}^n,
\end{equation}
\mathleft
\begin{equation}
\qquad \left( \begin{array}{c}\chi_{mn} \\[2mm]   \chi_{mn}^{eq} \end{array} \right)  = \sum_{\alpha = 0}^{8} \left( \begin{array}{c} h_{\alpha} \\[2mm]  h_{\alpha}^{eq} \end{array} \right) (e_{\alpha x}-u_x)^m  ( e_{\alpha y}-u_y)^n.
\end{equation}
\end{subequations}
For convenience, we list the components of the distribution function and its equilibrium, respectively, using $\mathbf{h}=(h_0,h_1,h_2,\ldots,h_8)^\dagger$ and $\mathbf{h}^{eq}=(h_0^{eq},h_1^{eq},h_2^{eq},\ldots,h_8^{eq})^\dagger$, and analogously for the raw moments and central moments via
\mathleft
\begin{subequations}
\begin{eqnarray}
\qquad \bm{\chi^{'}} \! \! \! &=& \! \! \! ( \chi_{00}^{'}, \chi_{10}^{'},\chi_{01}^{'}, \chi_{20}^{'}, \chi_{02}^{'}, \chi_{11}^{'},\chi_{21}^{'}, \chi_{12}^{'},\chi_{22}^{'} ),\label{eqn:4a} \\[3mm]
\qquad \bm{\chi} \! \! \! &=& \! \! \! ( \chi_{00},\chi_{10}, \chi_{01}, \chi_{20}, \chi_{02}, \chi_{11}, \chi_{21}, \chi_{12}, \chi_{22} ).
\end{eqnarray}
\end{subequations}
To construct a central moment-based collision model for solving the energy equation, similar to Sec.~\ref{Sec.4.1}, we obtain the discrete equilibrium central moments from the corresponding continuous counterpart of the Maxwellian by replacing the density $\rho$ with the temperature $T$, and the results read as
\mathleft
\begin{gather}
\qquad \chi_{00}^{eq} = T, \qquad
\chi_{10}^{eq} = 0,\qquad
\chi_{01}^{eq} = 0,\nonumber \\[2mm]
\qquad \chi_{20}^{eq} = c_{sT}^2 T,\qquad
\chi_{02}^{eq} = c_{sT}^2 T,\qquad
\chi_{11}^{eq} = 0,\nonumber  \\[2mm]
\qquad \chi_{21} = 0,\qquad
\chi_{12}^{eq} = 0,\qquad
\chi_{22}^{eq} = c_{sT}^4 T,
\end{gather}
where, typically, $c_{sT}^2 = 1/3$. Then, the computational procedure for solving the energy equation for a time step $\Delta t$ starting from $h_\alpha=h_\alpha(\bm{x},t)$ can be summarized as follows:
\begin{itemize}
  \item Compute pre-collision raw moments from distribution functions via $\bm{\chi^{'}} = \PP\mathbf{h}$ (see Eq.~(\ref{eq:tensorP}) in Appendix~\ref{App B} for $\tensr{P}$)
  \item Compute pre-collision central moments from raw moments via $\bm{\chi} = \F \bm{\chi^{'}}$ (see Eq.~(\ref{eq:tensorF}) in Appendix~\ref{App B} for $\tensr{F}$)
  \item Perform collision step via relaxation of central moments $\chi_{mn}$ to their equilibria $\chi_{mn}^{eq}$: \newline
        \begin{equation}\label{eq:centralmomentrelaxationenergyequation}
           \qquad \tilde{\chi}_{mn} = \chi_{mn} + \omega^T_{mn}  (\chi_{mn}^{eq} - \chi_{mn}),
        \end{equation}
        where $(mn)=(00),(10),(01),(20),(02),(11),(21),(12)$, and $(22)$, and $\omega^T_{mn}$ is the relaxation parameter for moment of order ($m+n$). The relaxation parameters of the first order moments $\omega_{10}^T$=$\omega_{01}^T$=$ \omega^T$ are related to the thermal diffusivity $\alpha=k/(\rho c_p)$ via $\alpha = c_{sT}^2 \left(1/\omega^{T} - 1/2\right)\Delta t$, and the rest of the relaxation parameters of higher central moments are typically set to unity. The results of Eq.~(\ref{eq:centralmomentrelaxationenergyequation}) are then grouped in $\bm{\tilde{\chi}}$.
  \item Compute post-collision raw moments from post-collision central moments via $\bm{\tilde{\chi}^{'}} = \F^{-1} \bm{\tilde{\chi}}$ (see Eq.~(\ref{eq:tensorFinverse}) in Appendix~\ref{App B} for $\tensr{F}^{-1}$)
  \item Compute post-collision distribution functions from post-collision raw moments via $\mathbf{\tilde{h}} = \PP^{-1}\bm{\tilde{\chi}^{'}}$ (see Eq.~(\ref{eq:tensorPinverse}) in Appendix~\ref{App B} for $\tensr{P}^{-1}$)
  \item Perform streaming step via $h_{\alpha}(\bm{x}, t+ \Delta t) = \tilde{h}_{\alpha}(\bm{x}-\bm{e}_{\alpha} \Delta t)$, where $\alpha = 0,1,2,...,8$.
  \item Update the temperature field $T$ is obtained from \newline
        \begin{equation}
           \qquad \qquad T = \sum_{\alpha=0}^{8} h_{\alpha}.
        \end{equation}
\end{itemize}
While the central moment LB schemes outlined here are applicable for a general class surface-tension driven flows with thermocapillary effects, in this work, they will be mainly applied, in conjunction with the analytical solution derived earlier, to study the effect of various characteristic parameters on the flow patterns and the intensity of thermocapillary convention in superimposed layers of two self-rewetting fluids (SRFs) bounded within a microchannel nonuniformly heated on one side. Before proceeding with this, we will first validate our numerical approach given above for some standard thermocapillary benchmark problems for normal fluids for which analytical solutions available in the literature in the next section.

\section{Numerical validation} \label{Sec.5}
\subsection{Thermocapillary migration of a droplet of a normal fluid under a temperature gradient}
The first validation problem that we consider is the thermocapillary migration of a droplet of a normal fluid in the field of a linear variation in temperature or a uniform gradient in temperature. Young \emph{et al.}~\cite{young1959motion} presented an analytical solution of the droplet migration velocity in the creeping flow limit and at small Marangoni numbers. Taking $\sigma (T) = \sigma_{0} + \sigma_T (T-T_{ref})$ for the surface tension variation, consider a droplet of radius $R$ with a density $\rho_b$, viscosity $\mu_b$, and thermal conductivity $k_b$ in the presence of a uniform temperature gradient $\nabla T_{\infty}$, then the characteristic velocity $U_*$ obtained under a balance of the thermocapillary force and the viscous force can be written as
\mathleft
\begin{equation}
\qquad U_* = -\frac{\sigma_{T}   |\nabla T_{\infty}| R} {\mu_{b}}.\label{eq:characteristicvelocitydroplet}
\end{equation}
Defining the Reynolds number and the Marangoni number as $\mbox{Re} = \rho_b U_*R/\mu_{b}$ and $\mbox{Ma} = U_*R/k_{b}$, respectively, in the limit $Re \ll 1$ and $Ma \ll 1$, Young \emph{et al.} performed a theoretical analysis and derived an expression for the terminal migration velocity of the droplet $U_{YGB}$ given by
\mathleft
\begin{equation}
\qquad U_{YGB}= \frac{2U_*}{(2+\tilde{k})(2+3 \tilde{\mu})},    \label{YGBmigrationvelocity}
\end{equation}
where $\tilde{k}$ and $\tilde{\mu}$ are the property ratios defined in Eq.~(\ref{three}).

For performing the numerical simulation using the LB schemes presented in the previous sections, here and in what follows for the rest of this paper, when required, the no-slip velocity boundary condition is prescribed using the standard half-way bounce-back condition, while the specification of the scalar field such as the temperature on the boundaries is carried out using the anti-bounce back scheme; the no-gradient conditions on any boundary are imposed using the free-slip condition; finally, as is typical for the LB method, all the values are specified in the lattice units. See Ref.~\cite{kruger2017lattice} for further details.

We consider a droplet of radius $R=20$ initially kept at the center of a 2D domain of size $8R \times 16R$. No slip boundary conditions are imposed on the top and bottom walls while periodic boundary conditions are used on the left and right walls. In the direction normal to the bottom and top walls, a linear variation in the temperature field with $T_{bot}=0$ on the bottom wall and $T_{top}=32$ on the top wall is imposed, resulting in a constant temperature gradient in the far field $\nabla T_{\infty} = 0.1$. For the fluid properties, we take $\rho_{a,b} = c_{p_{a,b}} = 1$, $\mu_{a,b} = k_{a,b} = 0.2$, $T_{ref} = 16$, $\sigma_{0} = 2.5\times10^{-3}$, and $\sigma_{T} = -10^{-4}$ and the values of the parameters are such that the assumption of the negligible convection in deriving the analytical solution is satisfied. For the above choice, the theoretically predicted value of the terminal velocity of the droplet is $U_{YGB}= 1.333 \times10^{-4}$, and both $\mbox{Re}$ and $\mbox{Ma}$ are $0.1$.

\begin{figure}[H]
\centering
\includegraphics[trim = 0 0 0 0,clip, width = 80mm]{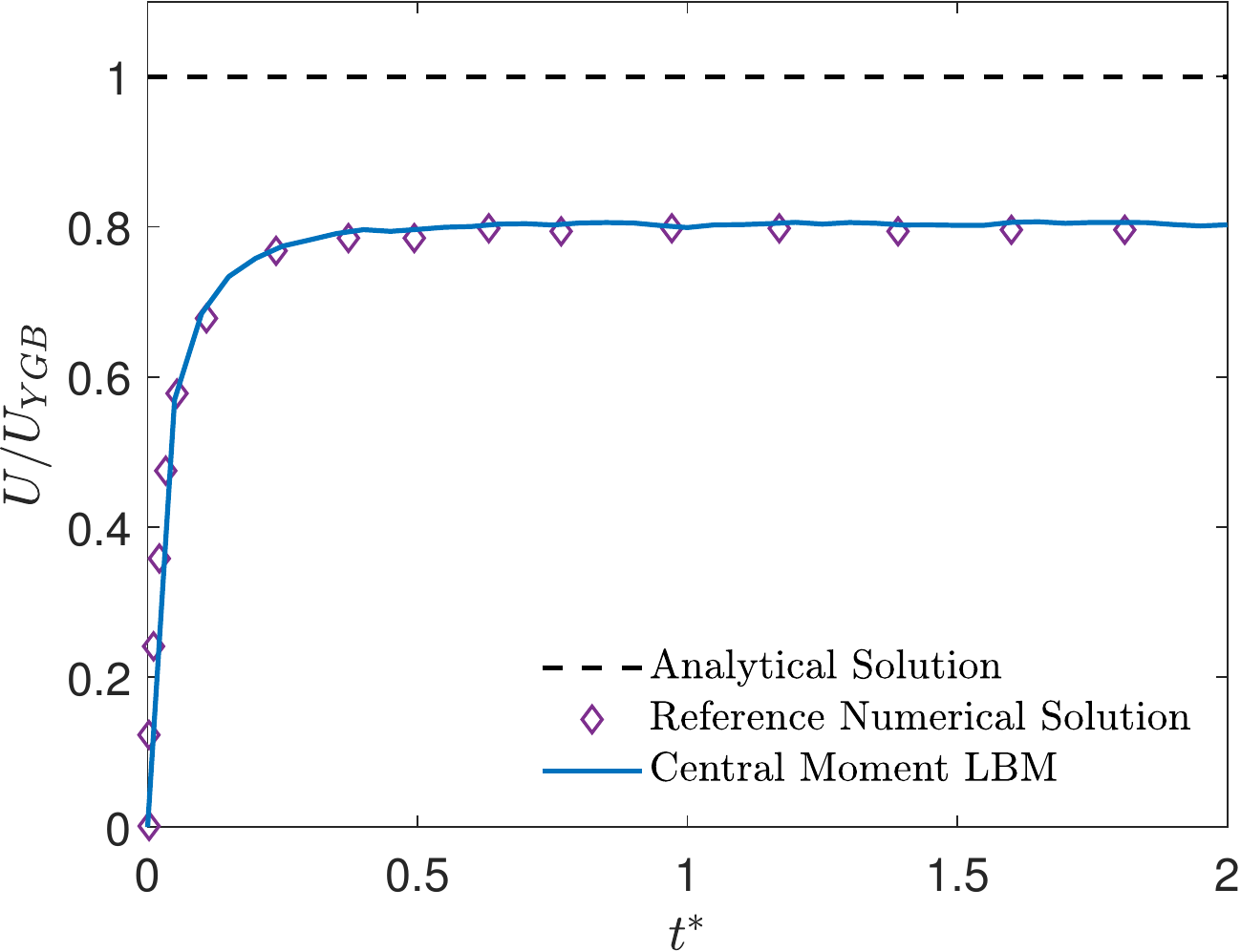}
\caption{The drop migration velocity for a 2D droplet at  $\mbox{Re}=\mbox{Ma}=0.1$ normalized by the analytical prediction velocity $U_{YGB}$ versus the dimensionless time $t^*=tU/R$. The other reference numerical solution for a 2D droplet is taken from Ref.~\cite{guo2015thermodynamically}.}
\label{Young_Velocity}
\end{figure}

Figure~\ref{Young_Velocity} shows the temporal variations of the normalized drop migration velocity $U / U_{YGB}$ as a function of the dimensionless time ${t}^{\star} = tU/R$ computed using the LB schemes presented earlier along with the theoretical prediction, as well as results from another reference numerical solution involving a 2D droplet~\cite{guo2015thermodynamically}. It should be noted that the theory assumed a 3D axisymmetric, non-deformable spherical droplet, while the present LB results as well as the reference numerical results are based on considering a 2D droplet. As a result, all the numerical schemes shown consistently attain $U/U_{YGB}\approx  0.80$ or about $80\%$ of the theoretical value. Nevertheless, the present results are in good quantitative agreement with the reference results given in~\cite{guo2015thermodynamically} for similar conditions. Moreover, this trend is also consistent with the results obtained by the use of different numerical methods for this problem involving a 2D droplet (see e.g.,~\cite{guo2015thermodynamically, zheng2016continuous, nabavizadeh2019effect}).

\subsection{Thermocapillary-driven flow in a heated microchannel with two superimposed normal fluids}
As a second benchmark problem, we will test our LB schemes for the simulation of the thermocapillary-driven flow in a sinusoidally heated microchannel which confines two supperimposed normal fluids (NFs)~\cite{pendse2010analytical}. The problem setup for this case is the same as the one presented in Sec.~\ref{Subsec.1.1} (see Fig.~\ref{Model_setup} for the geometric set up). The wall temperatures are applied according to Eqs.~(\ref{two}) and (\ref{one}). The dimensionless parameters $\mbox{Re}$, $\mbox{Ma}$, and $\mbox{Ca}$ for this case are defined in Eq.~(\ref{ReMaCa}) and the ratio of the material properties are given in Eq.~(\ref{three}). For  $Re\ll 1$, $Ma\ll 1$, and $Ca\ll 1$, and considering the flow is driven by  a surface tension gradient, where the surface tension decreases linearly with the increasing temperature for the NF as $\sigma (T) = \sigma_{0} + \sigma_T (T-T_{ref})$, Ref.~\cite{pendse2010analytical} derived analytical solutions for the temperature $T(x,y)$, stream function $\psi(x,y)$, and the components of the velocity field $u(x,y)$ and $v(x,y)$. It can also be obtained as a special case of the analytical solution derived in this work by setting the coefficient for the quadratic term for the surface tension variation with temperature to be zero, i.e., $\sigma_{TT}=0$.

We performed LB simulations by considering two normal fluids of equal thickness $a=b=50$ in a microchannel of length $l=200$. Periodic boundary conditions are used on the left and right sides of the domain, while no-slip boundary conditions are imposed on the top and bottom walls, and the wall temperatures are applied using Eqs.~(\ref{two}) and (\ref{one}) with $T_h=T_c=T_{ref}=\Delta T = 1.0$ for simplicity. The various fluid properties are chosen as follows: $\sigma_{0}= 1.0\times 10^{-2}$, $M_1=-5.0 \times 10^{-2}$,  $M_2=0$, $\tilde{k} = 1$, and $\tilde{\mu} = 1$; moreover, the dimensionless parameters are $\mbox{Re} = 1.59\times 10^{-1}$, $\mbox{Ma}=3.83\times 10^{-2}$, and $\mbox{Ca}=1.26\times 10^{-2}$, so that the assumptions made in deriving the analytical solution are satisfied. For the phase-field model, the values of the interface thickness and the mobility parameter are $W = 5$ and $M_{\phi} = 0.02$, respectively.

\begin{figure}[H]
\centering
\begin{subfigure}{0.475\textwidth}
\includegraphics[trim = 0 0 0 0,clip, width = 72mm]{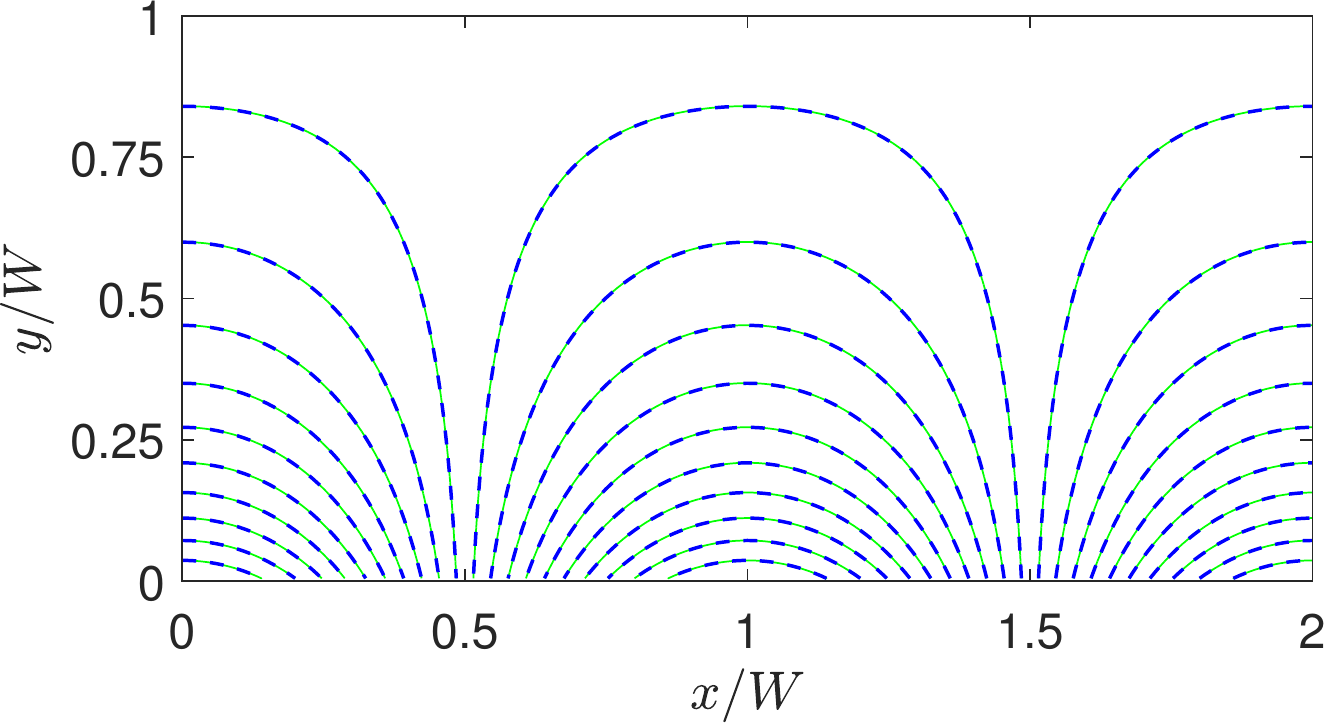}
\caption{Temperature contours}\label{temp_k1_NF}
\end{subfigure}
\begin{subfigure}{0.475\textwidth}
\includegraphics[trim = 0 0 0 0,clip, width = 72mm]{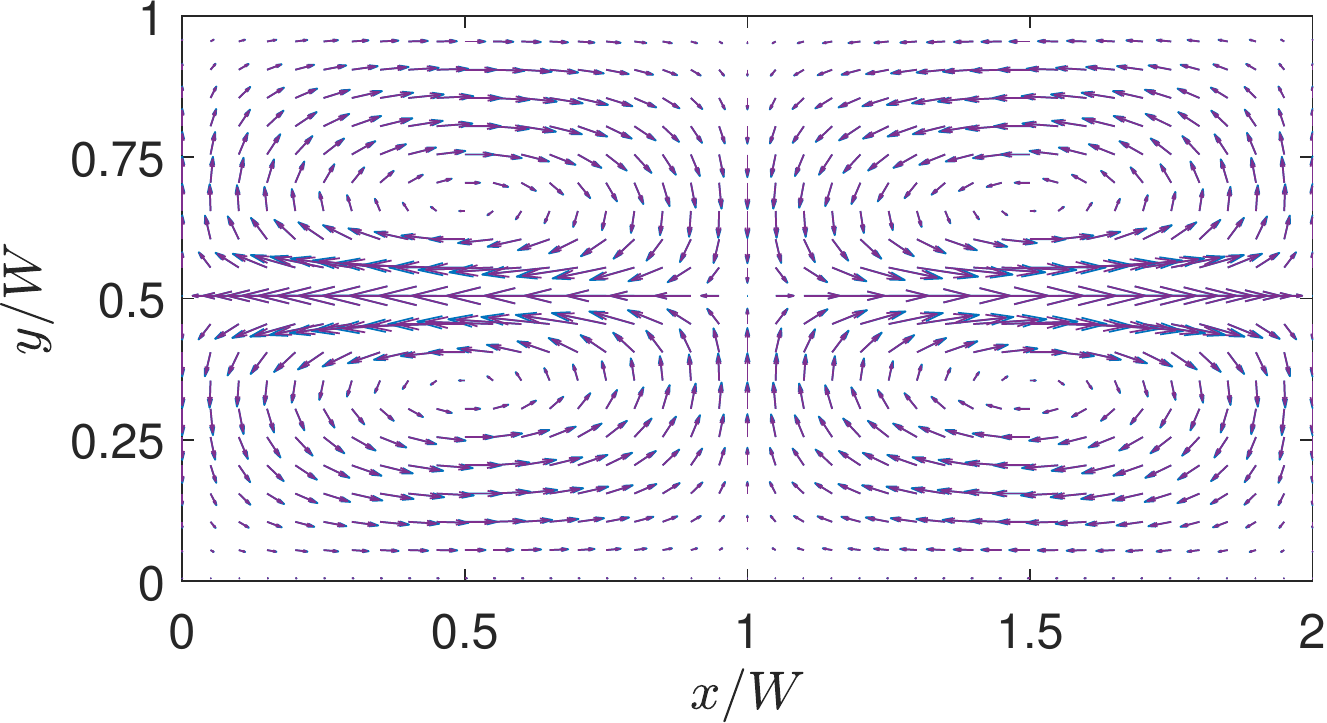}
\caption{Velocity vectors}\label{velocity_k1_NF}
\end{subfigure}
\caption{$(a)$ Temperature contours of the NF in thermocapillary flow within a heated microchannel with thermal conductivity ratio $\tilde{k} = 1$ and viscosity ratio $\tilde{\mu}=1$ obtained from the LB simulation results (solid green lines) and the analytical solution (dashed blue lines). $(b)$ Velocity vectors due to thermocapillary flow of the NF obtained from the LB simulation results (blue arrows) and the analytical solution (purple arrows).}
\label{}
\end{figure}

\begin{figure}[H]
\centering
\begin{subfigure}{0.475\textwidth}
\includegraphics[trim = 0 0 0 0,clip, width = 72mm]{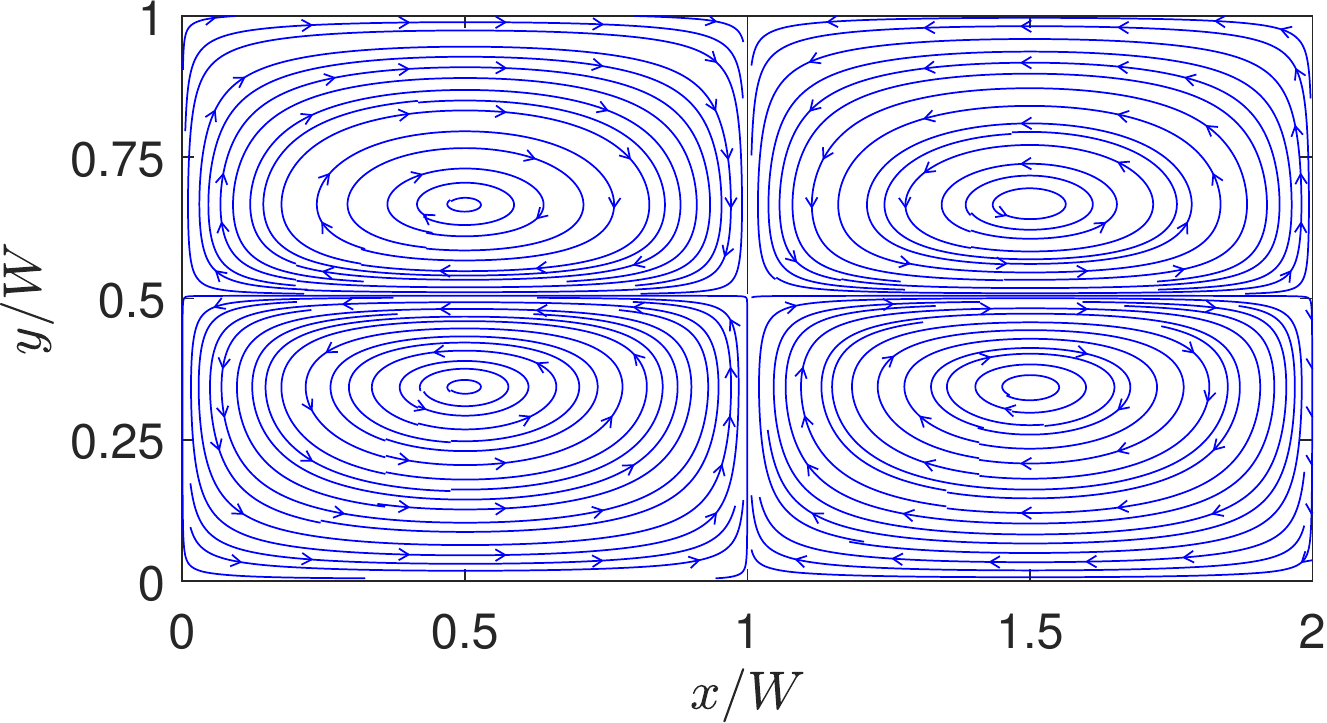}
\caption{Analytical Solution}
\end{subfigure}
\begin{subfigure}{0.475\textwidth}
\includegraphics[trim = 0 0 0 0,clip, width = 72mm]{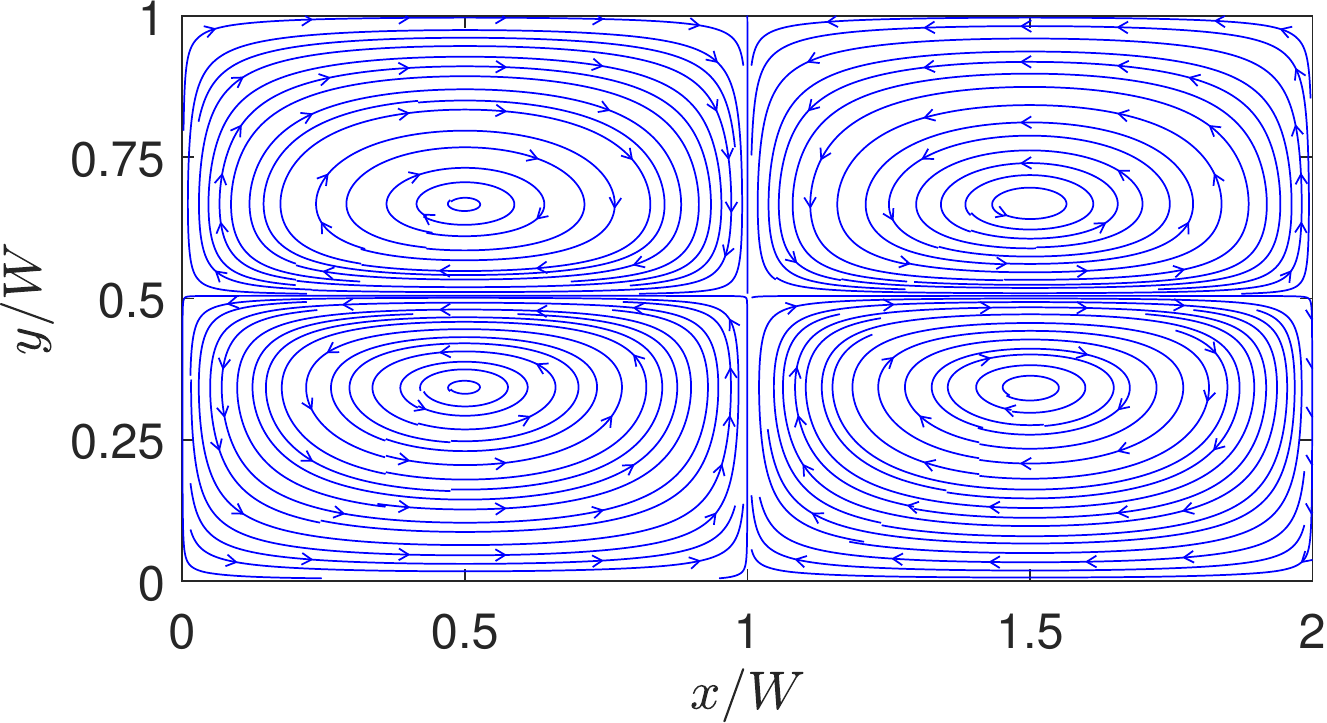}
\caption{LBM Simulation}
\end{subfigure}
\caption{Streamlines of the thermocapillary flow in NFs within a heated microchannel with thermal conductivity ratio $\tilde{k} = 1$ and viscosity ratio $\tilde{\mu}=1$ obtained from the analytical solution (left) and the LB simulation results (right). The arrows indicate the direction of the thermocapillary convection.}
\label{streamlines_k1_NF}
\end{figure}

\begin{figure}[H]
\centering
\begin{subfigure}{0.325\textwidth}
\includegraphics[trim = 0 0 0 0,clip, width = 50mm]{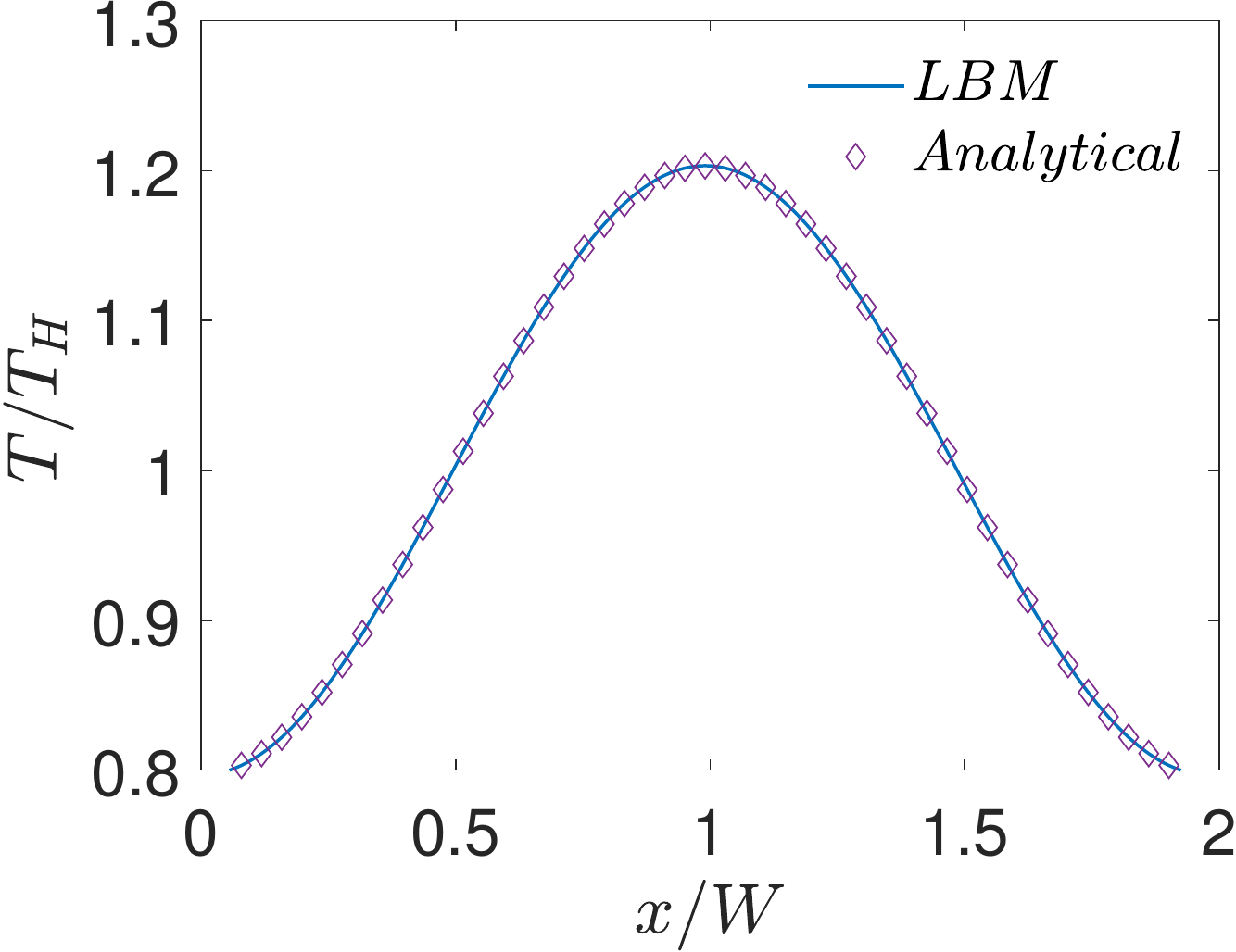}
\end{subfigure}
\begin{subfigure}{0.325\textwidth}
\includegraphics[trim = 0 0 0 0,clip, width = 50mm]{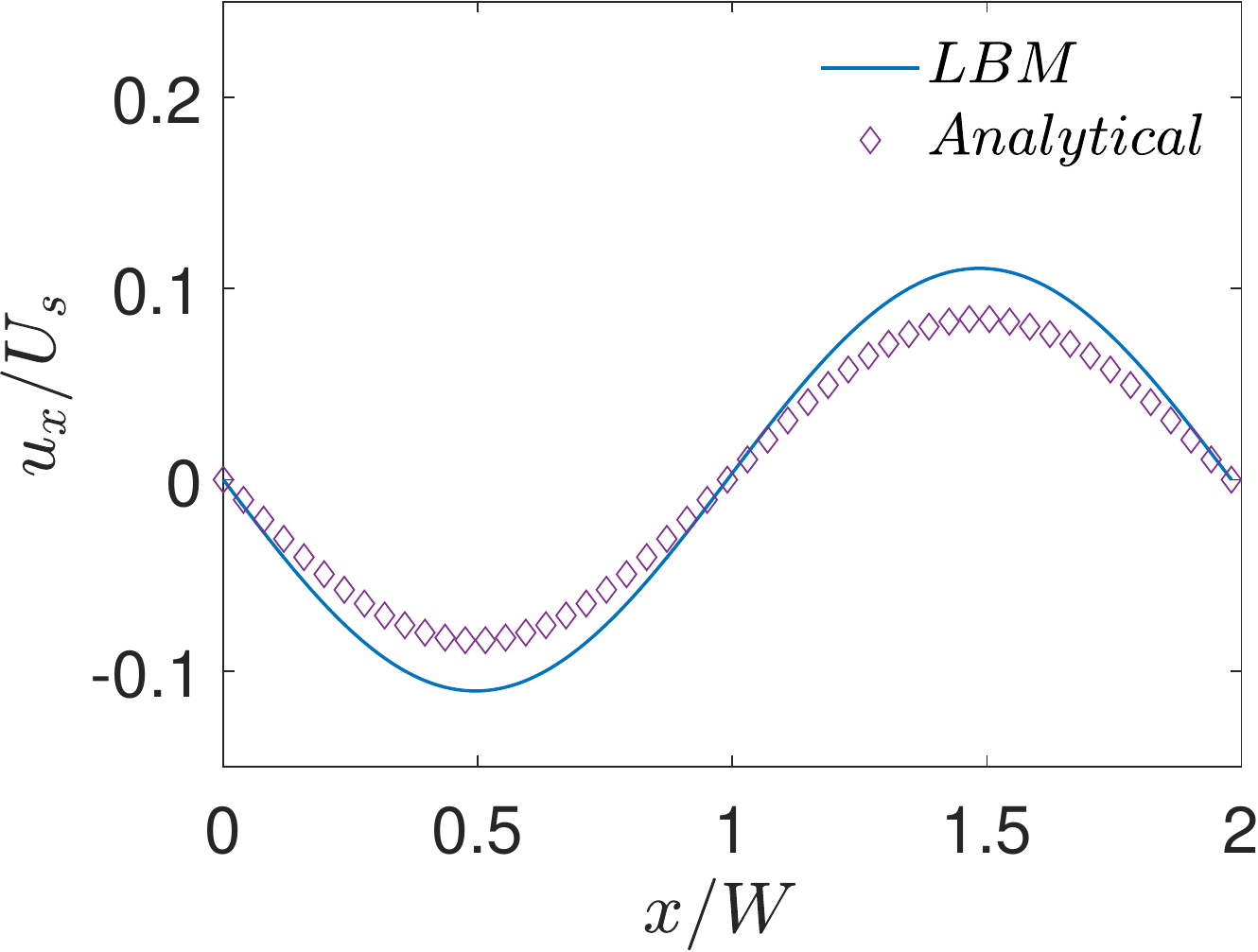}
\end{subfigure}
\begin{subfigure}{0.325\textwidth}
\includegraphics[trim = 0 0 0 0,clip, width = 50mm]{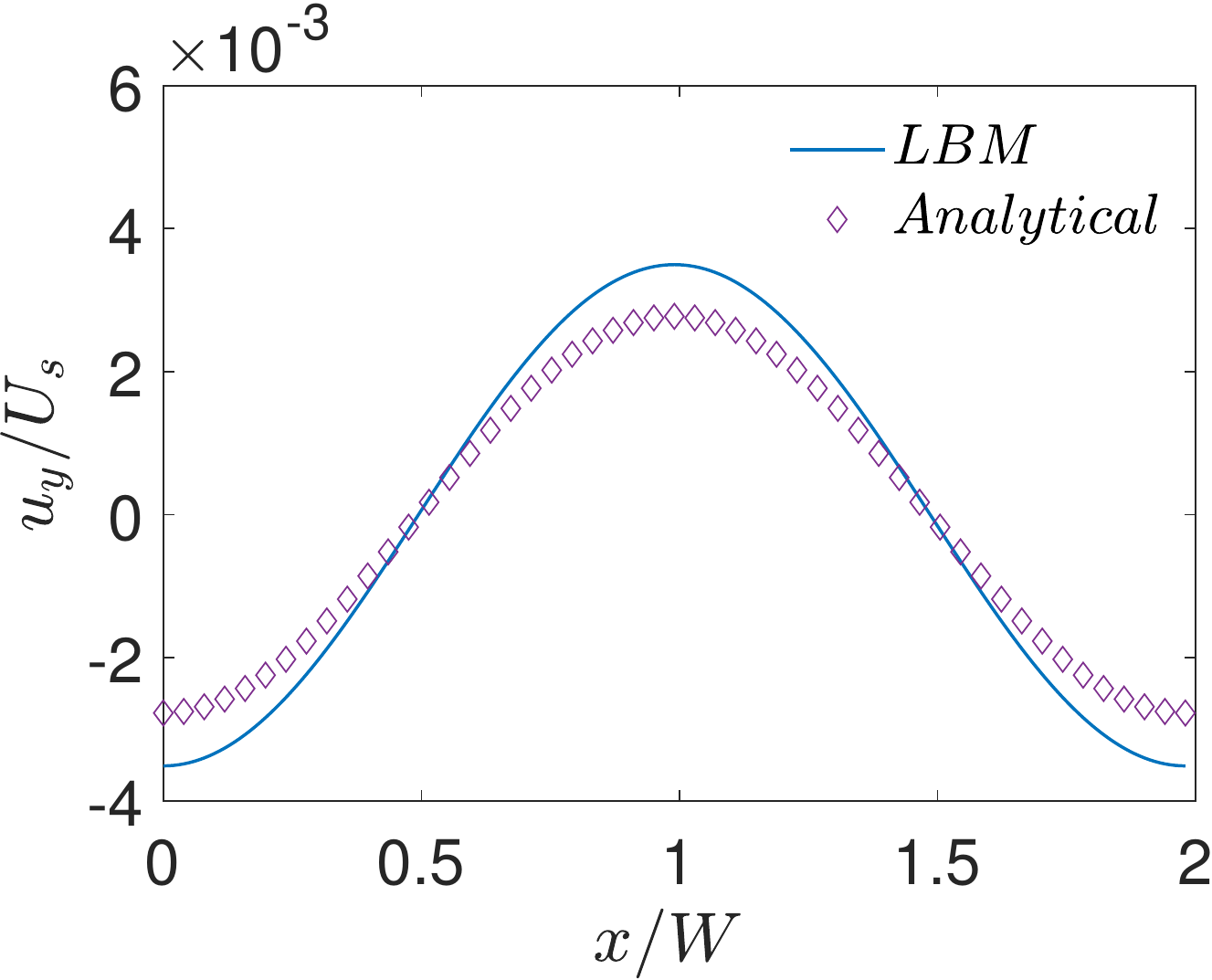}
\end{subfigure}
\caption{Profiles of the temperature and velocity components along the centerline of the domain in the $x$ direction for thermocapillary flow of a NF in a heated microchannel. The purple diamond symbols shown are obtained from the analytical solution given by~\cite{pendse2010analytical} and the blue lines are the LB simulation results.}
\label{u_v_T_x_k1_NF}
\end{figure}

\begin{figure}[H]
\centering
\begin{subfigure}{0.325\textwidth}
\includegraphics[trim = 0 0 0 0,clip, width = 50mm]{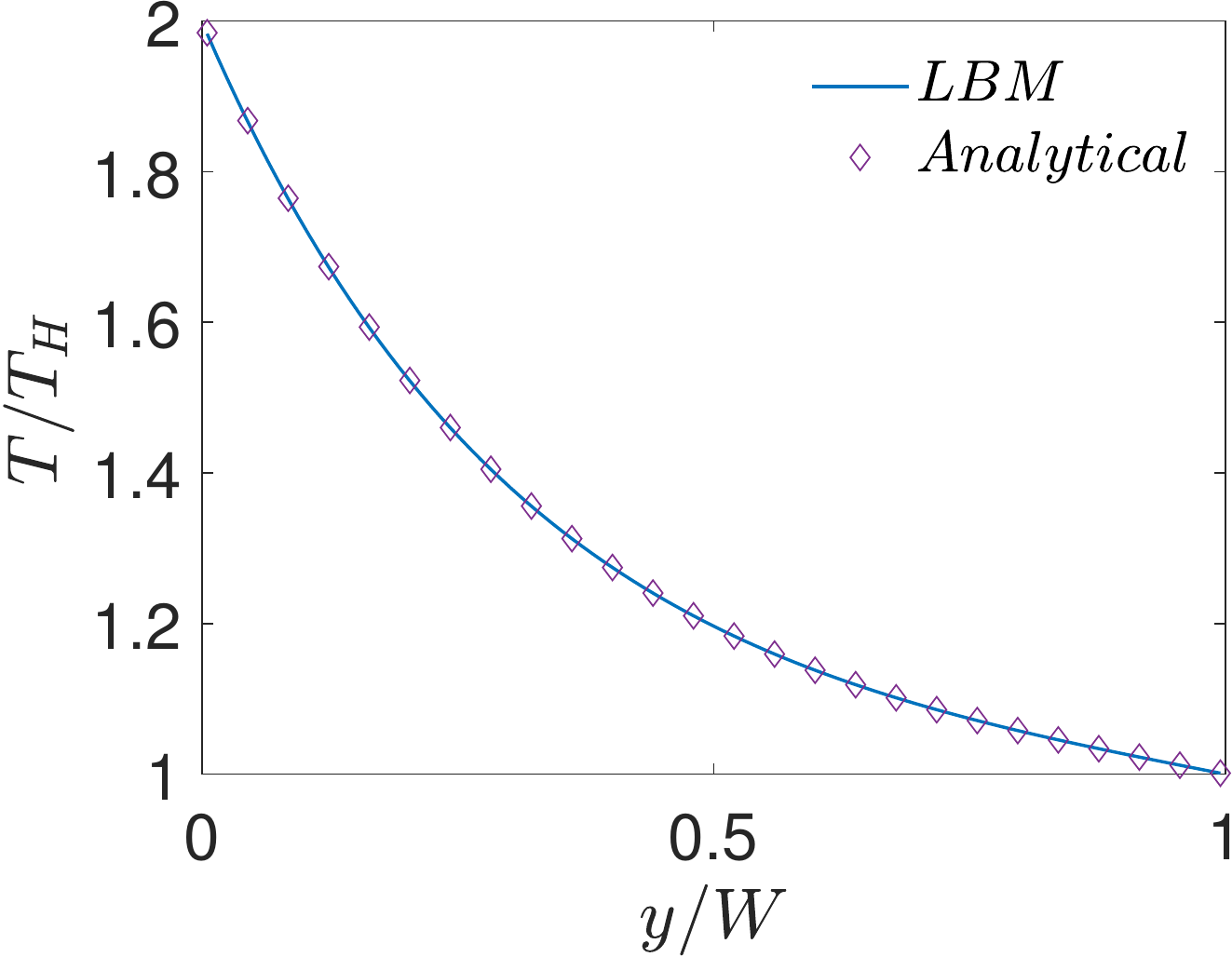}
\end{subfigure}
\begin{subfigure}{0.325\textwidth}
\includegraphics[trim = 0 0 0 0,clip, width = 50mm]{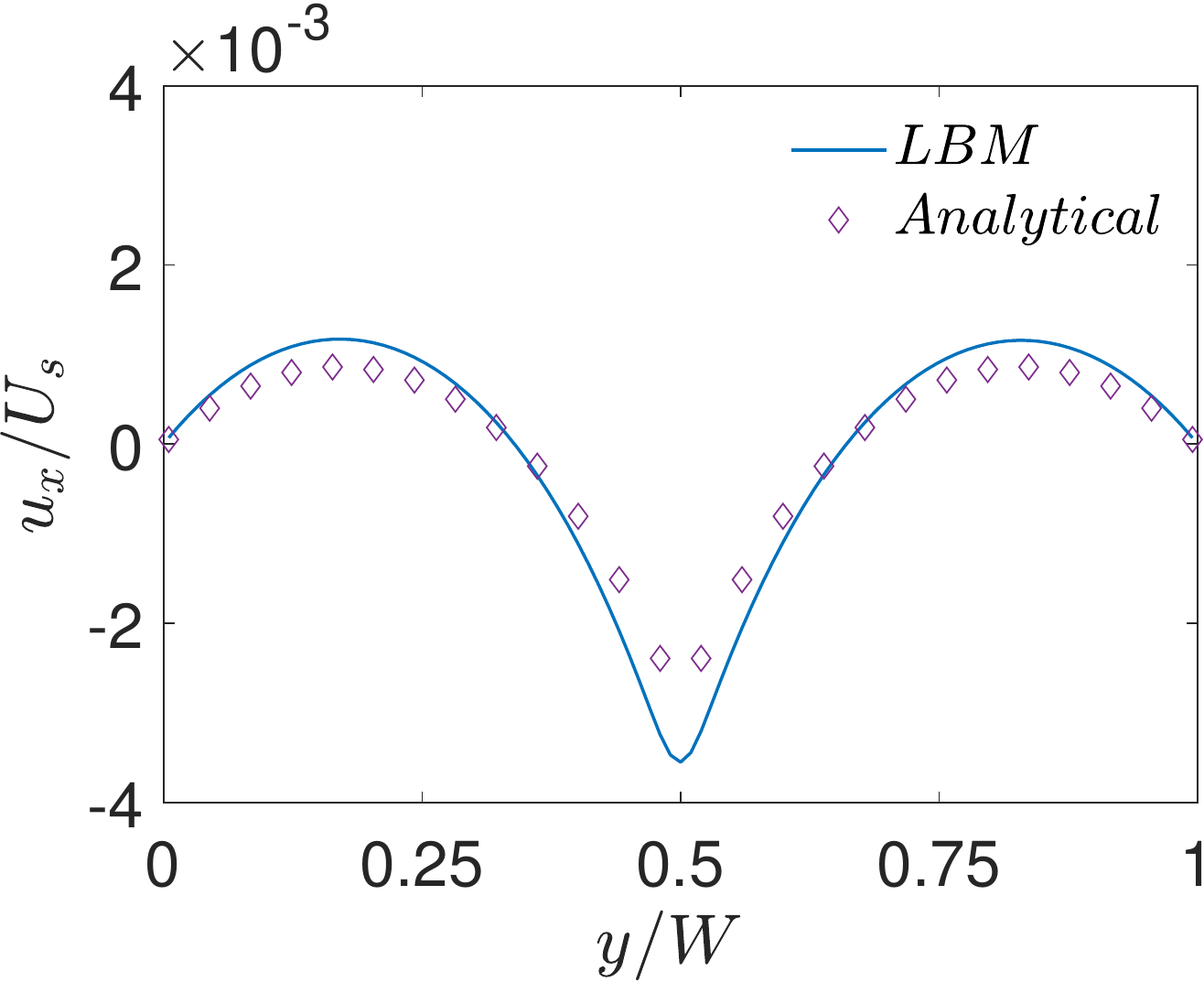}
\end{subfigure}
\begin{subfigure}{0.325\textwidth}
\includegraphics[trim = 0 0 0 0,clip, width = 50mm]{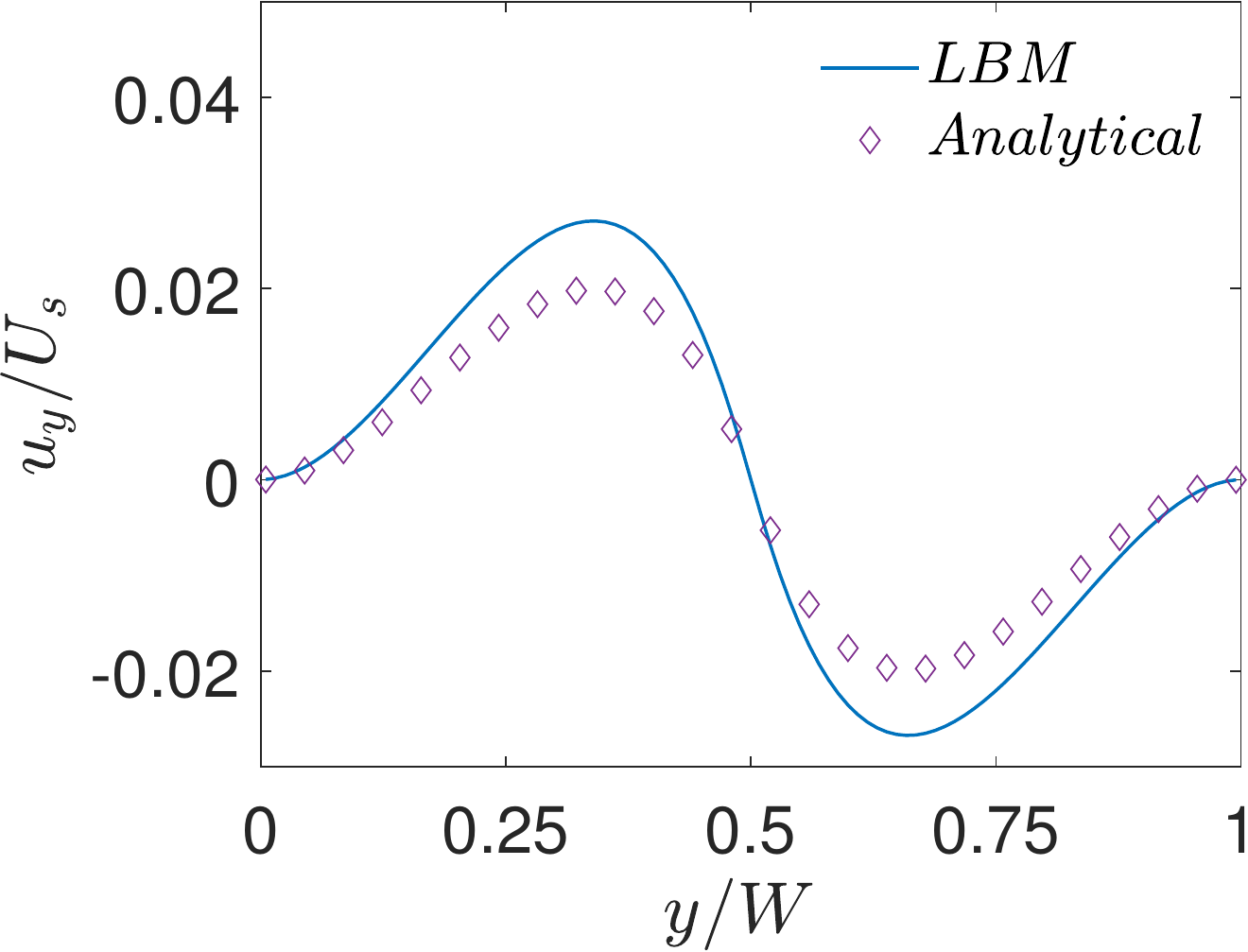}
\end{subfigure}
\caption{Profiles of the temperature and velocity components along the centerline of the domain in the $y$ direction for thermocapillary flow of a NF in a heated microchannel. The purple diamond symbols shown are obtained from the analytical solution given by~\cite{pendse2010analytical} and the blue lines are the LB simulation results.}
\label{u_v_T_y_k1_NF}
\end{figure}

Figure~\ref{temp_k1_NF} shows the equispaced contours of the temperature field for $\tilde{k} = 1$ and $\tilde{\mu} = 1$ obtained by the LB simulation as well as from the analytical solution~\cite{pendse2010analytical}; Moreover, Fig.~\ref{velocity_k1_NF} provides a similar comparison of the thermocapillary velocity vectors which shows that the fluid motion occurring in the direction away from the higher temperature zones on the interface as would be expected for NFs. Clearly, the simulation results agree well with the analytical solution.
The overall flow pattern for the thermocapillary convection in NFs is shown in Fig.~\ref{streamlines_k1_NF}, which consists of four periodic counter-rotating vortices in the two superimposed fluids. The numerical results based on the LB schemes are seen to be qualitatively consistent with the analytical solution~\cite{pendse2010analytical} for the streamline contours.
Finally, Figs.~\ref{u_v_T_x_k1_NF} and~\ref{u_v_T_y_k1_NF} present quantitative comparisons between our numerical approach and the analytical solution for the profiles of the temperature and the components of the thermocapillary velocity field along the centerlines of the domain in both the horizontal ($x$) and vertical ($y$) directions, respectively. In these figures, the temperature profiles are non-dimensionalized using the bottom wall temperature, while the velocity profiles are normalized using a characteristic velocity scale $U_s$ given in Eq.~(\ref{U_s}) by setting $\sigma_{TT}=0$ which is appropriate for NFs considered here. Again, they are fairly in good agreement with each other, thereby validating the implementation of our LB schemes presented earlier.

\section{Results and Discussion} \label{Results and Discussion}
We will now study the effect of various characteristic parameters on the physics of thermocapillary convection in superimposed layers of two self-rewetting fluids (SRFs) confined with a microchannel, where the bottom wall is nonuniformly heated by imposing a sinusoidal variation in temperature, while the top wall is maintained at a lower, but uniform temperature (see Fig.~\ref{Model_setup}). In this regard, we will utilize the new analytical solution developed in Sec.~\ref{Sec.2} and consider cases, where the quadratic coefficient of the surface tension variation with the temperature is non-zero, i.e., $\sigma_{TT}\neq 0$ or $M_2\neq 0$ to demonstrate the role of SRFs, and compare both the qualitative and quantitative differences in their behavior when compared with the normal fluids, where only the linear coefficient exists. Also, the LB schemes, which were validated in the previous section, will be used in conjunction with the analytical solution for providing additional confirmation of the applicability as well as for ensuring quantitative accuracy of the latter via numerical simulations of thermocapillary-driven flows in SRFs. For clarification, it suffices to mention that three distribution functions are used in our LB formulation to compute the two-fluid motion, interface capturing, as well as the transport of the energy within the SRFs. The temperature-dependent surface tension, which is used as a regularized volumetric body force in a single-field formulation for the fluid motion is used to couple the latter with the temperature field. The values of the bulk fluid properties such as the viscosity and the thermal conductivity are chosen that the resulting flow and transport occurs in the creeping regime and at small Marangoni and capillary numbers (i.e., $\mbox{Re}\ll 1$, $\mbox{Ma}\ll 1$, and $\mbox{Ca}\ll 1$) in numerical simulations, where the latter also ensures that the interface is naturally maintained as flat, which is valid in situations in microchannel flows.

We perform simulations in a 2D computational domain with $200 \times 100$, thereby the length $l$ of the microchannel is $200$ and the total thickness of both the SRFs ($a+b$) is $100$. Periodic boundary conditions are used in the horizontal direction, while no-slip boundary conditions are imposed on the top and bottom walls, and the wall temperatures are applied from Eqs.~(\ref{two}) and (\ref{one}) where we choose $T_h=T_c=T_{ref}=\Delta T=1.0$ for simplicity. The reference surface tension is taken is $\sigma_0=1\times 10^{-2}$. Thermocapillary flow patterns and their strengths are determined by the choice of the dimensionless linear and quadratic coefficients of the surface tension variation with temperature, i.e., $M_1$ and $M_2$, respectively. For the model parameters in the conservative ACE for interface tracking, we chose $W = 5$ and $M_{\phi} = 0.02$.

\begin{figure}[H]
\centering
\begin{subfigure}{0.475\textwidth}
\includegraphics[trim = 0 0 0 0,clip, width = 72mm]{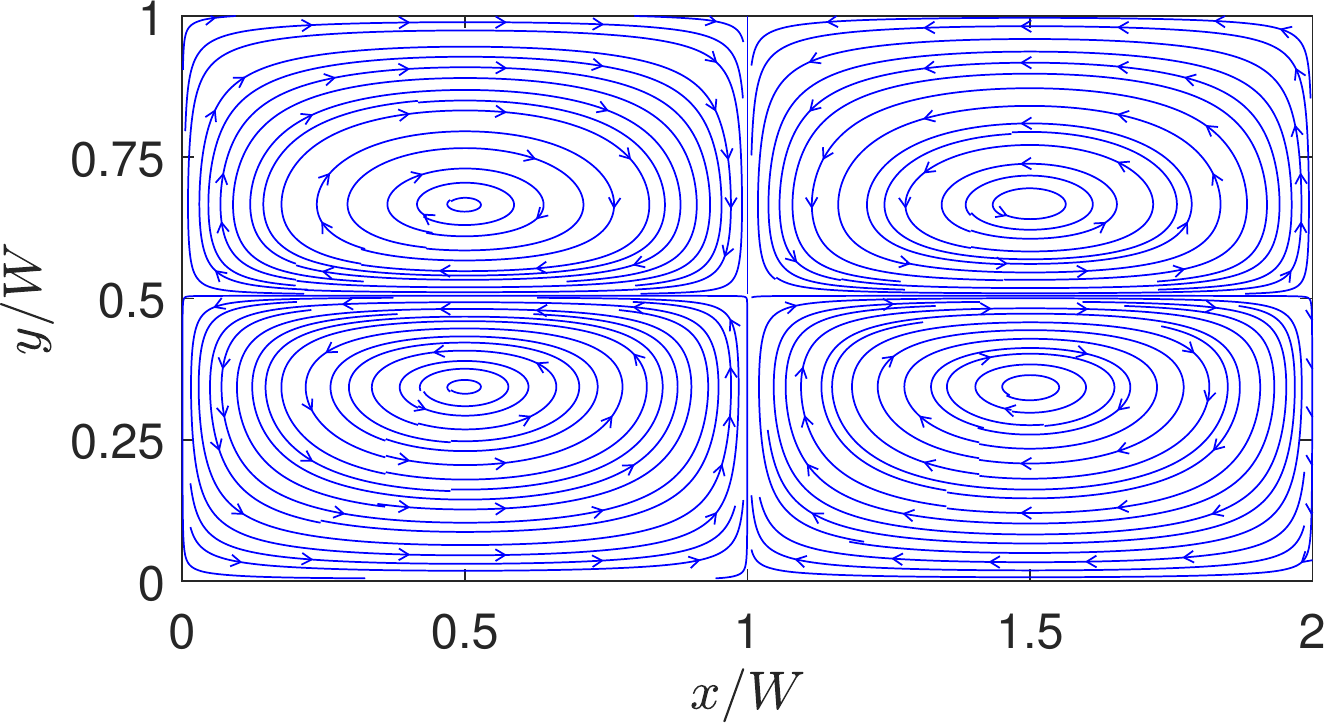}
\caption{Analytical Solution}
\end{subfigure}
\begin{subfigure}{0.475\textwidth}
\includegraphics[trim = 0 0 0 0,clip, width = 72mm]{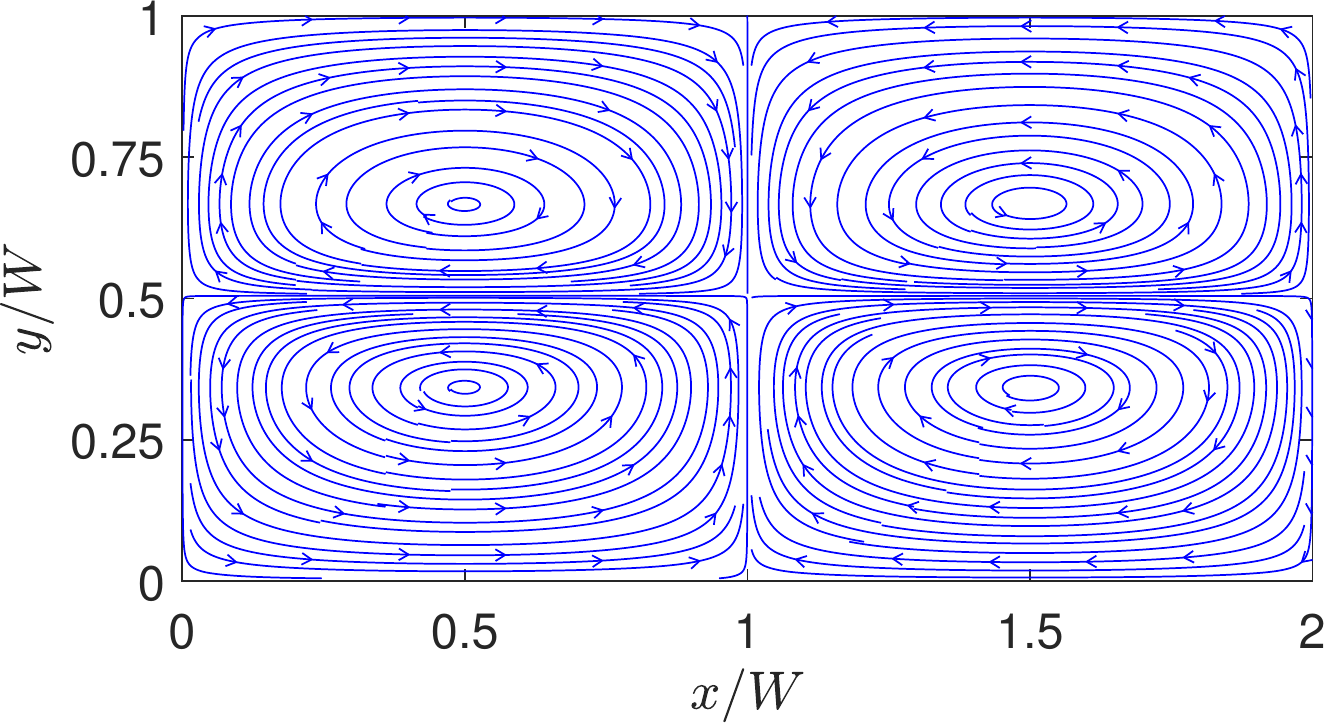}
\caption{LBM Simulation}
\end{subfigure}
\caption{Comparison of the streamlines between the analytical solution with the LBM simulation of thermocapillary convection in NFs for the case of aspect ratio of $a/b = 1$, thermal conductivity ratio of $\tilde{k} = 1$, and viscosity ratio of $\tilde{\mu} = 1$. Here, the dimensionless linear and quadratic coefficients of surface tension variation with temperature are $M_1=-5\times10^{-2}$ and $M_2=0$, respectively.}
\label{streamlines_NF}
\end{figure}

First, we consider cases with two superimposed fluids having the same thickness or $a/b=1$ and with property ratios $\tilde{k}=1$ and $\tilde{\mu}=1$. To provide a perspective and a basis for comparison, we will first show the streamlines a case with NFs in Fig.~\ref{streamlines_NF} by considering $M_1=-5\times 10^{-2}$ and $M_2=0.0$. We treat these choices for the dimensionless surface tension coefficients as the baseline case for NFs. Moreover, the choices of the other fluid properties are such that $\mbox{Re} = 1.59\times 10^{-1}$, $\mbox{Ma}=3.83\times 10^{-2}$, and $\mbox{Ca}=1.26\times 10^{-2}$. In defining these dimensionless parameters here and in what follows, a characteristic velocity $U_s$ derived in Appendix~\ref{AppendixC} is used. Clearly, if the quadratic coefficient for the surface tension is absent (i.e., $\sigma_{TT}=0$ or $M_2=0$), then \emph{four periodic counterrotating vortices} are induced, where the fluids \emph{move away from the hotter region on the interface} at the center of the domain.

\begin{figure}[H]
\centering
\begin{subfigure}{0.475\textwidth}
\includegraphics[trim = 0 0 0 0,clip, width = 72mm]{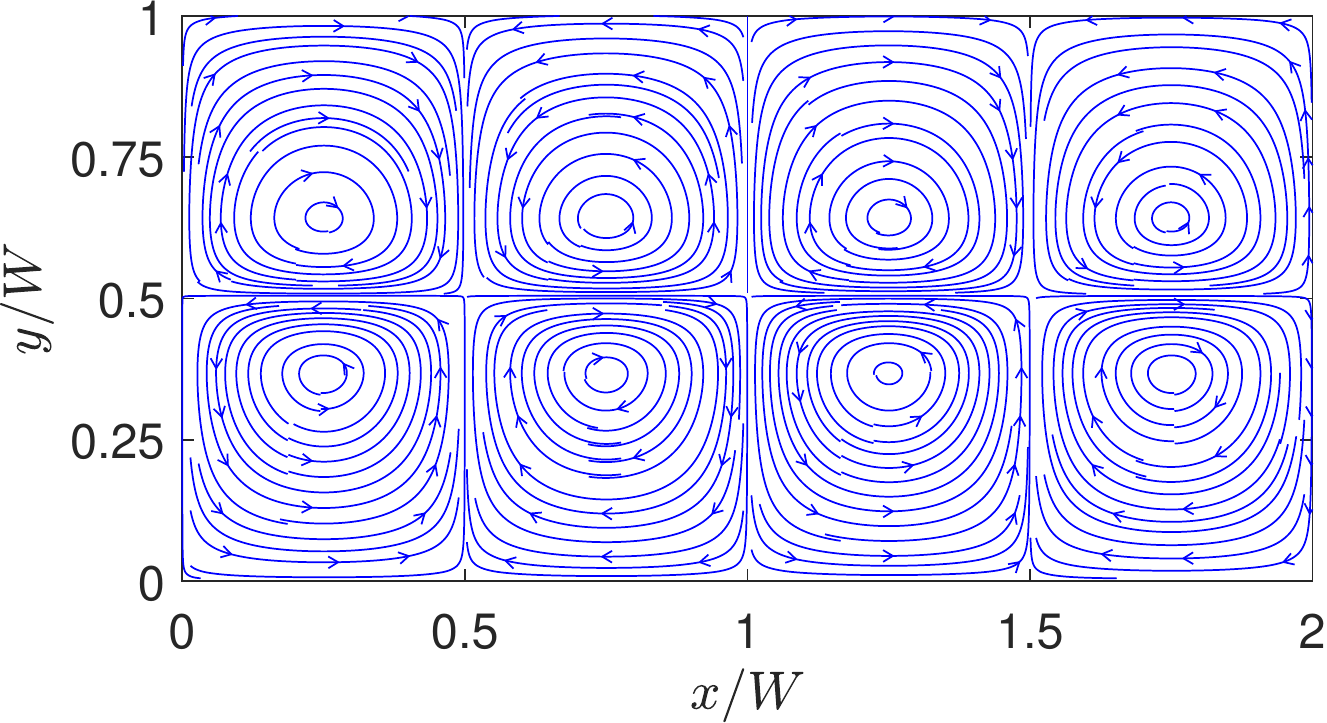}
\caption{Analytical Solution}
\end{subfigure}
\begin{subfigure}{0.475\textwidth}
\includegraphics[trim = 0 0 0 0,clip, width = 72mm]{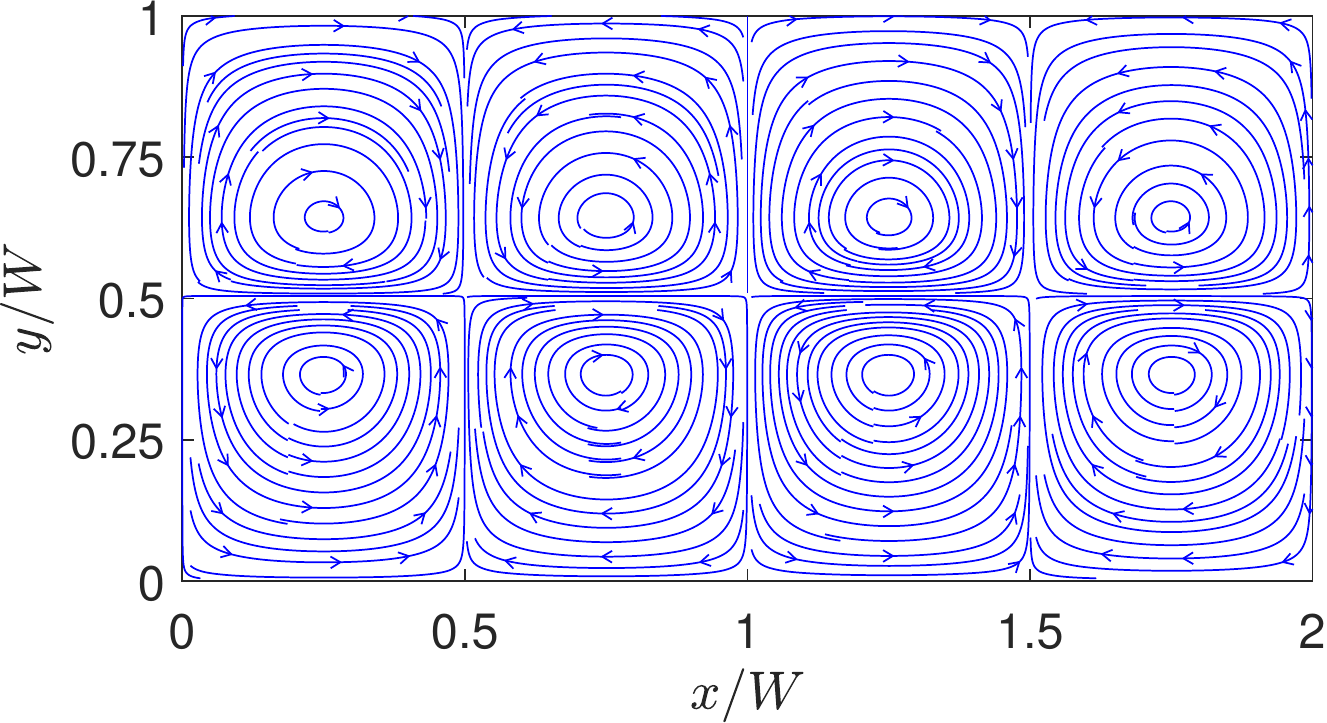}
\caption{LBM Simulation}
\end{subfigure}
\caption{Comparison of the streamlines between the analytical solution with the LBM simulation of thermocapillary convection in SRFs for the case of aspect ratio of $a/b = 1$, thermal conductivity ratio of $\tilde{k} = 1$, and viscosity ratio of $\tilde{\mu} = 1$. Here, the dimensionless linear and quadratic coefficients of surface tension variation with temperature are $M_1=0$ and $M_2=1\times 10^{-1}$, respectively.}
\label{streamlines_SRF}
\end{figure}

\begin{figure}[H]
\centering
\begin{subfigure}{0.475\textwidth}
\includegraphics[trim = 0 0 0 0,clip, width = 72mm]{NF_VF_Test1}
\caption{Normal Fluid}
\end{subfigure}
\begin{subfigure}{0.475\textwidth}
\includegraphics[trim = 0 0 0 0,clip, width = 72mm]{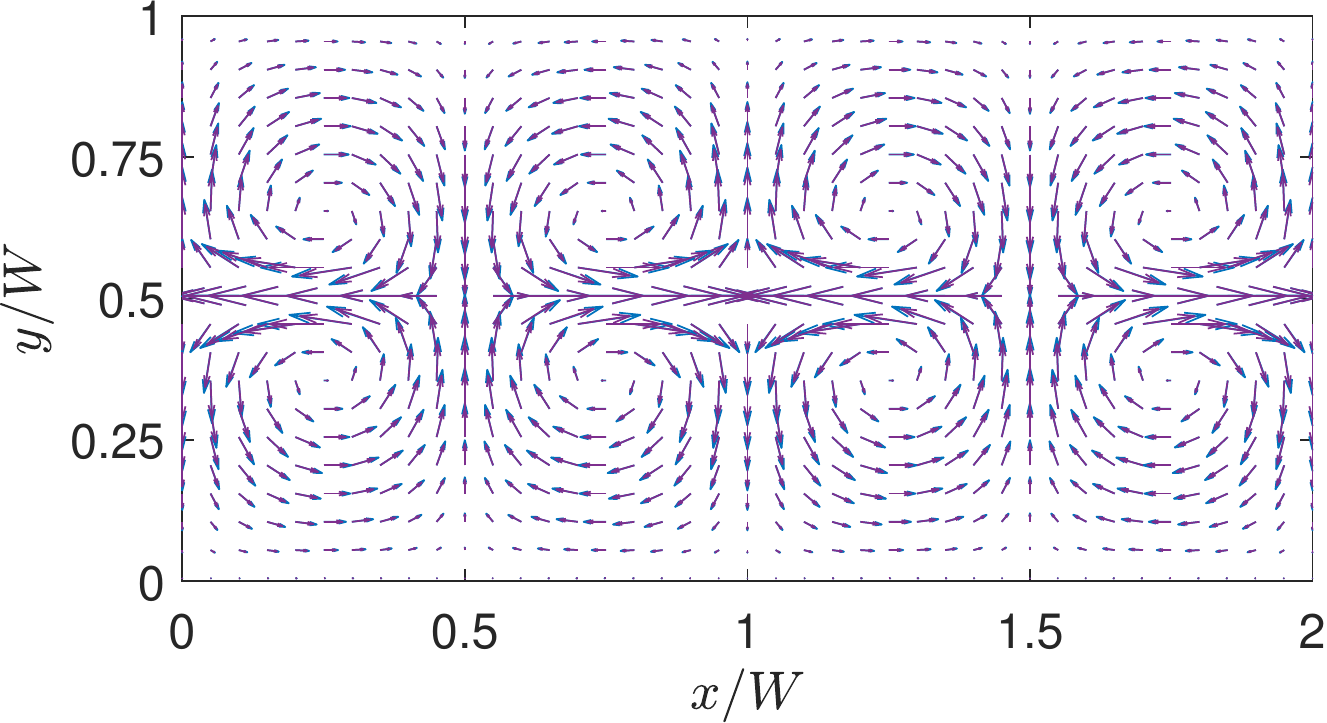}
\caption{Self Re-wetting Fluid}
\end{subfigure}
\caption{Velocity vectors due to thermocapillary convection for the case of (a) NFs ($M_1=-5\times10^{-2}$ and $M_2=0$) and (b) SRFs ($M_1=0$ and $M_2=1\times 10^{-1}$). Here, the aspect ratio is $a/b = 1$, thermal conductivity ratio is $\tilde{k} = 1$, and the viscosity ratio is $\tilde{\mu} = 1$. The blue arrows are for the LBM simulation while the purple arrows are for the analytical solution.}
\label{Vel_Vectors}
\end{figure}

On the other hand, by turning off the linear coefficient of surface tension (i.e., $\sigma_T=0$) and keeping only the quadratic coefficient non-zero, i.e., $\sigma_{TT}\neq 0$, for otherwise the same property ratios, we simulate the thermocapillary convection in SRFs. In dimensionless form, we take $M_1=0$ and $M_2=1\times 10^{-1}$, which we consider as the choices for the baseline case for SRFs; the rest of the dimensionless parameters resulting from specifying the other fluid properties are $\mbox{Re} = 1\times 10^{-1}$, $\mbox{Ma}=3\times 10^{-2}$, and $\mbox{Ca}=9.9\times 10^{-3}$. The results given in terms of the streamlines are plotted in Fig.~\ref{streamlines_SRF}. It is evident that the thermocapillary flow pattern in SRFs is strikingly different from that in NFs: First, \emph{eight periodic counterrotating vortices} are generated in SRFs, which is double the number of the convection rolls in NFs. Second, the fluids on the interface seek to \emph{move towards the hotter region on the interface} at the center of the domain. Such differences in direction of the thermocapillary flow fields between the NFs and SRFs are more explicit in the velocity vector diagrams shown in Fig.~\ref{Vel_Vectors}, which is a manifestation of flow arising from the Marangoni stress generated due to a positive (negative) surface tension gradient on the interface for SRFs (NFs). The doubling of the vortical structures in the case of the SRFs can be interpreted from an earlier and simpler form of the analytical solution given in terms of the streamfunction in Eq.~(\ref{eq:formofstreamfunction}): this equation contains the `fundamental solution' related to $\sin(\omega x)$ arising from the linear part surface tension coefficient $\sigma_T$ and a `first order harmonic solution' related to $\sin(\omega x)\cos(\omega x)$ generated from the quadratic part surface tension coefficient $\sigma_{TT}$. The latter is of the form $\sin(2\omega x)/2$, which has double the wavenumber compared to the former case. Thus, fluids with surface tension such that $\sigma_{TT}\neq 0$ (or SRFs) would result in double the number of thermocapillary convection rolls when compared to fluids with only linear variations in the surface tension, i.e., only $\sigma_T\neq 0$ (or NFs).
Finally, we note that in all cases, the side-by-side comparisons between the analytical solution and the numerical results based on the LB schemes show very good agreement with each other.

\subsection{Effect of relative magnitudes of dimensionless linear $M_1$ and quadratic $M_2$ surface tension coefficients of SRF layers}

\begin{figure}[H]
\centering
\begin{subfigure}{0.475\textwidth}
\includegraphics[trim = 0 0 0 0,clip, width = 72mm]{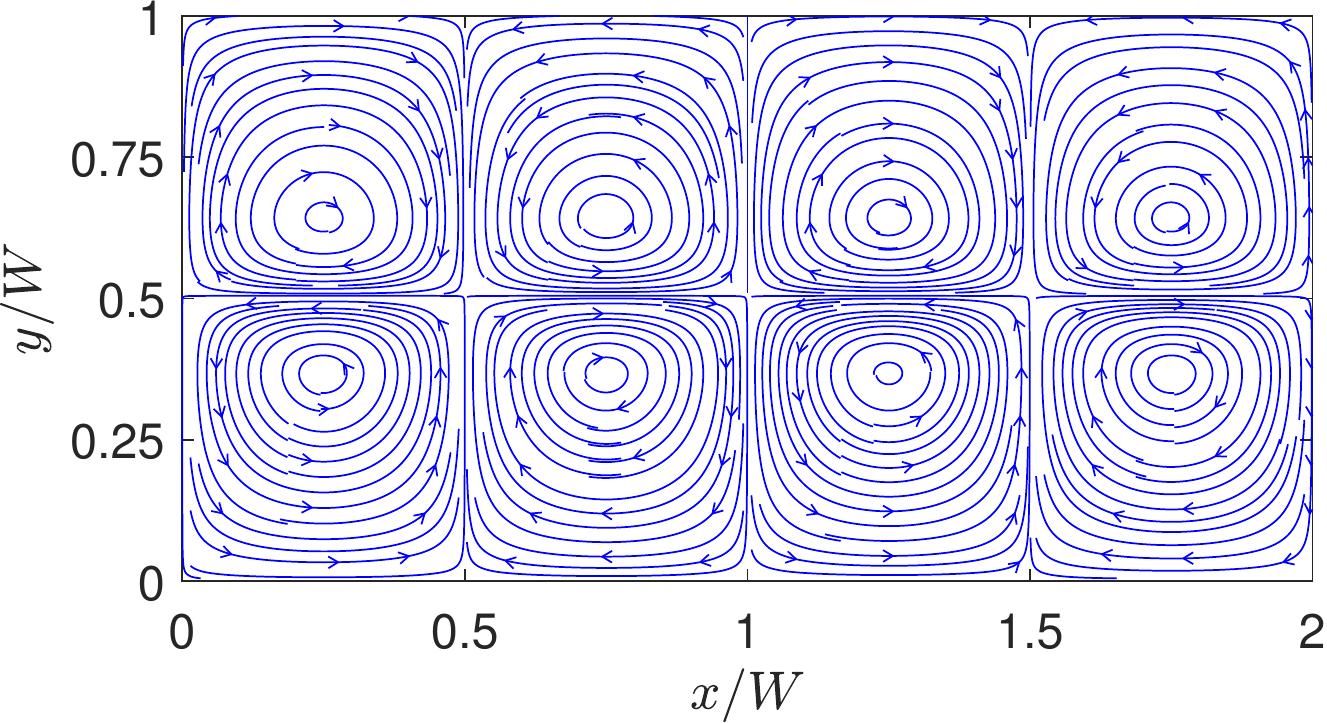}
\caption{Analytical Solution}
\end{subfigure}
\begin{subfigure}{0.475\textwidth}
\includegraphics[trim = 0 0 0 0,clip, width = 72mm]{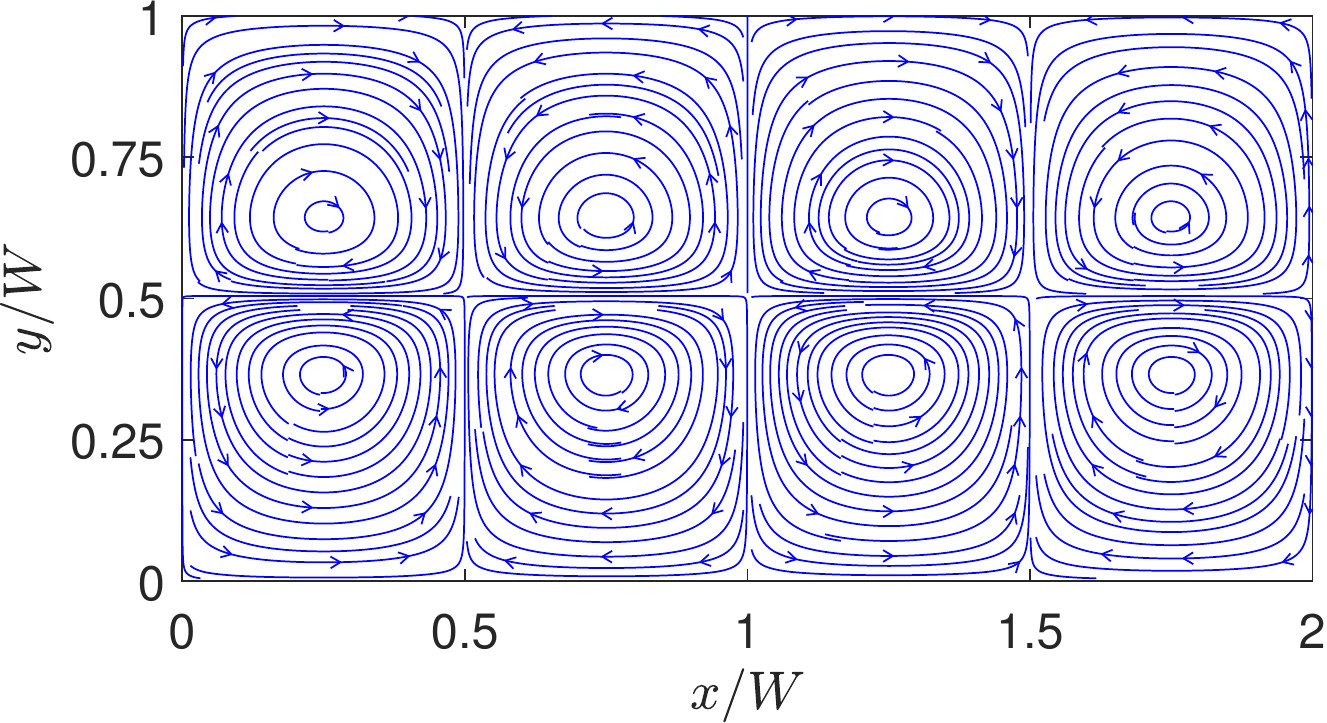}
\caption{LBM Simulation}
\end{subfigure}
\caption{Comparison of the streamlines between the analytical solution with the LBM simulation of thermocapillary convection in SRFs for the case of aspect ratio of $a/b = 1$, thermal conductivity ratio of $\tilde{k} = 1$, and viscosity ratio of $\tilde{\mu} = 1$. Here, the dimensionless linear and quadratic coefficients of surface tension variation with temperature are $M_1=1 \times 10^{-5}$ and $M_2=5 \times 10^{-1}$, respectively.}
\label{Real_Fluid_SRF}
\end{figure}

In the previous case, we considered a particular type of SRF for which only the quadratic coefficient is non-zero, while the linear part of the coefficient is absent, i.e., $M_2\neq 0$, but $M_1 = 0$. While this is a plausible assumption, it doesn't encompass all types of SRFs, for which it is possible to have both $M_2\neq 0$ and $M_1\neq 0$, and the unique nature of the flow patterns associated with the SRFs can still be manifested provided that the overall surface tension gradient is positive in the flow domain of interest. To test this hypothesis, we used the following parameters for SRFs with both the linear and quadratic coefficients: $\sigma_{0} = 1\times 10^{-3}$, $M_1=1 \times 10^{-5}$, and $M_2=5 \times 10^{-1}$. Based on such more general forms of surface tension coefficients, results from the analytical solution and well as the LB simulations are obtained and the corresponding thermocapillary convection patterns given in terms of the streamlines are presented in Fig.~\ref{Real_Fluid_SRF}. Again, we notice here that eight counterrotating  convection rolls are generated, where the fluid motion along the interface is directed towards the higher temperatures, which confirms our hypothesis mentioned above. Moreover, the theoretical prediction is consistent with the numerical results based on LB schemes.

\begin{figure}[H]
\centering
\begin{subfigure}{0.475\textwidth}
\includegraphics[trim = 0 0 0 0,clip, width = 72mm]{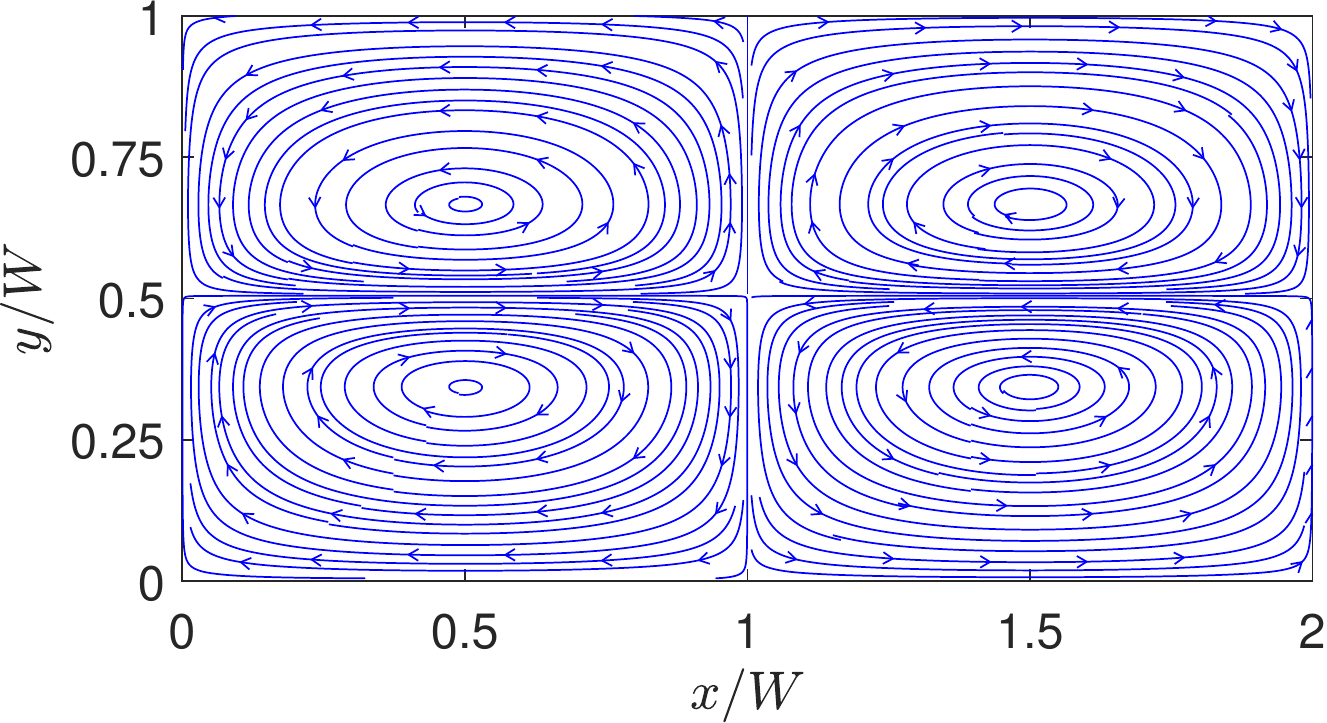}
\caption{Analytical Solution}
\end{subfigure}
\begin{subfigure}{0.475\textwidth}
\includegraphics[trim = 0 0 0 0,clip, width = 72mm]{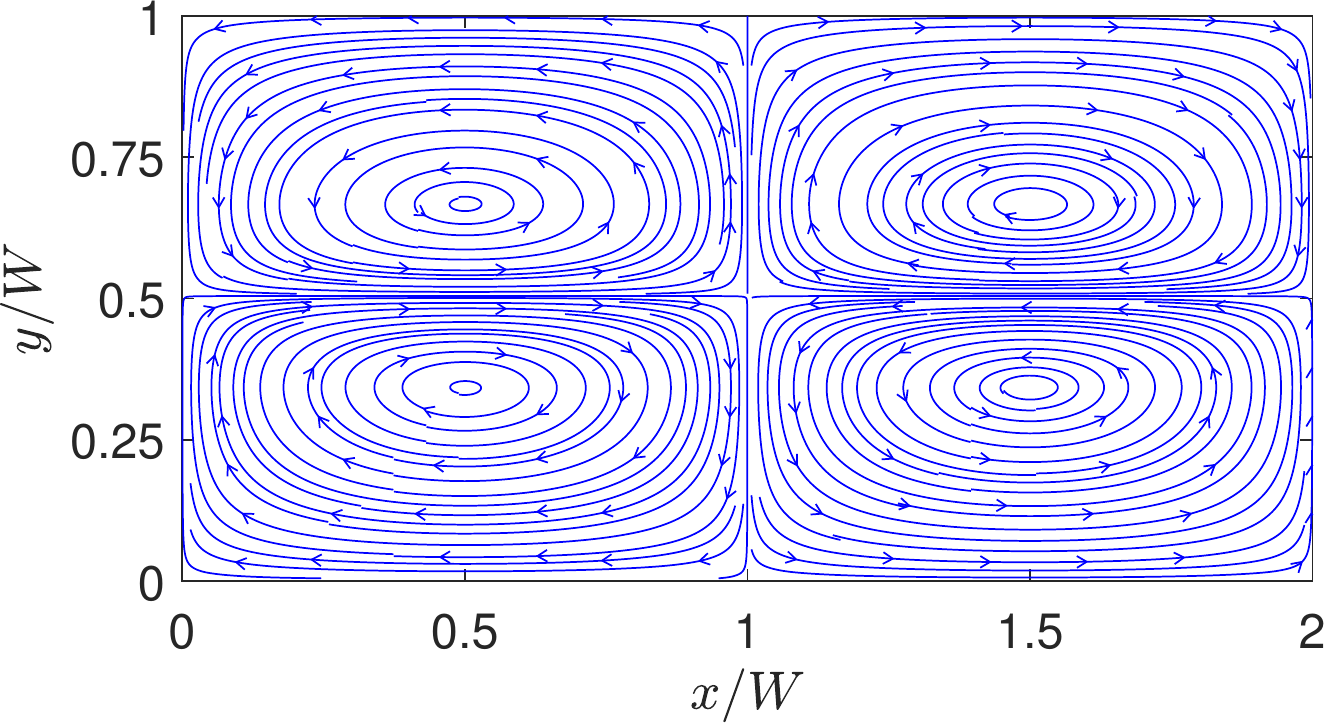}
\caption{LBM Simulation}
\end{subfigure}
\caption{Comparison of the streamlines between the analytical solution with the LBM simulation of thermocapillary convection in SRFs for the case of aspect ratio of $a/b = 1$, thermal conductivity ratio of $\tilde{k} = 1$, and viscosity ratio of $\tilde{\mu} = 1$. Here, the dimensionless linear and quadratic coefficients of surface tension variation with temperature are $M_1=1 \times 10^{-1}$ and $M_2=1 \times 10^{-4}$, respectively.}
\label{Real_Fluid_SRF-case2}
\end{figure}

This last example concerned a situation where the dimensionless linear coefficient of surface tension is much smaller than that of the quadratic coefficient. Let's now explore another case by inverting this situation where the linear coefficient is much larger than the quadratic coefficient in SRF layers. In particular, we take $M_1=1 \times 10^{-1}$ and $M_2=1 \times 10^{-4}$, and the streamline patterns based on the analytical solution and the LBM simulation results are shown in Fig.~\ref{Real_Fluid_SRF-case2}. Interestingly, in this case only four periodic counterrotating vortices are generated; however, unlike those observed for the NFs in the previous section where the fluids on the interface move away from the center (see Fig.~\ref{streamlines_NF}), here the thermocapillary motion along the interface is directed towards the higher temperature zones at the center of the microchannel, which is consistent withe expected behavior of SRFs. Now, the presence of four vortices for the present case where $M_2\ll M_1$ and eight vortices for the previous case where $M_2\gg M_1$ can be explained as follows. The analytical solution derived in a previous section consists of the superposition of two results: one that arises from the linear coefficient of the surface tension $M_1$ and the other is generated from the quadratic coefficient $M_2$, which contains contribution of thermocapillary flow with a wavenumber that is twice as the former case. Moreover, the resulting magnitudes of the flow in each case is proportional to the magnitude of the respective coefficient of the surface tension. Thus, the overall solution, in terms of the dominant flow pattern, is then dictated by the contribution of the part of the solution which has the largest magnitude arising between the two surface tension coefficients.

\begin{figure}[H]
\centering
\includegraphics[trim = 0 0 0 0,clip, width = 100mm]{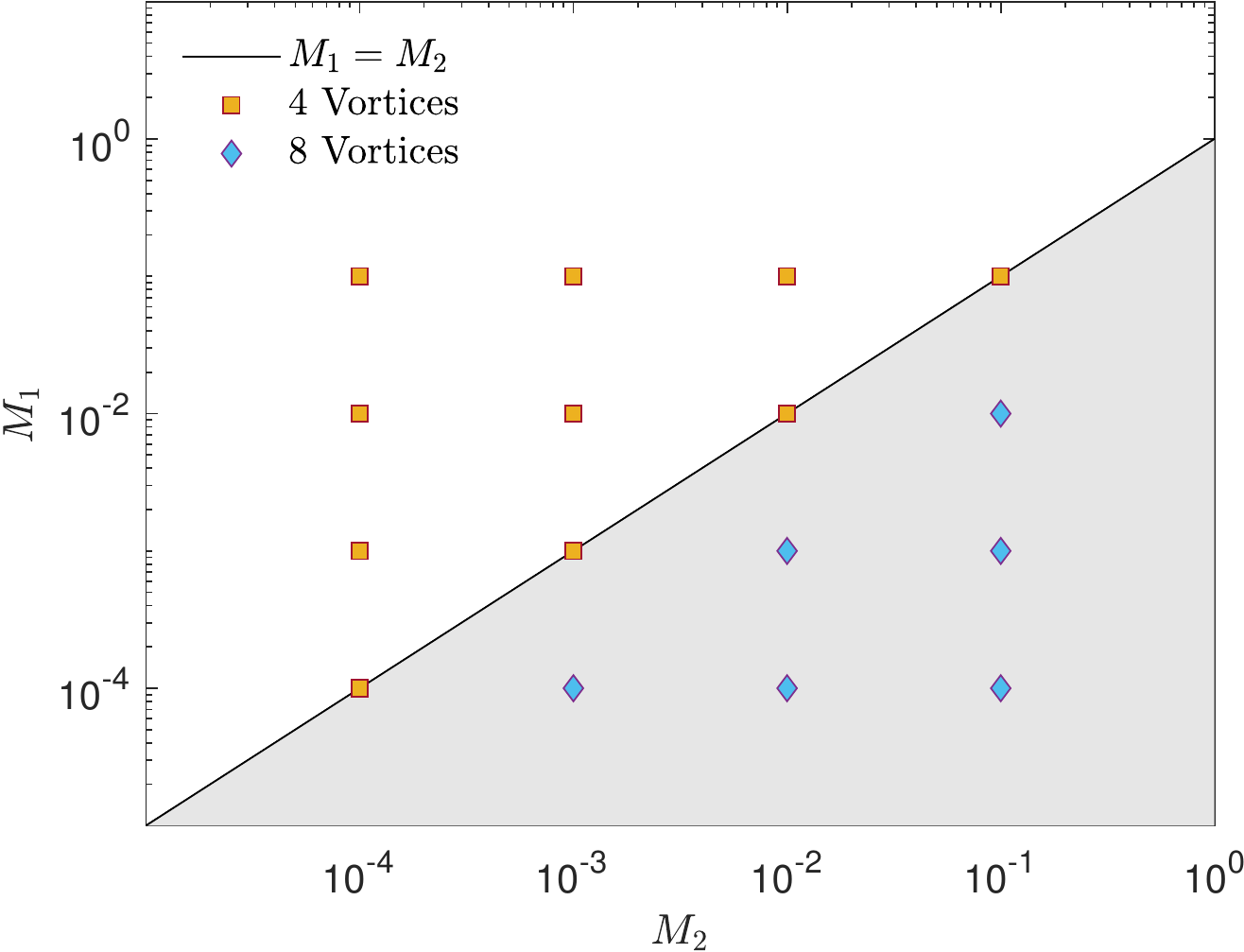}
\caption{Parametric regime map given in terms of the dimensionless linear $M_1$ and quadratic $M_2$ surface tension coefficients for the four and eight vortex convection roll cases induced by thermocapillary effects in SRFs in a nonuniformly heated microchannel. Here, the aspect ratio is $a/b = 1$, the thermal conductivity ratio is $\tilde{k} = 1$, and viscosity ratio is $\tilde{\mu} = 1$. The symbols correspond to our analytical prediction, with the shaded region encompassing the eight vortex cases.}
\label{fig:parametericregimemapM1M2}
\end{figure}

Indeed, in view of the above considerations, we performed a systematic study to deduce the parameter space $M_1 - M_2$ that delineates the cases with four vortices with those of eight vortices in SRFs. Figure~\ref{fig:parametericregimemapM1M2} presents a parametric regime map in terms of the linear and quadratic surface tension coefficients, when all the other characteristic parameters are fixed as follows: $a/b = 1$, $\tilde{k} = 1$, and $\tilde{\mu} = 1$. We find for all cases where $M_2 > M_1$ with the above parametric choices, the thermocapillary convection in SRFs manifests in the form of eight counterrotating vortex cells as shown by the shaded region in Fig.~\ref{fig:parametericregimemapM1M2}; otherwise the SRFs exhibit four vortex cells. Moreover, unlike in NFs, the SRFs, regardless of the choice of $M_1$ and $M_2$, always seek to move towards the hotter regions at the center on the interface. These findings may be exploited in creating new pathways to specifically promote certain targeted mixing patterns in microfluidic channels subjected to nonuniform wall heating by tuning surface tension coefficients, i.e., by synthesizing SRFs with appropriate interfacial properties $\sigma_T$ and $\sigma_{TT}$ (or equivalently, $M_1$ and $M_2$), e.g., by selecting appropriate number of carbon atoms in the molecular chain arrangements in aqueous solutions of alcohols.

\subsection{Effect of relative thickess ratio $a/b$ of SRF layers}
\begin{figure}[H]
\centering
\begin{subfigure}{0.475\textwidth}
\includegraphics[trim = 0 0 0 0,clip, width = 72mm]{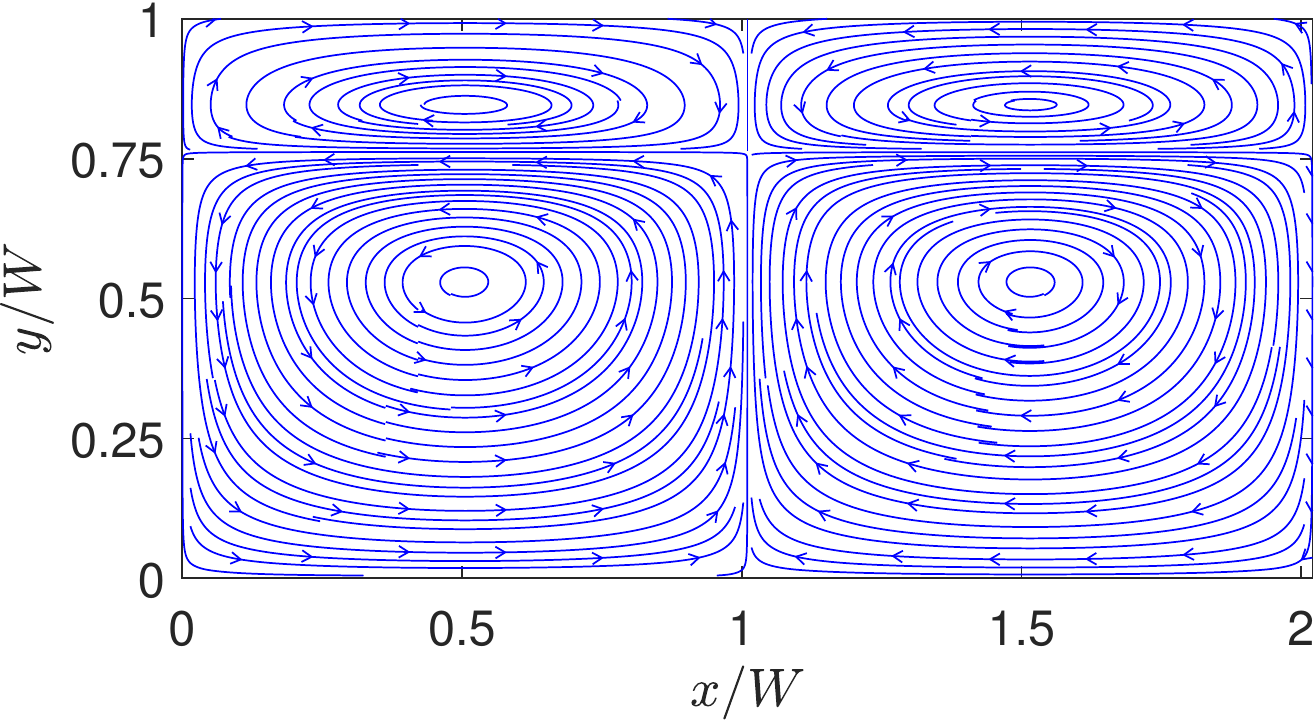}
\caption{Analytical Solution}
\end{subfigure}
\begin{subfigure}{0.475\textwidth}
\includegraphics[trim = 0 0 0 0,clip, width = 72mm]{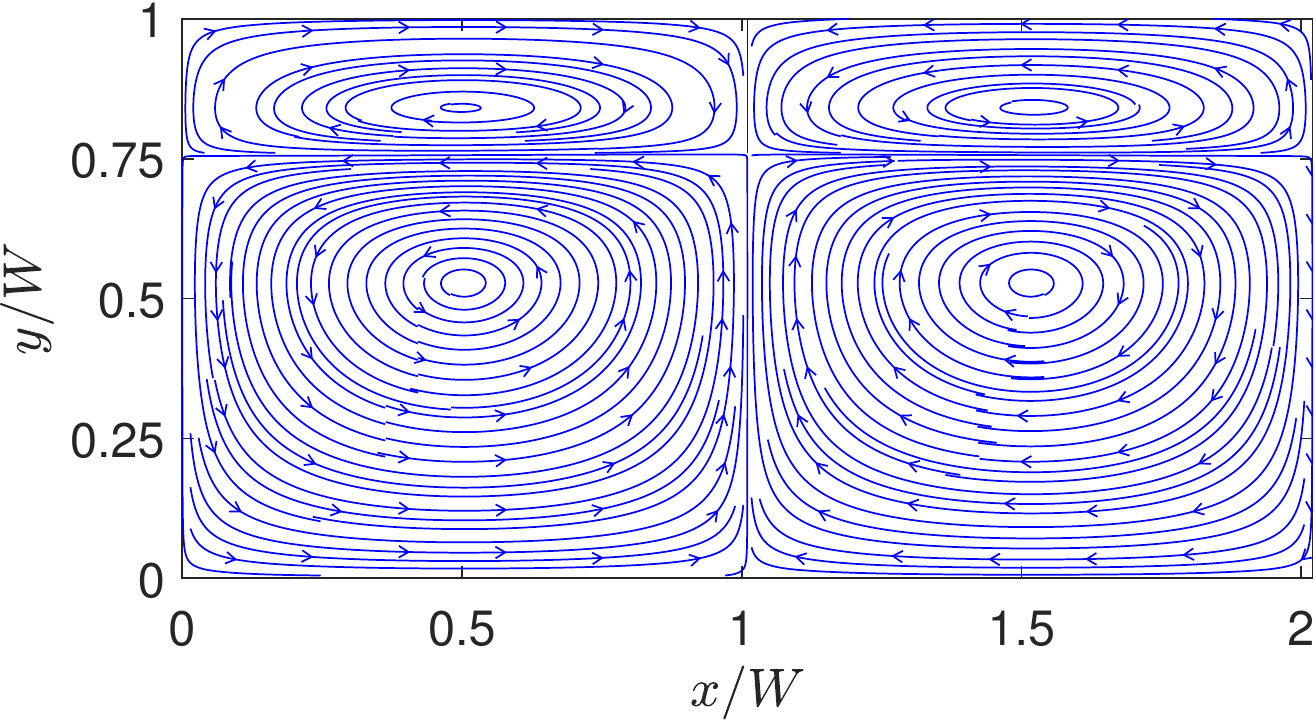}
\caption{LBM Simulation}
\end{subfigure}
\caption{Comparison of the streamlines between the analytical solution with the LBM simulation of thermocapillary convection in NFs for the case of aspect ratio of $a/b = 1/3$, thermal conductivity ratio of $\tilde{k} = 1$, and viscosity ratio of $\tilde{\mu} = 1$. Here, the dimensionless linear and quadratic coefficients of surface tension variation with temperature are $M_1=-5\times10^{-2}$ and $M_2=0$, respectively.}
\label{streamlines_NF_ab_1_3}
\end{figure}

\begin{figure}[H]
\centering
\begin{subfigure}{0.475\textwidth}
\includegraphics[trim = 0 0 0 0,clip, width = 72mm]{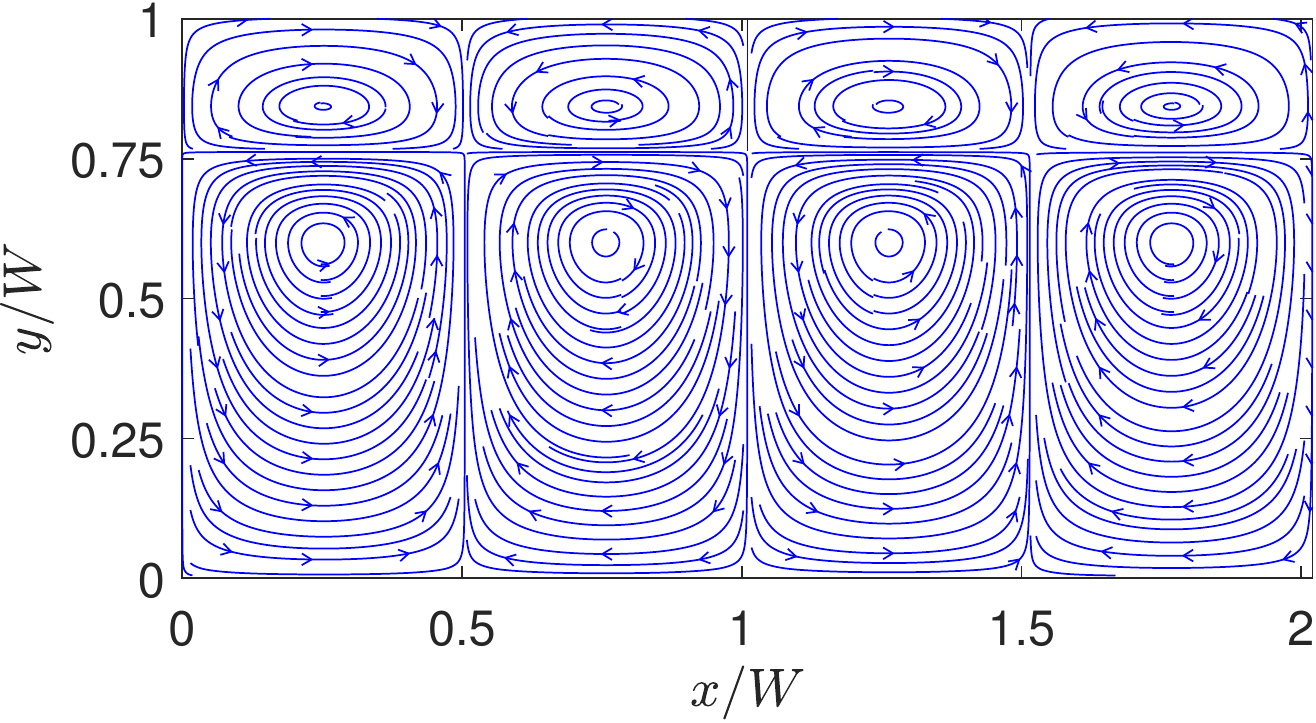}
\caption{Analytical Solution}
\end{subfigure}
\begin{subfigure}{0.475\textwidth}
\includegraphics[trim = 0 0 0 0,clip, width = 72mm]{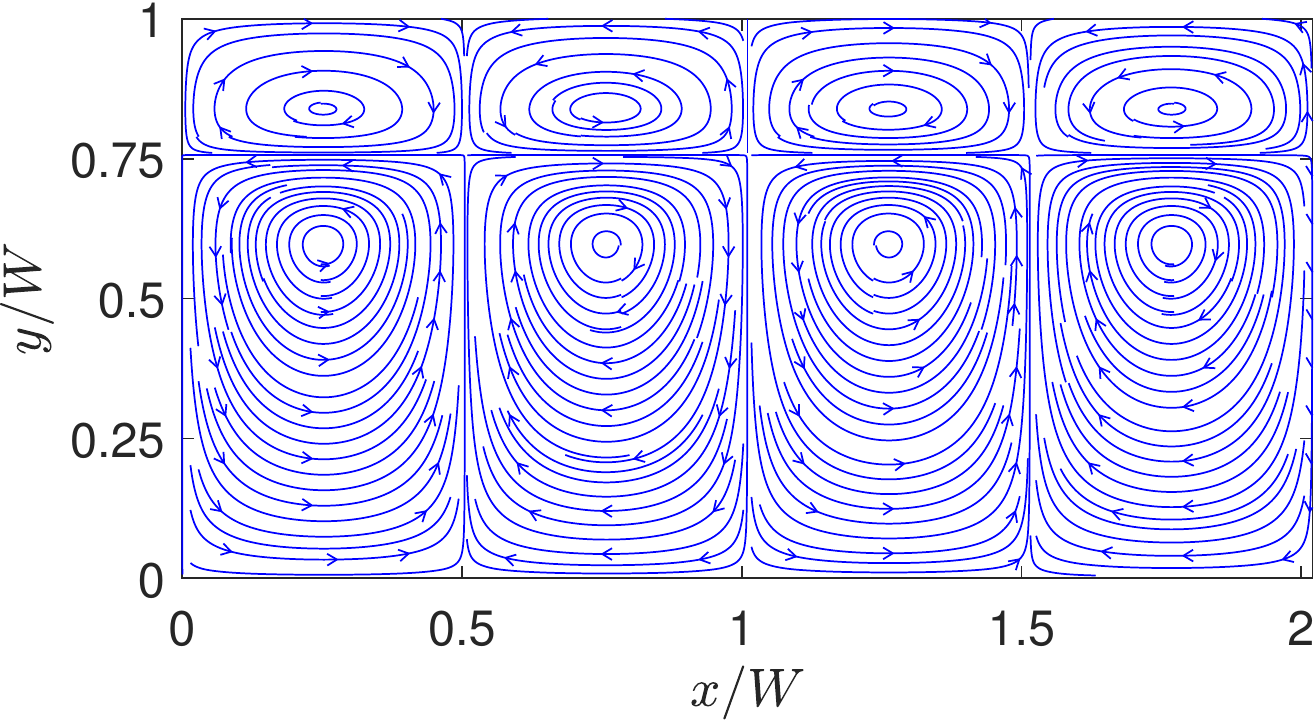}
\caption{LBM Simulation}
\end{subfigure}
\caption{Comparison of the streamlines between the analytical solution with the LBM simulation of thermocapillary convection in SRFs for the case of aspect ratio of $a/b = 1/3$, thermal conductivity ratio of $\tilde{k} = 1$, and viscosity ratio of $\tilde{\mu} = 1$. Here, the dimensionless linear and quadratic coefficients of surface tension variation with temperature are $M_1=0$ and $M_2=1\times 10^{-1}$, respectively.}
\label{streamlines_SRF_ab_1_3}
\end{figure}

Next, let's examine the effect of changing the relative thicknesses $a$ and $b$ of the top and bottom fluids or the aspect ratio $a/b$ on thermocapillary flow patterns for both NFs and SRFs. Figure~\ref{streamlines_NF_ab_1_3} shows the streamlines in NFs when the aspect ratio $a/b=1/3$, while the corresponding result for the SRFs is presented in Fig.~\ref{streamlines_SRF_ab_1_3}.

\begin{figure}[H]
\centering
\begin{subfigure}{0.475\textwidth}
\includegraphics[trim = 0 0 0 0,clip, width = 72mm]{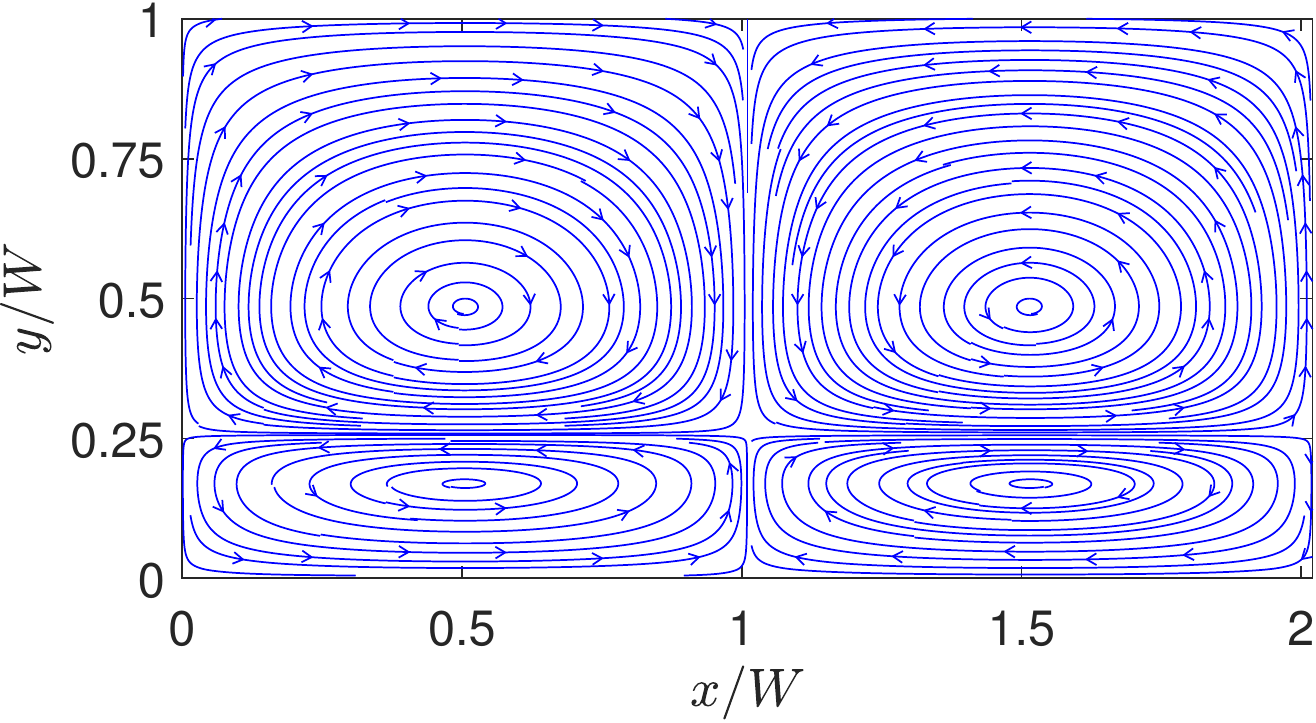}
\caption{Analytical Solution}
\end{subfigure}
\begin{subfigure}{0.475\textwidth}
\includegraphics[trim = 0 0 0 0,clip, width = 72mm]{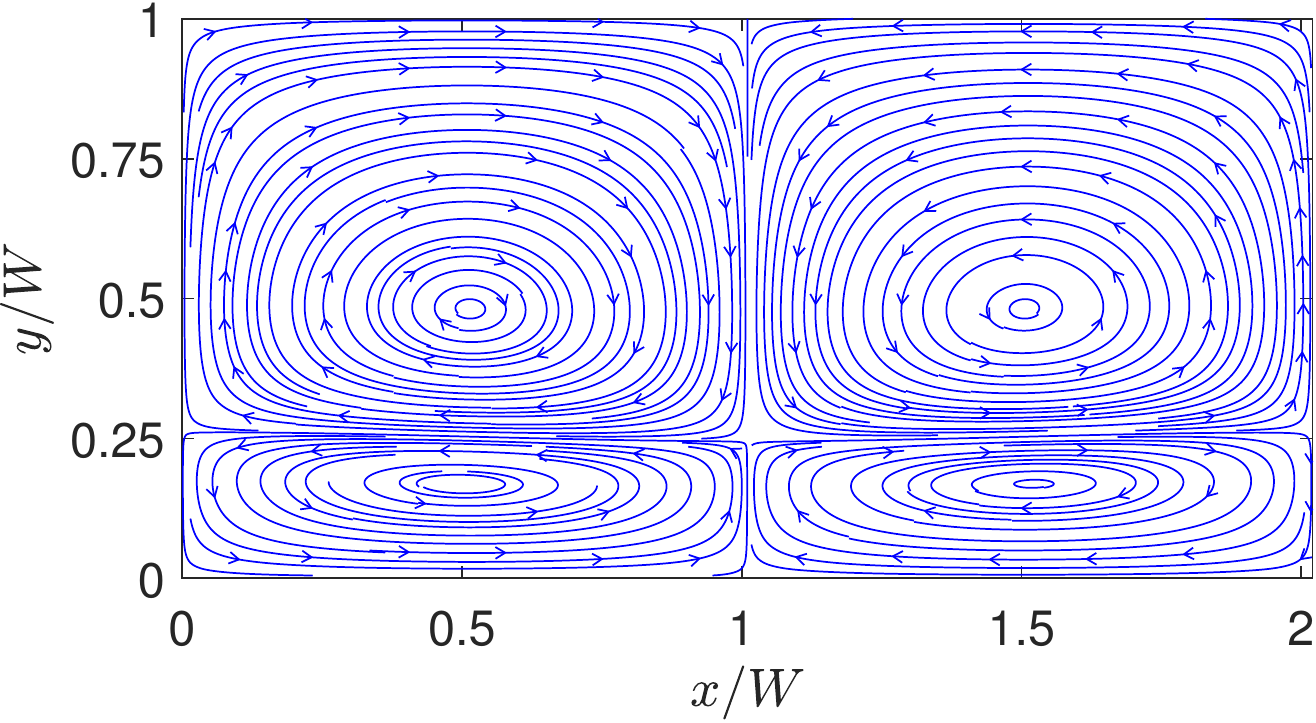}
\caption{LBM Simulation}
\end{subfigure}
\caption{Comparison of the streamlines between the analytical solution with the LBM simulation of thermocapillary convection in NFs for the case of aspect ratio of $a/b = 3$, thermal conductivity ratio of $\tilde{k} = 1$, and viscosity ratio of $\tilde{\mu} = 1$. Here, the dimensionless linear and quadratic coefficients of surface tension variation with temperature are $M_1=-5\times10^{-2}$ and $M_2=0$, respectively.}
\label{streamlines_NF_ab_3}
\end{figure}

\begin{figure}[H]
\centering
\begin{subfigure}{0.475\textwidth}
\includegraphics[trim = 0 0 0 0,clip, width = 72mm]{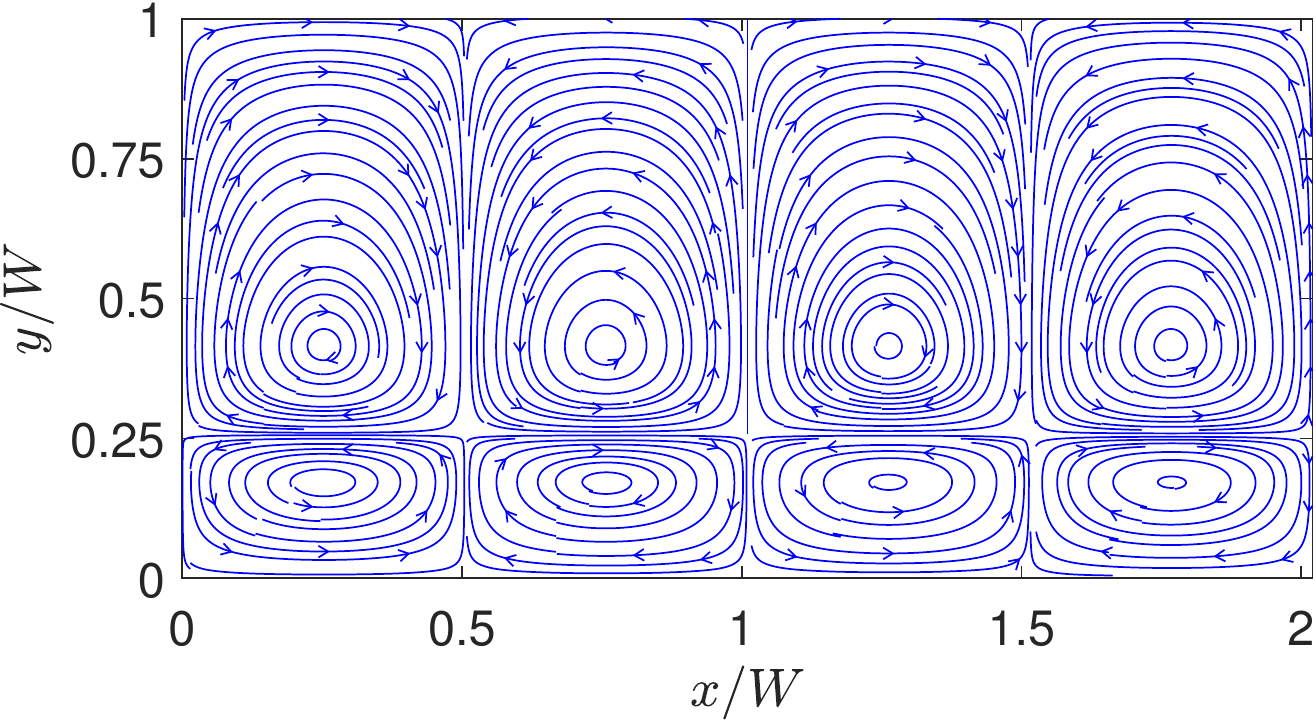}
\caption{Analytical Solution}
\end{subfigure}
\begin{subfigure}{0.475\textwidth}
\includegraphics[trim = 0 0 0 0,clip, width = 72mm]{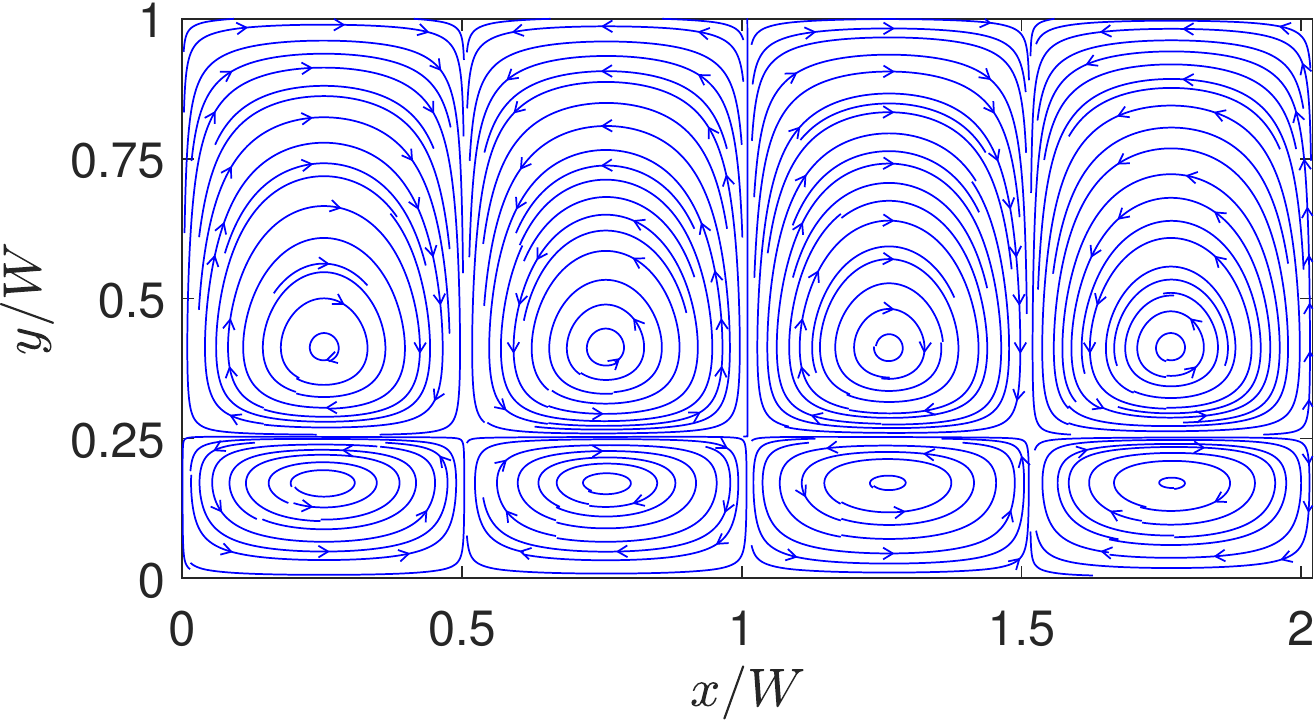}
\caption{LBM Simulation}
\end{subfigure}
\caption{Comparison of the streamlines between the analytical solution with the LBM simulation of thermocapillary convection in SRFs for the case of aspect ratio of $a/b = 3$, thermal conductivity ratio of $\tilde{k} = 1$, and viscosity ratio of $\tilde{\mu} = 1$. Here, the dimensionless linear and quadratic coefficients of surface tension variation with temperature are $M_1=0$ and $M_2=1\times 10^{-1}$, respectively.}
\label{streamlines_SRF_ab_3}
\end{figure}

\begin{figure}[H]
\centering
\begin{subfigure}{0.325\textwidth}
\includegraphics[trim = 0 0 0 0,clip, width = 50mm]{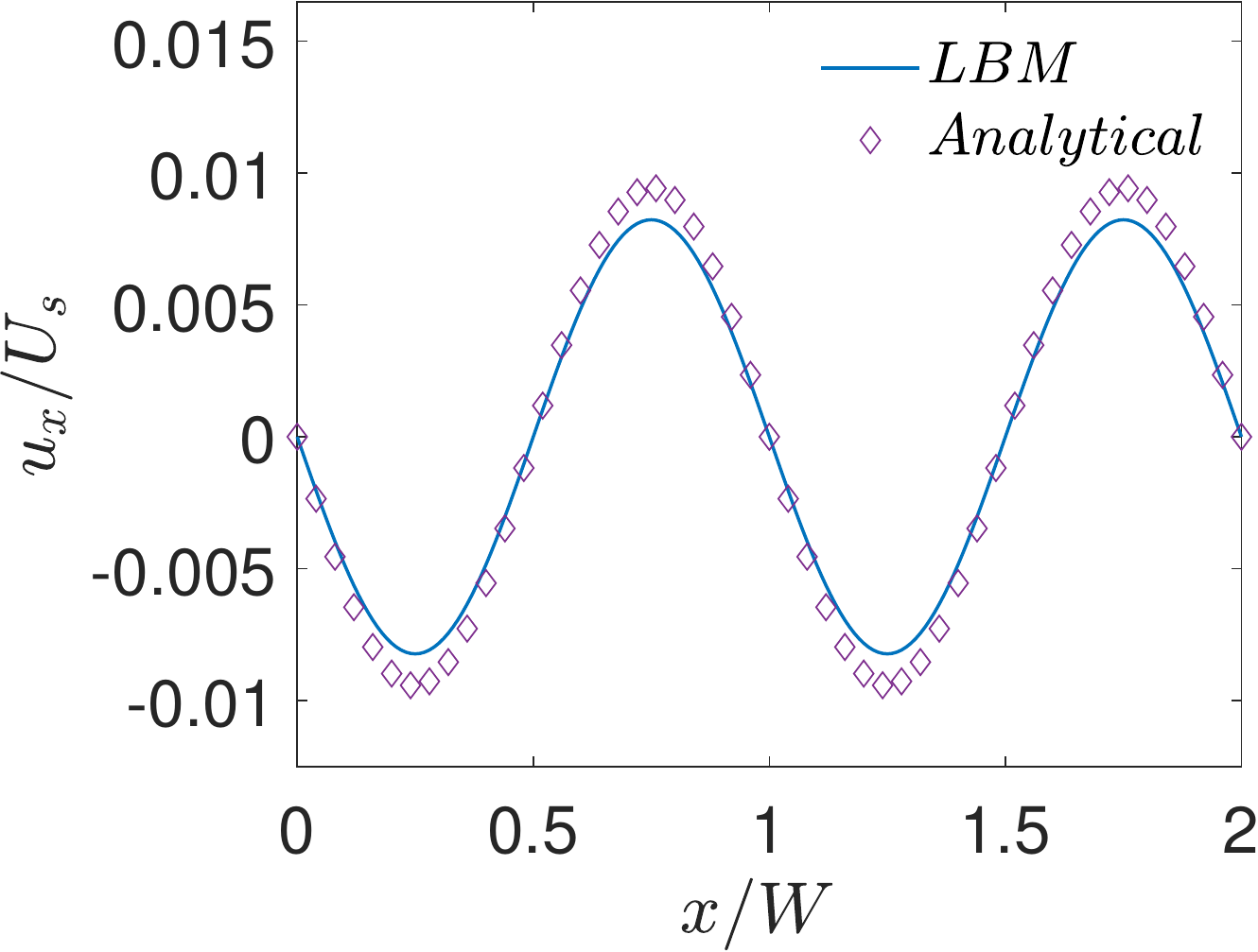}
\end{subfigure}
\begin{subfigure}{0.325\textwidth}
\includegraphics[trim = 0 0 0 0,clip, width = 50mm]{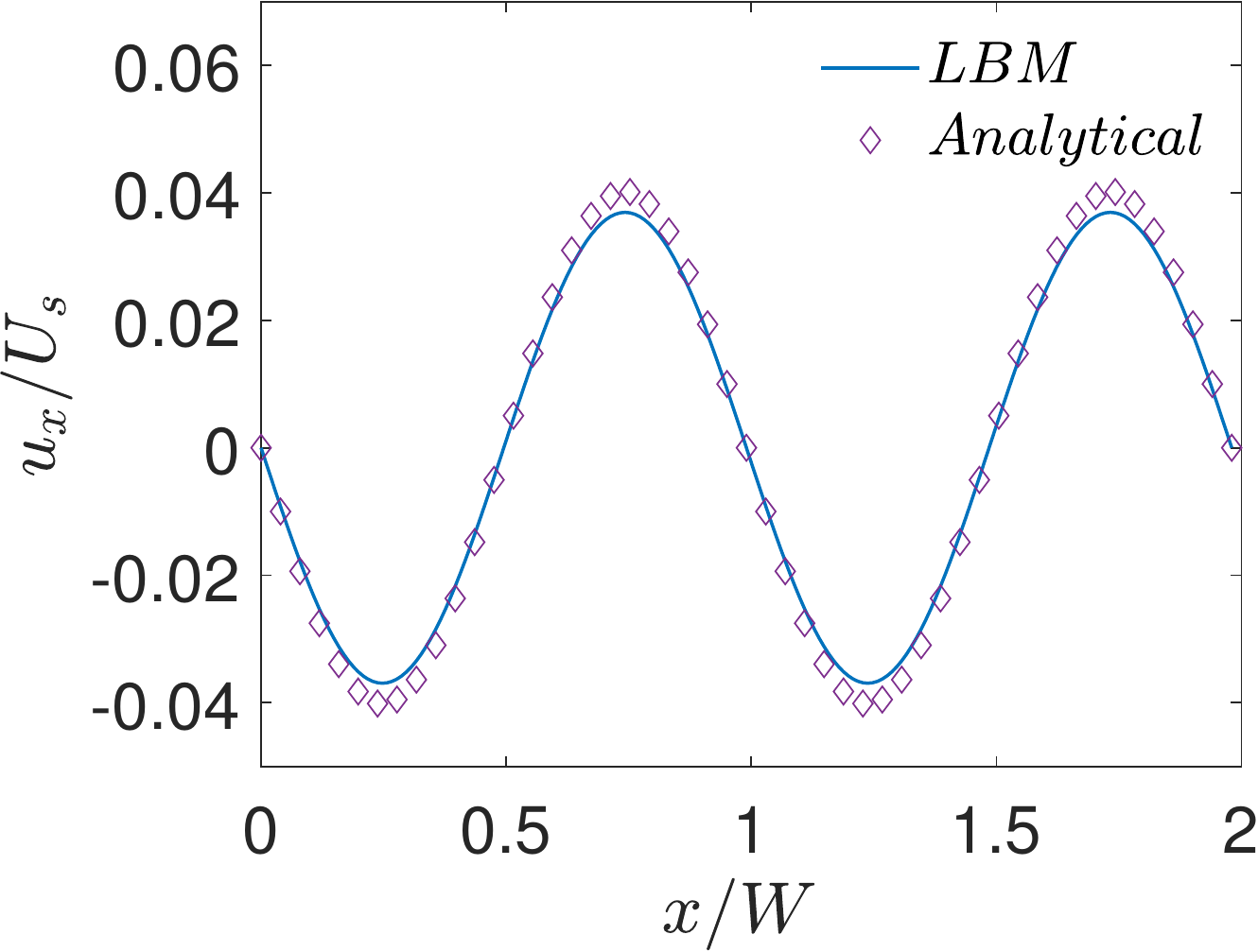}
\end{subfigure}
\begin{subfigure}{0.325\textwidth}
\includegraphics[trim = 0 0 0 0,clip, width = 50mm]{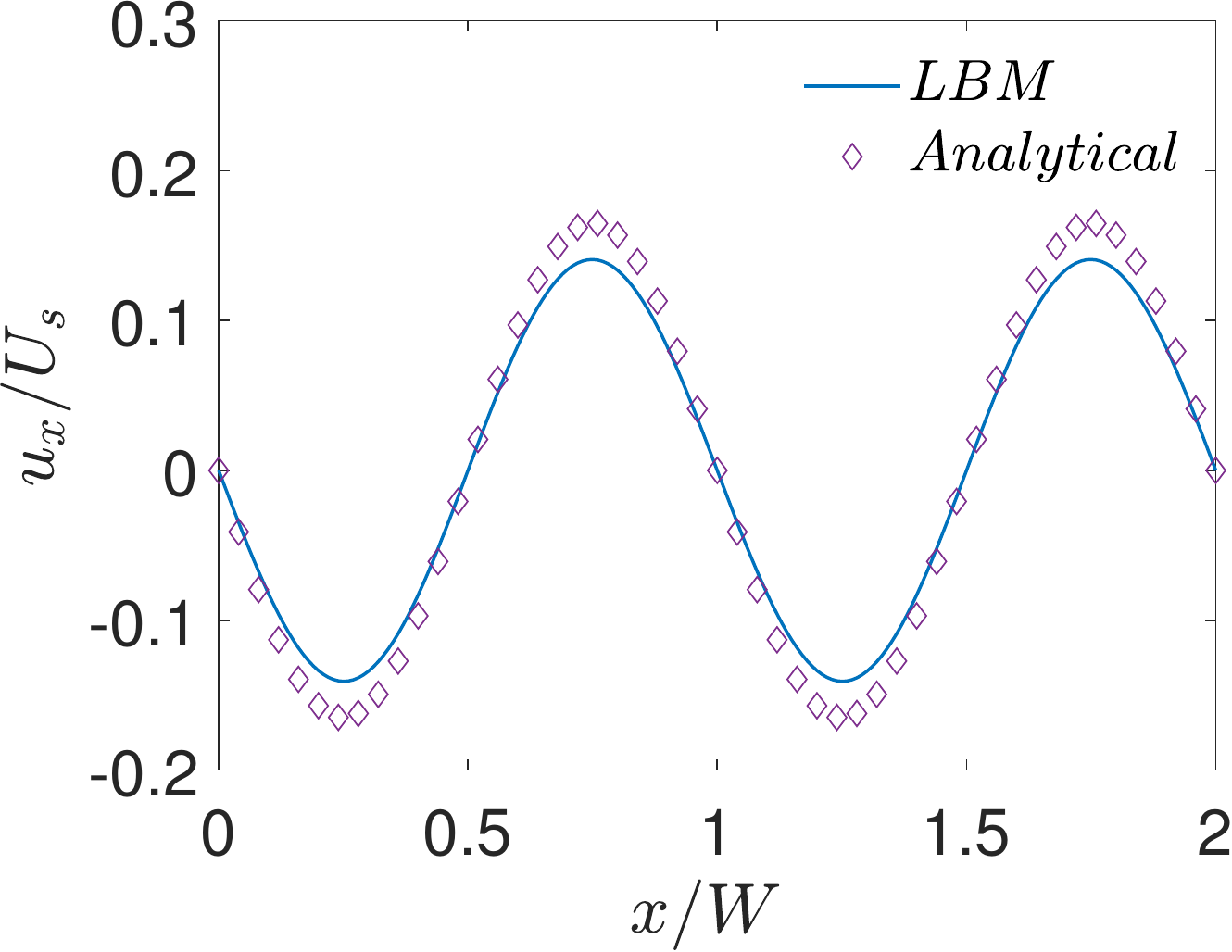}
\end{subfigure}
\caption{Profiles of the horizontal velocity component along the interface in the $x$ direction for thermocapillary flow in SRFs for three different values of the aspect ratio $a/b$: $a/b=1/3$ (left), $a/b=1$ (middle), and $a/b=3$ (right). The purple diamond symbols shown are obtained from the analytical solution and the lines are the LB simulation results. Here, the thermal conductivity ratio is $\tilde{k} = 1$, viscosity ratio is $\tilde{\mu}  = 1$, and the dimensionless surface tension coefficients are $M_1 = 0$ and $M_2 = 1\times 10^{-1}$.}
\label{u_x_abr}
\end{figure}

Again, four and eight counterrotating standing vortical cells are generated in the case of NFs and SRFs, respectively, though with the bottom fluid being thicker, asymmetrical patterns are generated, with the top fluid motion being more squished. By contrast, Figs.~\ref{streamlines_NF_ab_3} and \ref{streamlines_SRF_ab_3} illustrate the streamlines in the thermocapillary-driven flow in NFs and SRFs, respectively, when the aspect ratio $a/b=3$, where the bottom fluid motion appears more squished while retaining the overall distributions of flow in terms of their asymmetrical patterns in the respective cases. However, the streamlines, by themselves, do not necessarily provide a complete information, for example, the variations in the magnitude of the thermal convection currents due to changes in the aspect ratio or the confinement effects of the superimposed fluids.
Hence, by focusing on the interface, let's now investigate the variations in the distribution of the horizontal component of the thermocapillary velocity field $u_x$ in SRFs due to changes in the aspect ratio $a/b$, which we will normalize by a suitable characteristic velocity arising from the surface tension gradient. Based on the scaling argument given below Eq.~(\ref{three-prime}) involving a balance of the Marangoni stress and the viscous stress and using the average temperature on the interface in estimating the attendant temperature gradient, we can obtain the following characteristic velocity for thermocapillary convection in SRFs (see Appendix~\ref{AppendixC} for details):
\mathleft
\begin{equation}\label{U_s}
U_s \sim  \frac{\Delta T}{\mu^b}\left(\frac{b}{l}\right)  \left[  \sigma_T + 2 \sigma_{TT}\left(\frac{T_h \left(\frac{a}{b}\right) +T_c \tilde{k}}{\left(\frac{a}{b}\right) + \tilde{k}} + \frac{\Delta T \sinh(\tilde{a})}{\tilde{k} \cosh (\tilde{a}) \sinh (\tilde{b}) + \cosh (\tilde{b}) \sinh (\tilde{a})} - T_{ref} \right) \right].
\end{equation}
Then, taking $W=a+b$ as the width of the microchannel, Fig.~\ref{u_x_abr} presents the dimensionless horizontal velocity component $u_x/U_s$ on the interface in SRFs as a function of the dimensionless coordinate $x/W$ for three different aspect ratios $a/b=1/3, 1$ and $3$. It can be seen that while the velocity profiles are qualitatively similar, there are dramatic differences in the strength of the Marangoni convection currents in the interface depending on the aspect ratio. When the interface is far from the nonuniformly heated bottom wall, which occurs for the case $a/b=1/3$, the magnitude of the thermocapillary convection is found to be relatively weak; by contrast, when the interface is closer to the interface at $a/b=3$ than the other two cases, the Marangoni velocities are much larger, by at least an order of magnitude. This is due to the fact that the closer the interface is to the heated wall side, the greater is the heat transport by diffusion from the latter to the former, which in turn intensifies the generation of thermocapillary velocity currents via the surface tension gradient resulting from a nonuniform temperature distribution on the interface. Thus, the aspect ratio $a/b$ of the superimposed layers of SRFs has a major effect on not just in setting up asymmetrical thermocapillary convection roll cells, but also, and more importantly, in determining the resulting magnitude of the velocities of the fluids around interfaces. For the purpose of clarification, it should be noted that the length scale $b$ used in determining the shear stress used in obtaining a scale for the characteristic velocity $U_s$ (see Appendix~\ref{AppendixC} for its derivation), while a consistent scaling definition following~\cite{pendse2010analytical}, is an overestimate. Hence, the dimensionless velocity profiles $u_x/U_s$ shown are generally significantly smaller than unity.

\subsection{Effect of thermal conductivity ratio $\tilde{k}$ of SRF layers}
\begin{figure}[H]
\centering
\begin{subfigure}{0.325\textwidth}
\includegraphics[trim = 0 0 0 0,clip, width = 50mm]{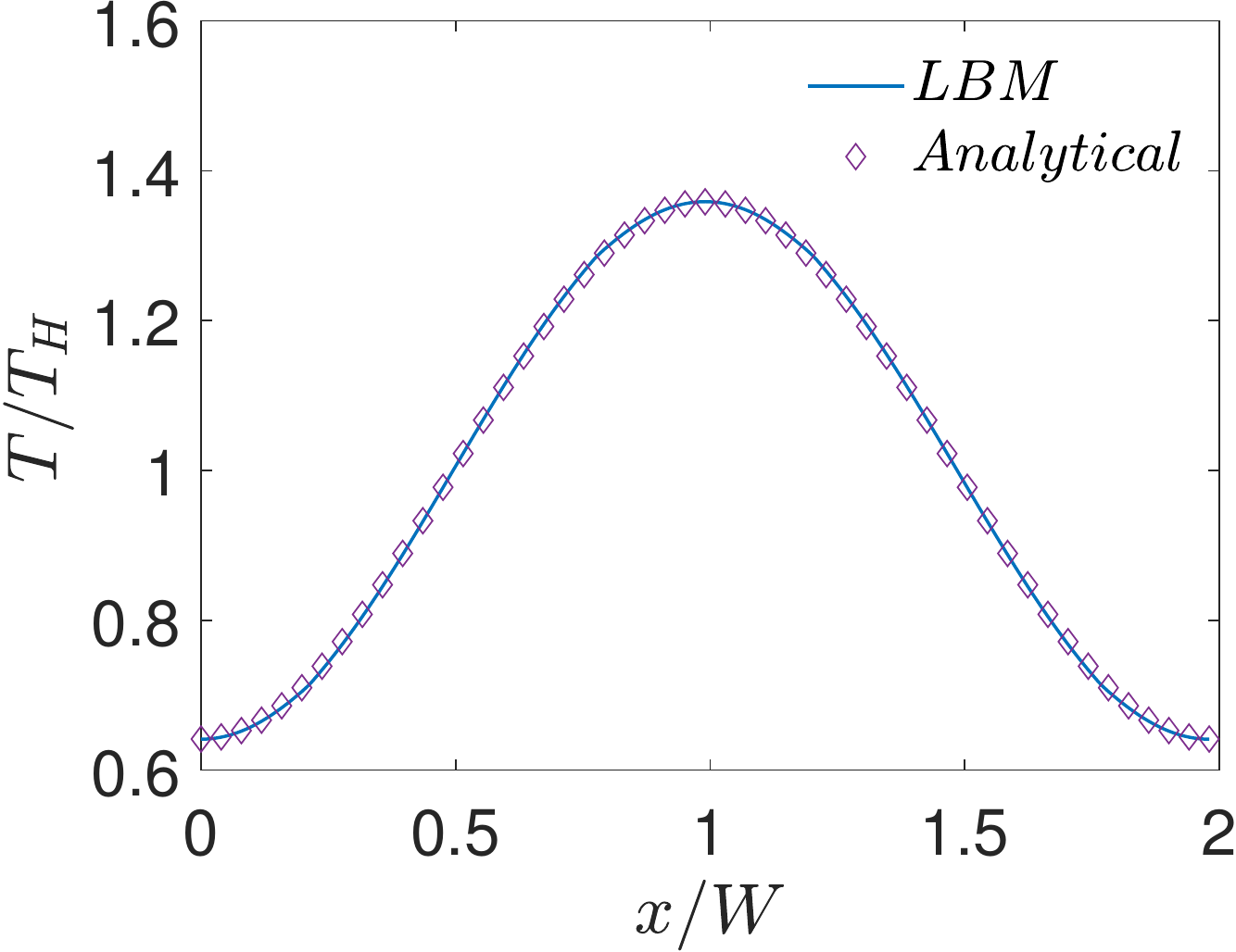}
\end{subfigure}
\begin{subfigure}{0.325\textwidth}
\includegraphics[trim = 0 0 0 0,clip, width = 50mm]{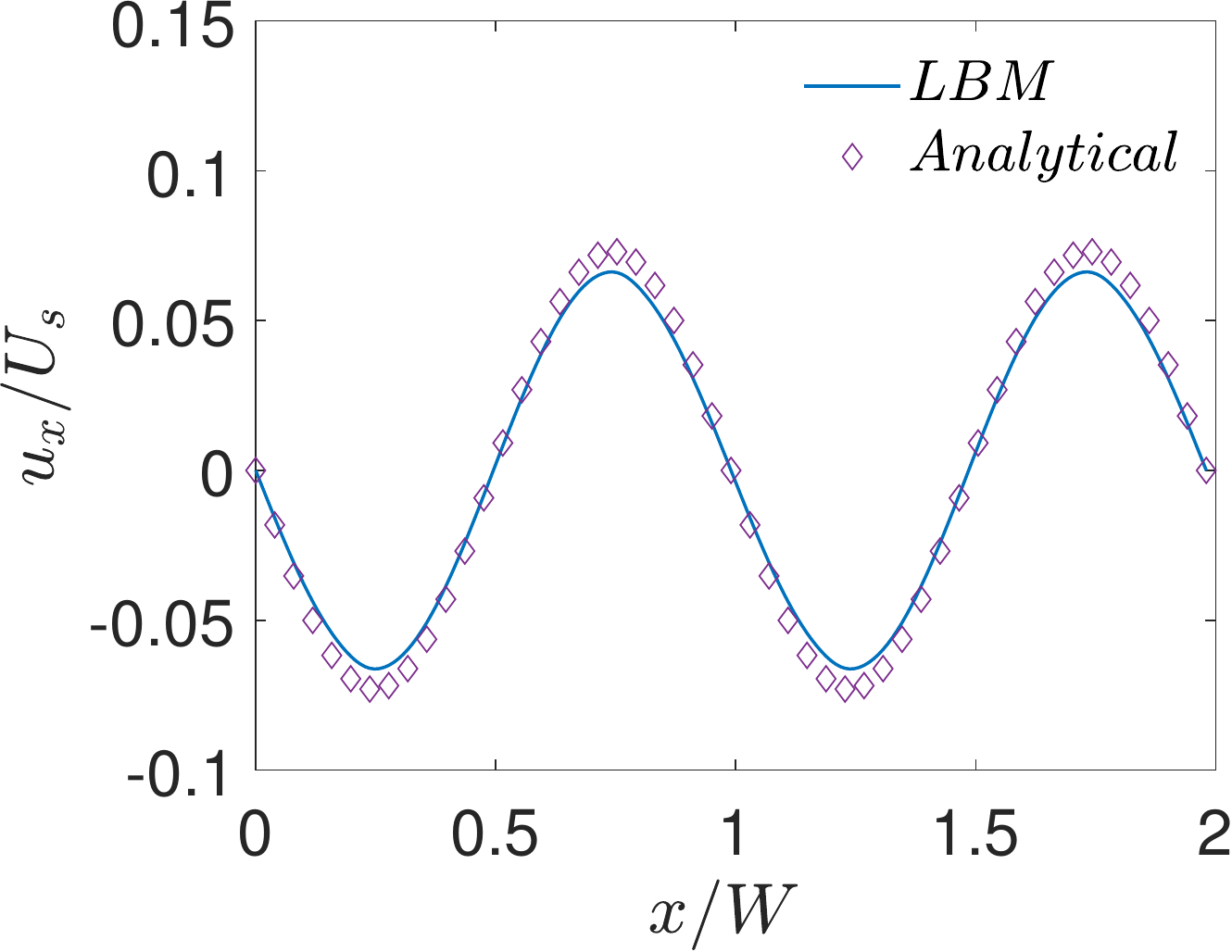}
\end{subfigure}
\begin{subfigure}{0.325\textwidth}
\includegraphics[trim = 0 0 0 0,clip, width = 50mm]{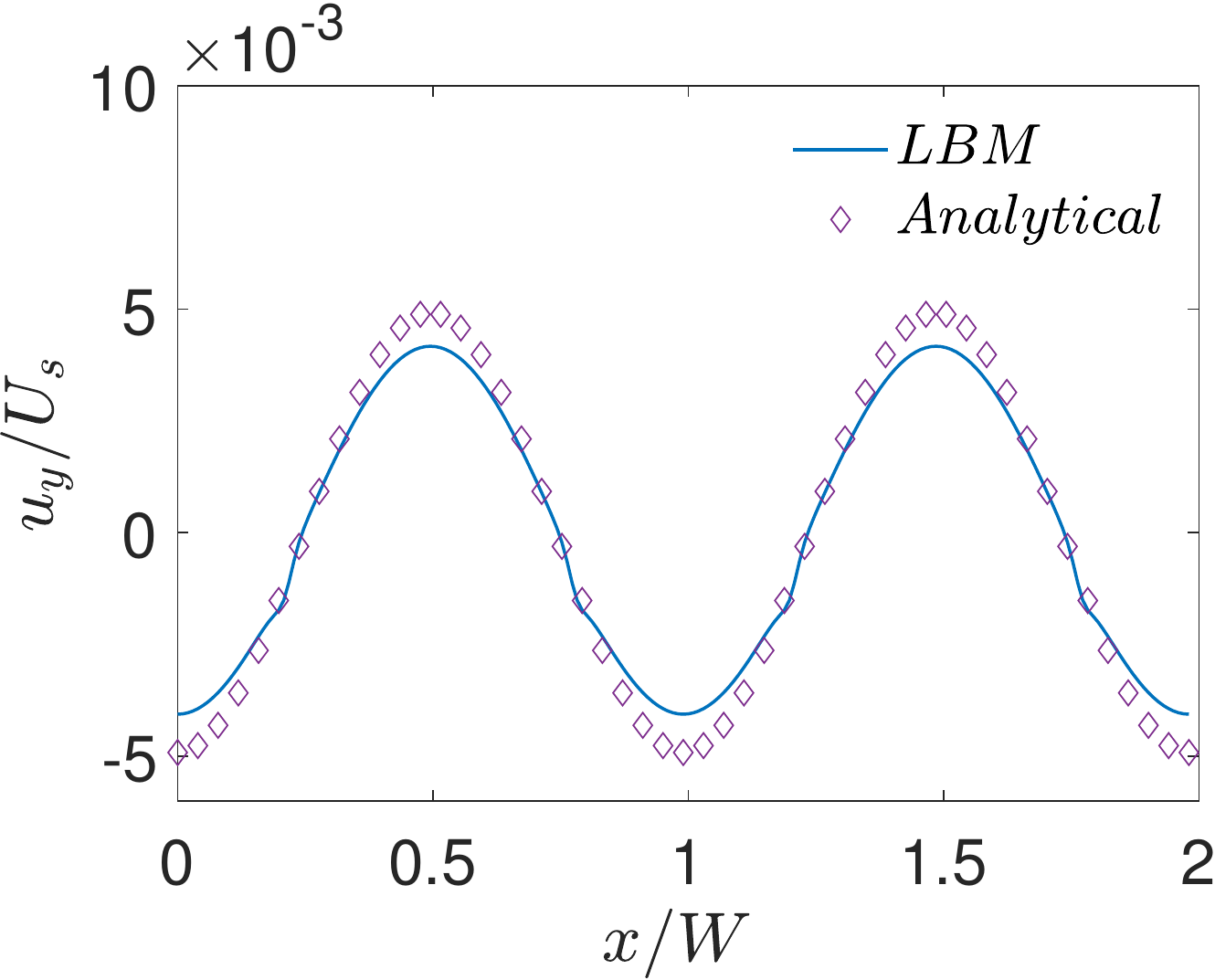}
\end{subfigure}
\caption{Profiles of the temperature and velocity components along the centerline of the domain in the $x$ direction for thermocapillary flow in SRFs for thermal conductivity ratio $\tilde{k} = 0.1$. The purple symbols shown are obtained from the analytical solution and the lines are the LB simulation results. Here, the aspect ratio is $a/b = 1$, viscosity ratio is $\tilde{\mu}  = 1$, and the dimensionless surface tension coefficients are $M_1 = 0$ and $M_2 = 1\times 10^{-1}$.}
\label{u_v_T_x_k13}
\end{figure}

\begin{figure}[H]
\centering
\begin{subfigure}{0.325\textwidth}
\includegraphics[trim = 0 0 0 0,clip, width = 50mm]{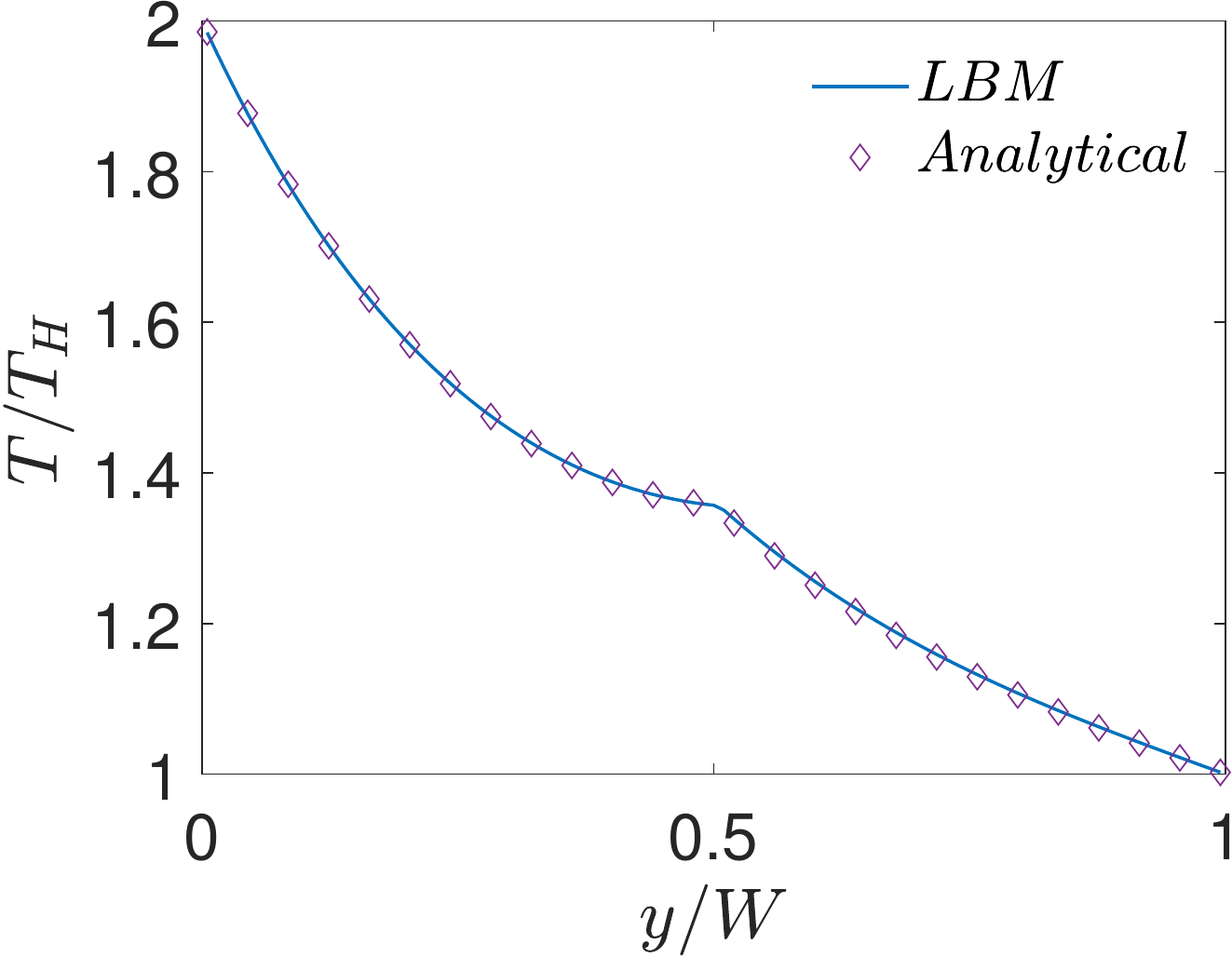}
\end{subfigure}
\begin{subfigure}{0.325\textwidth}
\includegraphics[trim = 0 0 0 0,clip, width = 50mm]{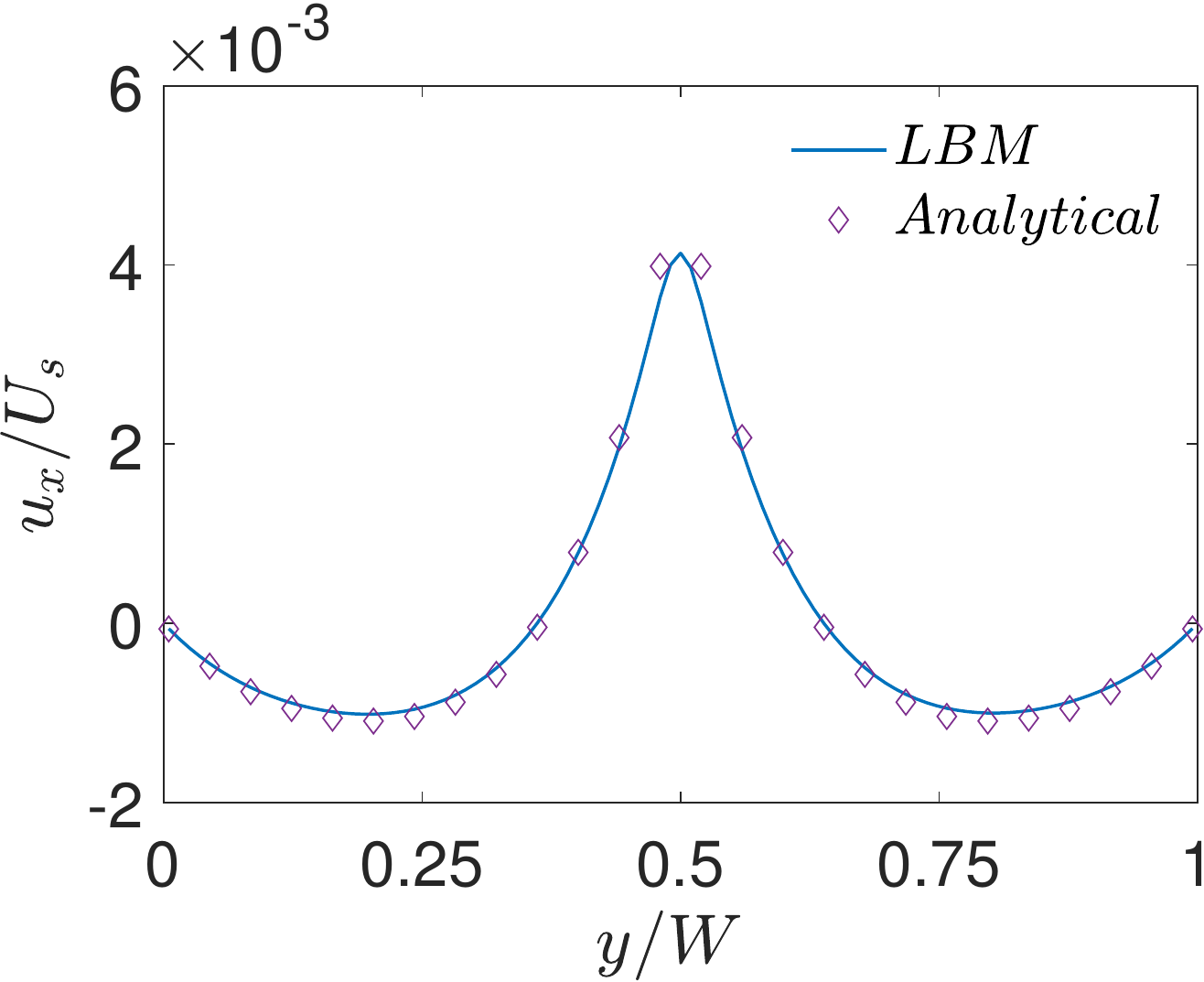}
\end{subfigure}
\begin{subfigure}{0.325\textwidth}
\includegraphics[trim = 0 0 0 0,clip, width = 50mm]{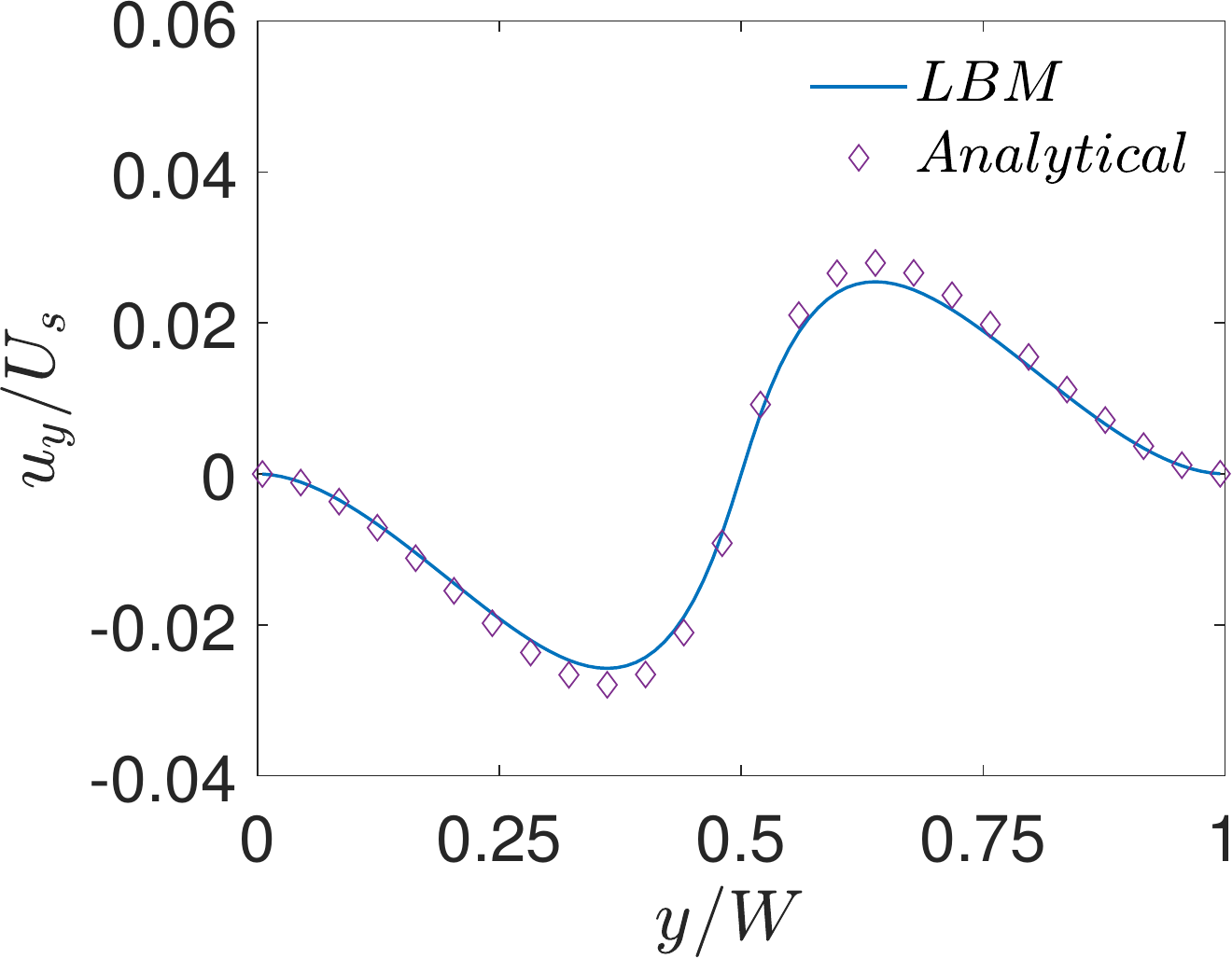}
\end{subfigure}
\caption{Profiles of the temperature and velocity components along the centerline of the domain in the $y$ direction for thermocapillary flow in SRFs for thermal conductivity ratio $\tilde{k} = 0.1$. The purple symbols shown are obtained from the analytical solution and the lines are the LB simulation results. Here, the aspect ratio is $a/b = 1$, viscosity ratio is $\tilde{\mu}  = 1$, and the dimensionless surface tension coefficients are $M_1 = 0$ and $M_2 = 1\times 10^{-1}$.}
\label{u_v_T_y_k13}
\end{figure}

\begin{figure}[H]
\centering
\begin{subfigure}{0.325\textwidth}
\includegraphics[trim = 0 0 0 0,clip, width = 50mm]{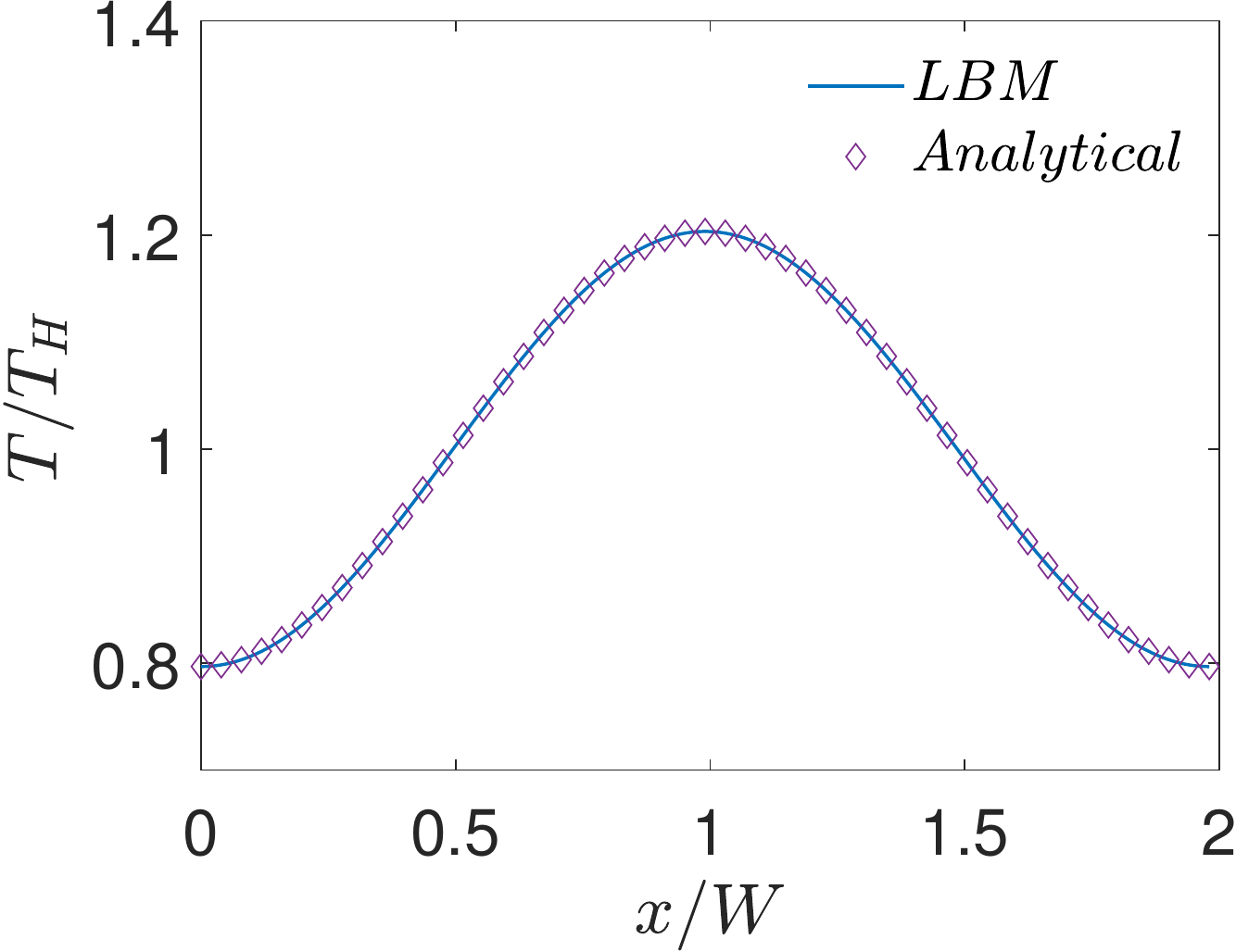}
\end{subfigure}
\begin{subfigure}{0.325\textwidth}
\includegraphics[trim = 0 0 0 0,clip, width = 50mm]{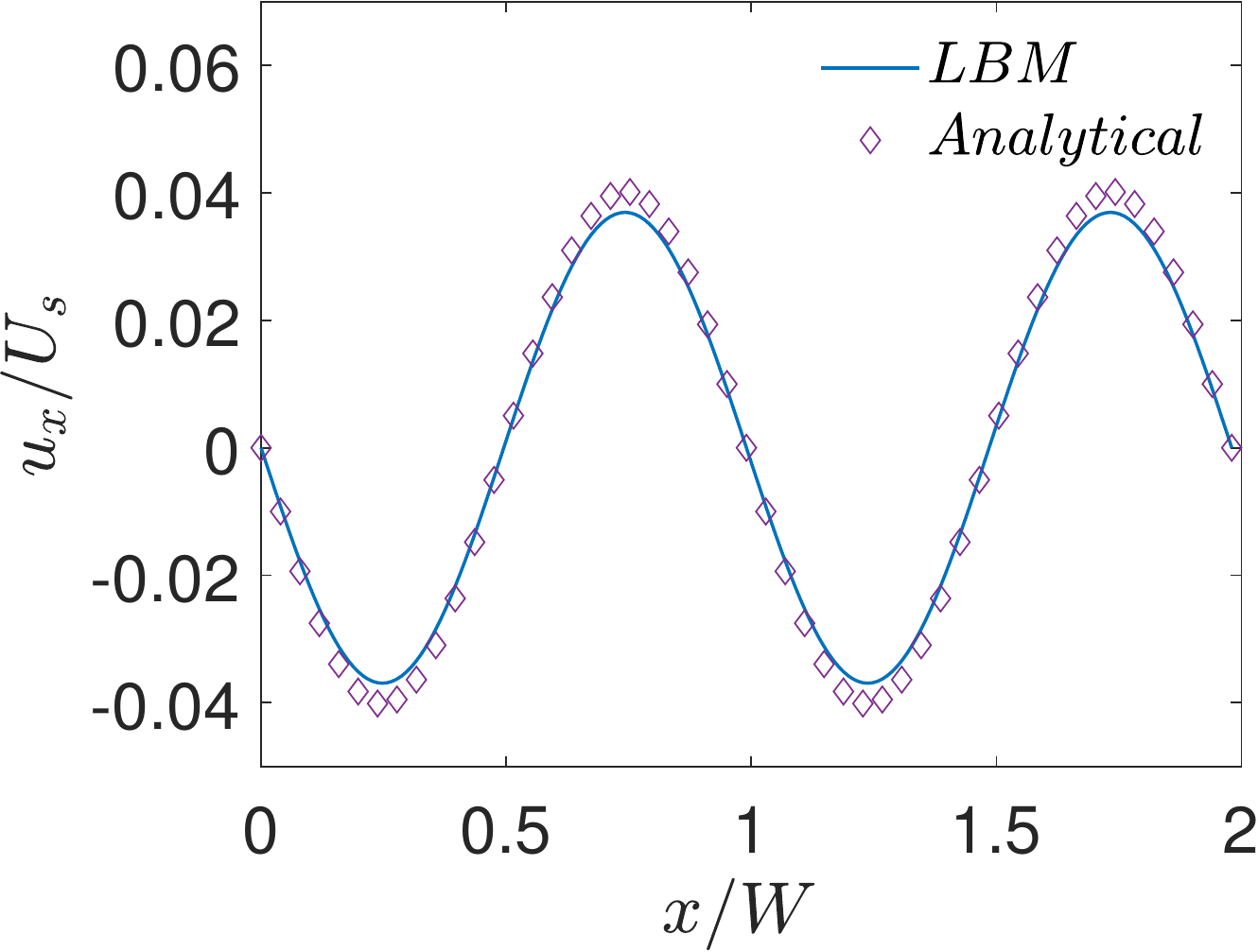}
\end{subfigure}
\begin{subfigure}{0.325\textwidth}
\includegraphics[trim = 0 0 0 0,clip, width = 50mm]{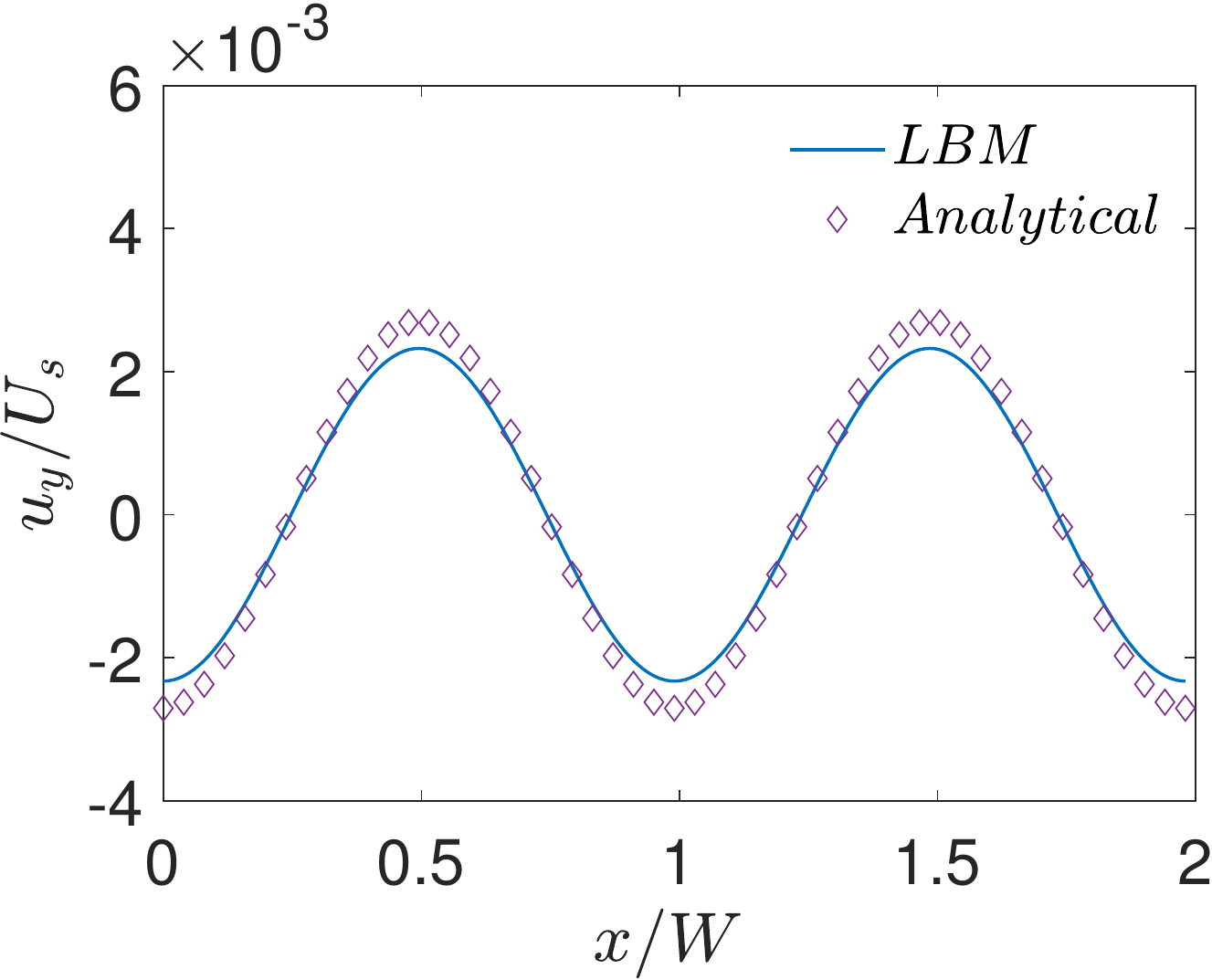}
\end{subfigure}
\caption{Profiles of the temperature and velocity components along the centerline of the domain in the $x$ direction for thermocapillary flow in SRFs for thermal conductivity ratio $\tilde{k} = 1.0$. The purple symbols shown are obtained from the analytical solution and the lines are the LB simulation results. Here, the aspect ratio is $a/b = 1$, viscosity ratio is $\tilde{\mu} = 1$, and the dimensionless surface tension coefficients are $M_1 = 0$ and $M_2 = 1\times 10^{-1}$.}
\label{u_v_T_x_k1}
\end{figure}

\begin{figure}[H]
\centering
\begin{subfigure}{0.325\textwidth}
\includegraphics[trim = 0 0 0 0,clip, width = 50mm]{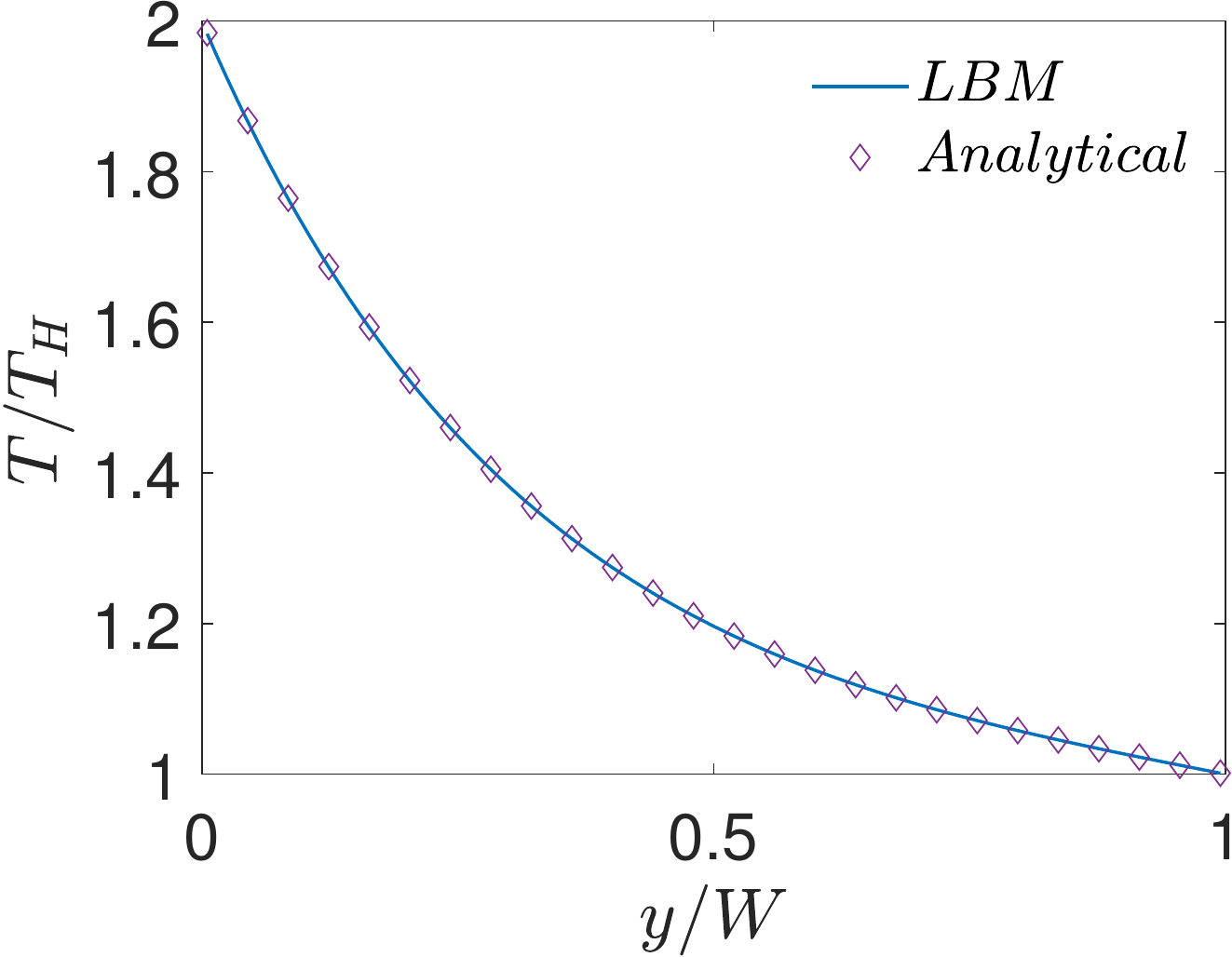}
\end{subfigure}
\begin{subfigure}{0.325\textwidth}
\includegraphics[trim = 0 0 0 0,clip, width = 50mm]{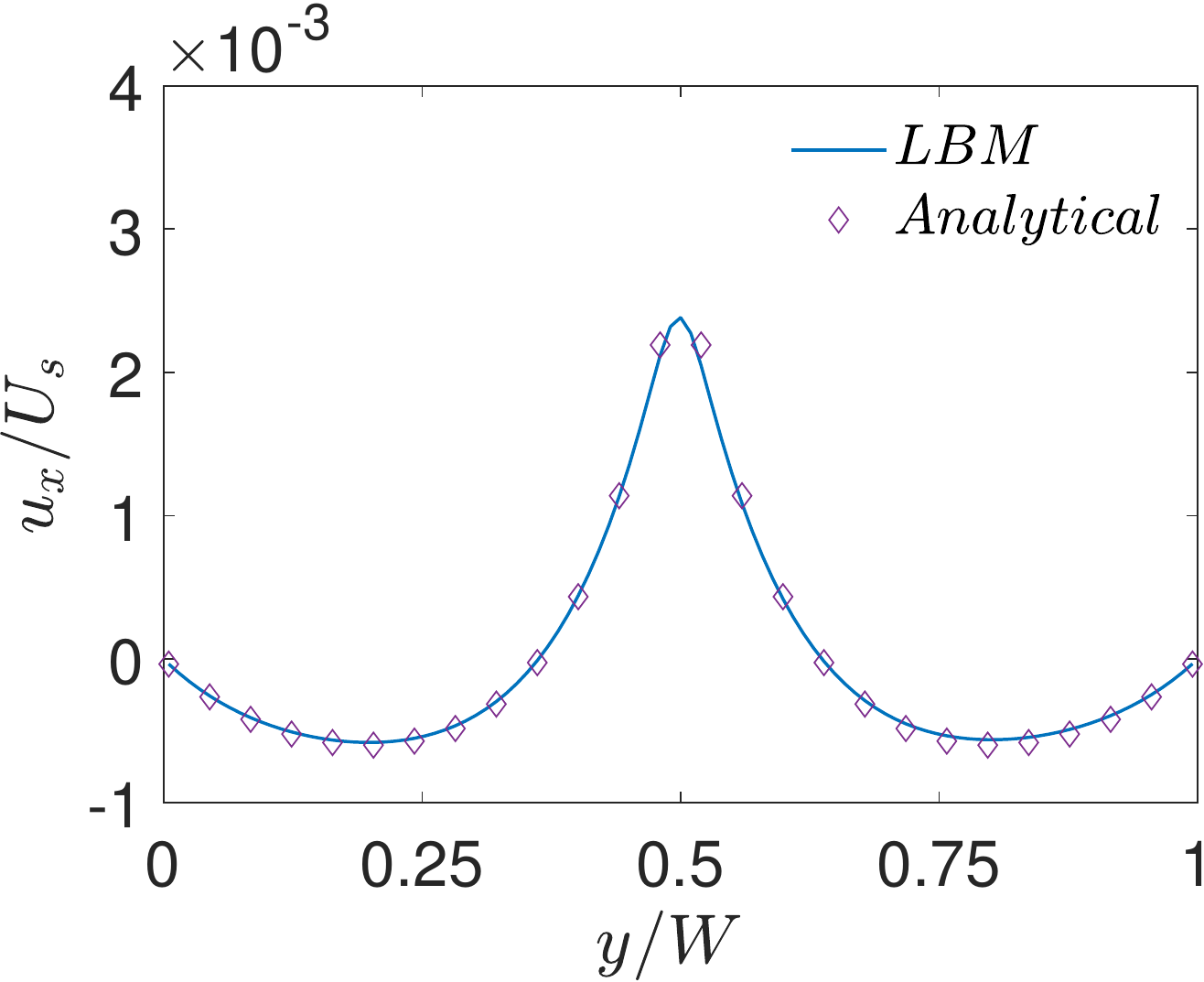}
\end{subfigure}
\begin{subfigure}{0.325\textwidth}
\includegraphics[trim = 0 0 0 0,clip, width = 50mm]{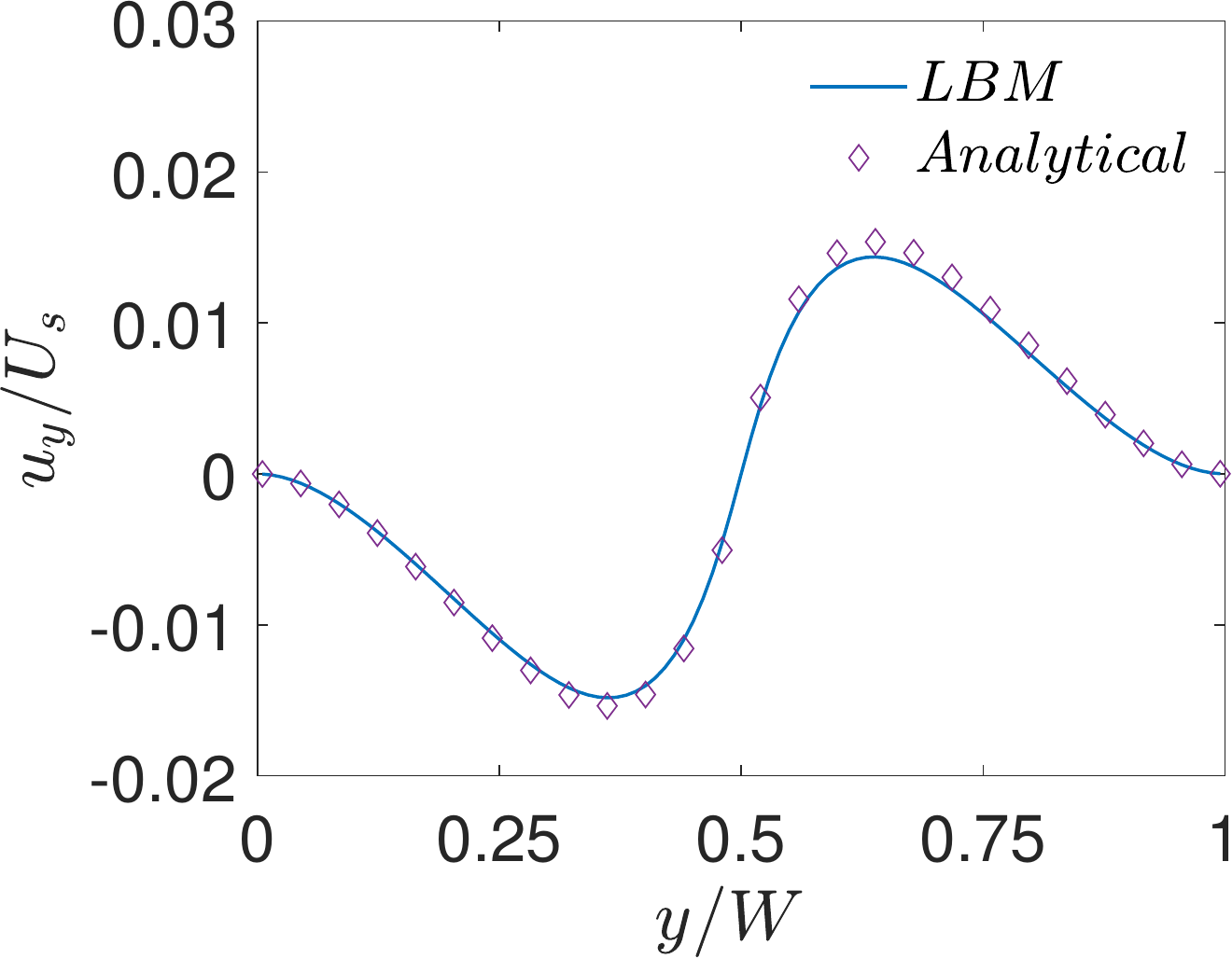}
\end{subfigure}
\caption{Profiles of the temperature and velocity components along the centerline of the domain in the $y$ direction for thermocapillary flow in SRFs for thermal conductivity ratio $\tilde{k} = 1.0$. The purple symbols shown are obtained from the analytical solution and the lines are the LB simulation results. Here, the aspect ratio is $a/b = 1$, viscosity ratio is $\tilde{\mu}  = 1$, and the dimensionless surface tension coefficients are $M_1 = 0$ and $M_2 = 1\times 10^{-1}$.}
\label{u_v_T_y_k1}
\end{figure}

\begin{figure}[H]
\centering
\begin{subfigure}{0.325\textwidth}
\includegraphics[trim = 0 0 0 0,clip, width = 50mm]{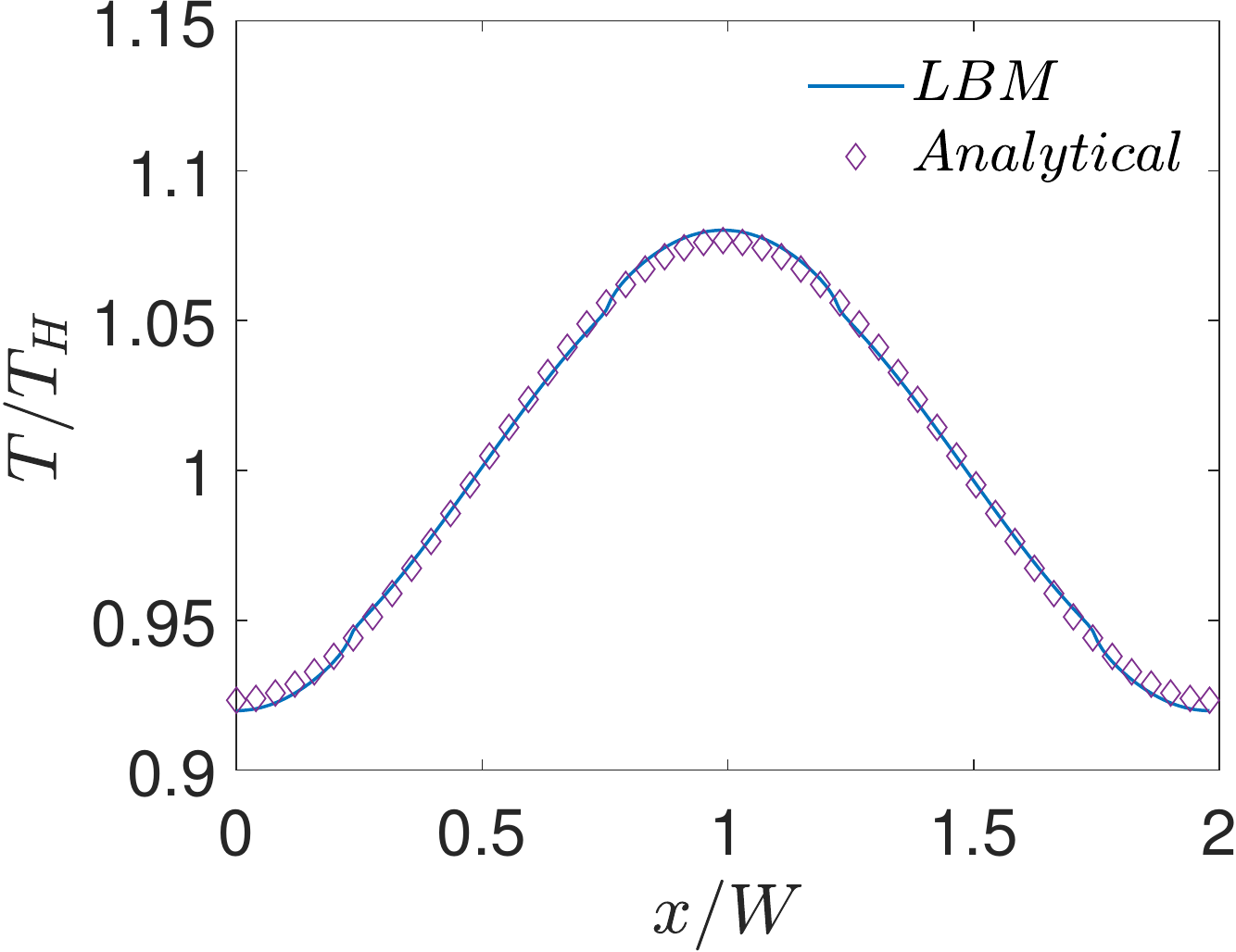}
\end{subfigure}
\begin{subfigure}{0.325\textwidth}
\includegraphics[trim = 0 0 0 0,clip, width = 50mm]{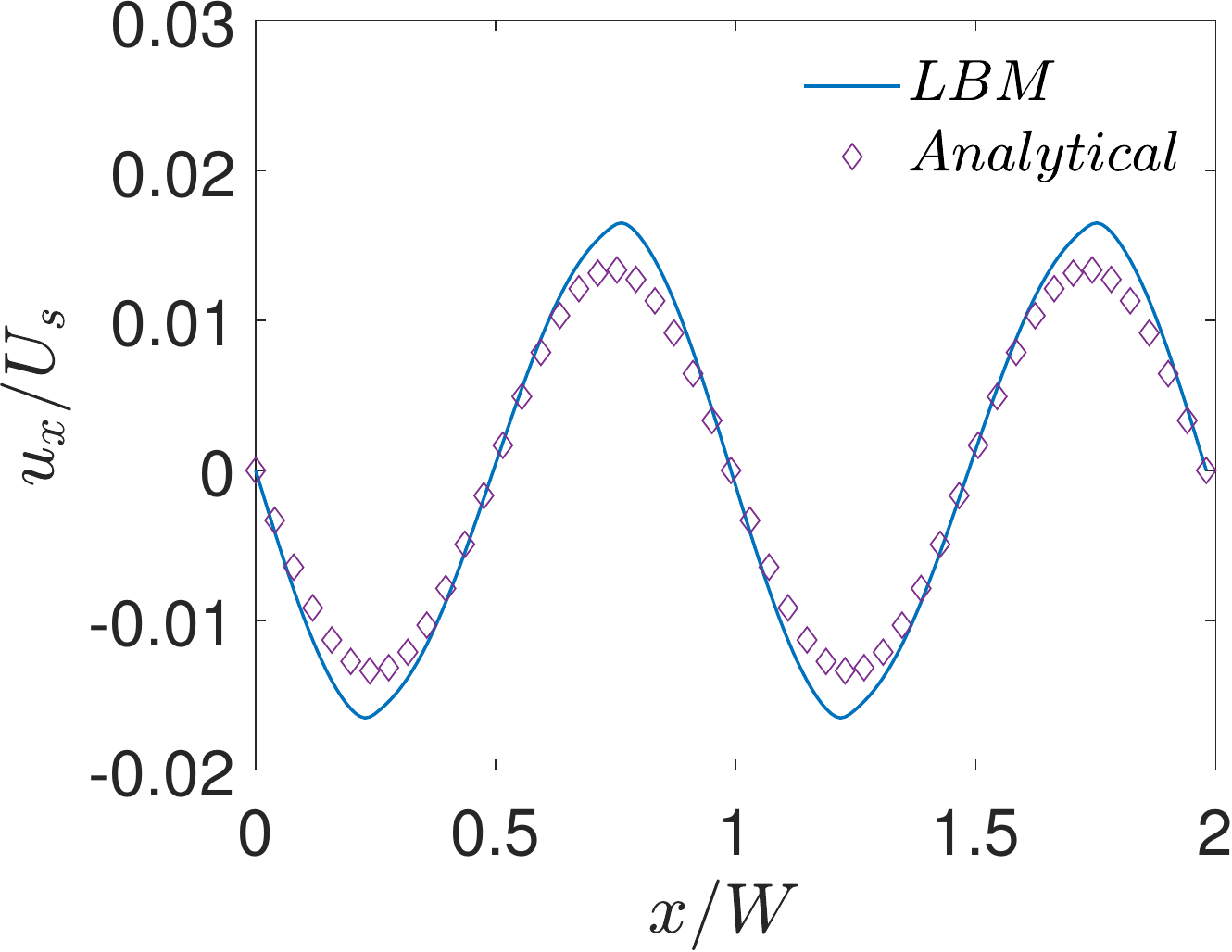}
\end{subfigure}
\begin{subfigure}{0.325\textwidth}
\includegraphics[trim = 0 0 0 0,clip, width = 50mm]{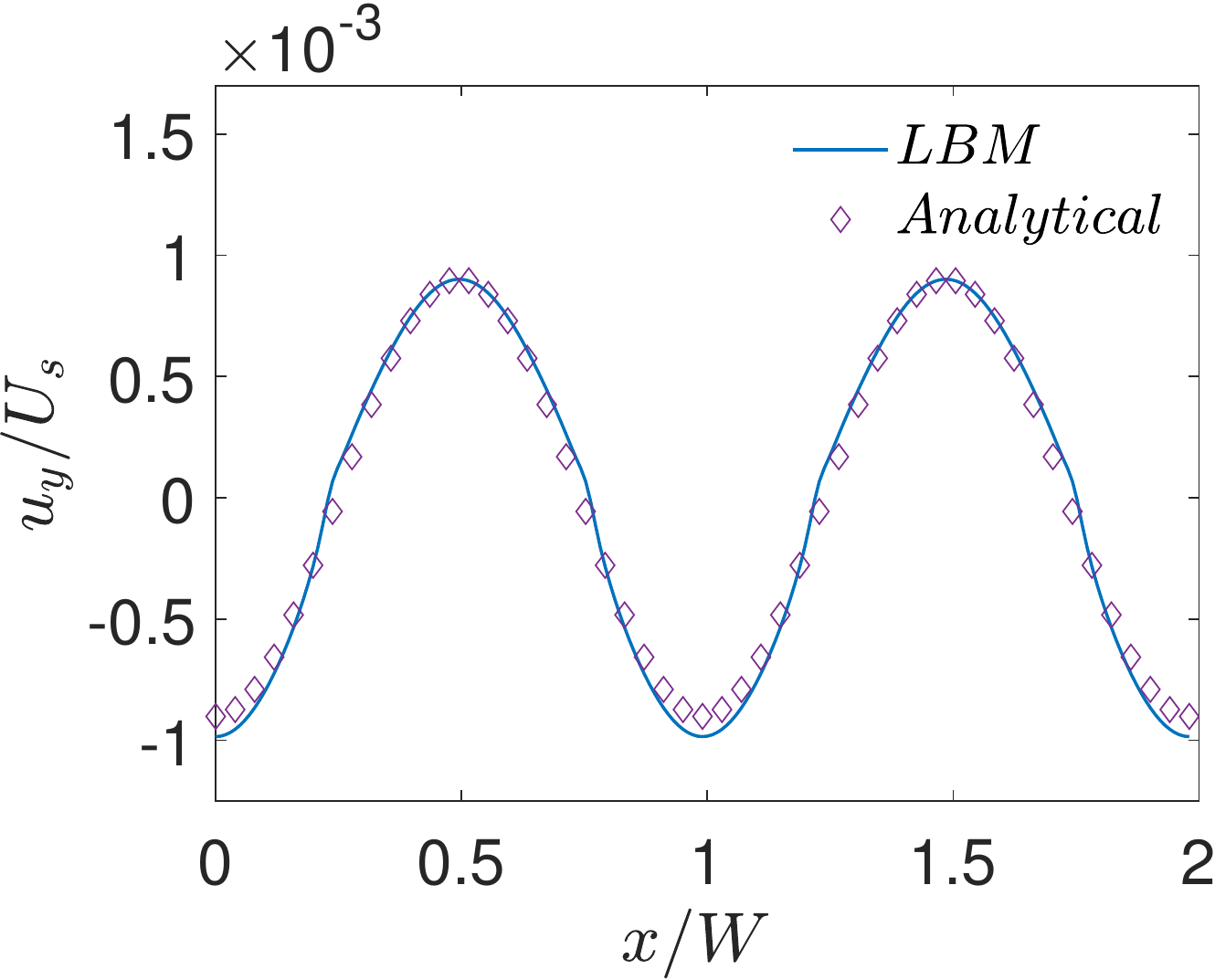}
\end{subfigure}
\caption{Profiles of the temperature and velocity components along the centerline of the domain in the $x$ direction for thermocapillary flow in SRFs for thermal conductivity ratio $\tilde{k} = 5.0$. The purple symbols shown are obtained from the analytical solution and the lines are the LB simulation results. Here, the aspect ratio is $a/b = 1$, viscosity ratio is $\tilde{\mu} = 1$, and the dimensionless surface tension coefficients are $M_1 = 0$ and $M_2 = 1\times 10^{-1}$.}
\label{u_v_T_x_k3}
\end{figure}

\begin{figure}[H]
\centering
\begin{subfigure}{0.325\textwidth}
\includegraphics[trim = 0 0 0 0,clip, width = 50mm]{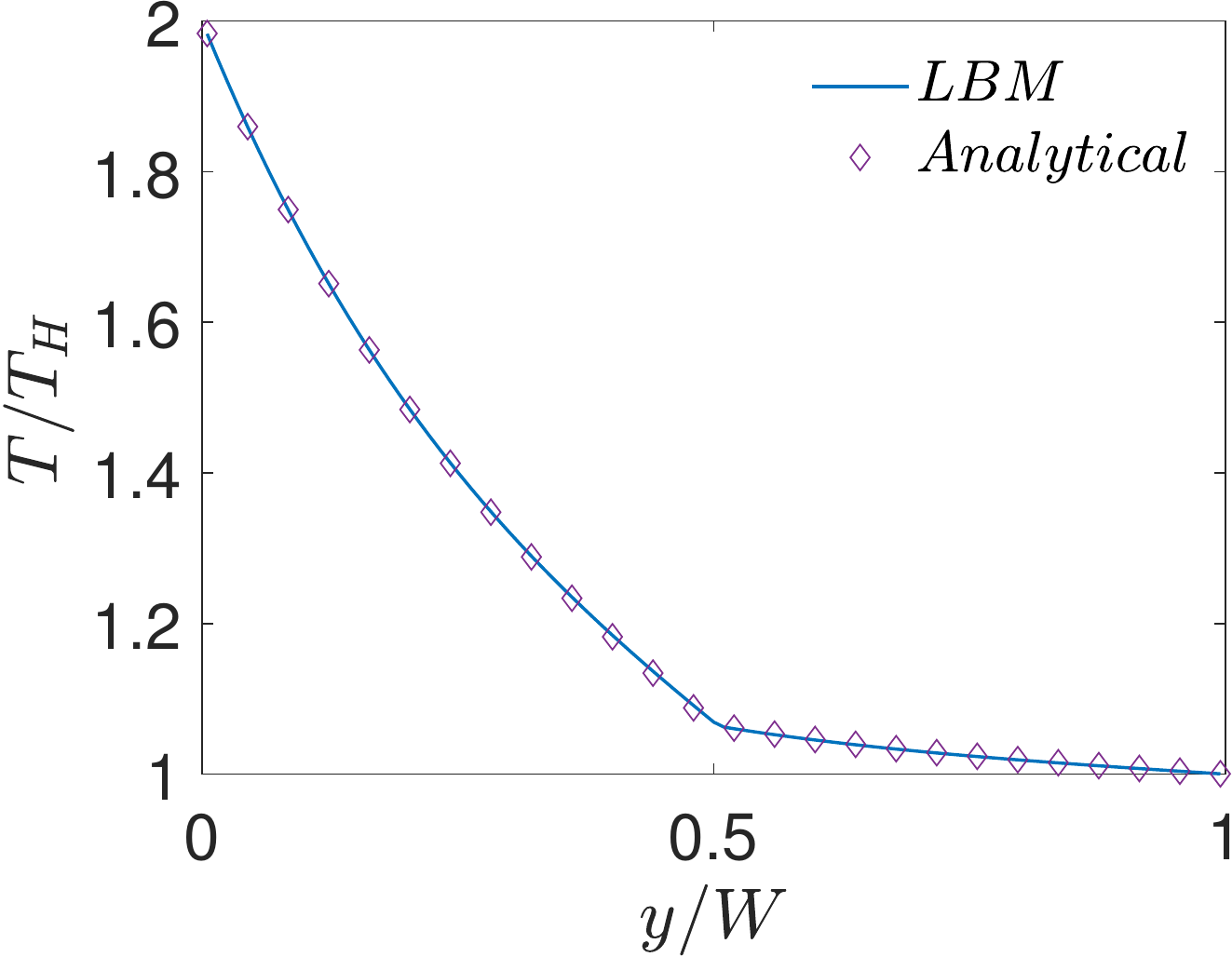}
\end{subfigure}
\begin{subfigure}{0.325\textwidth}
\includegraphics[trim = 0 0 0 0,clip, width = 50mm]{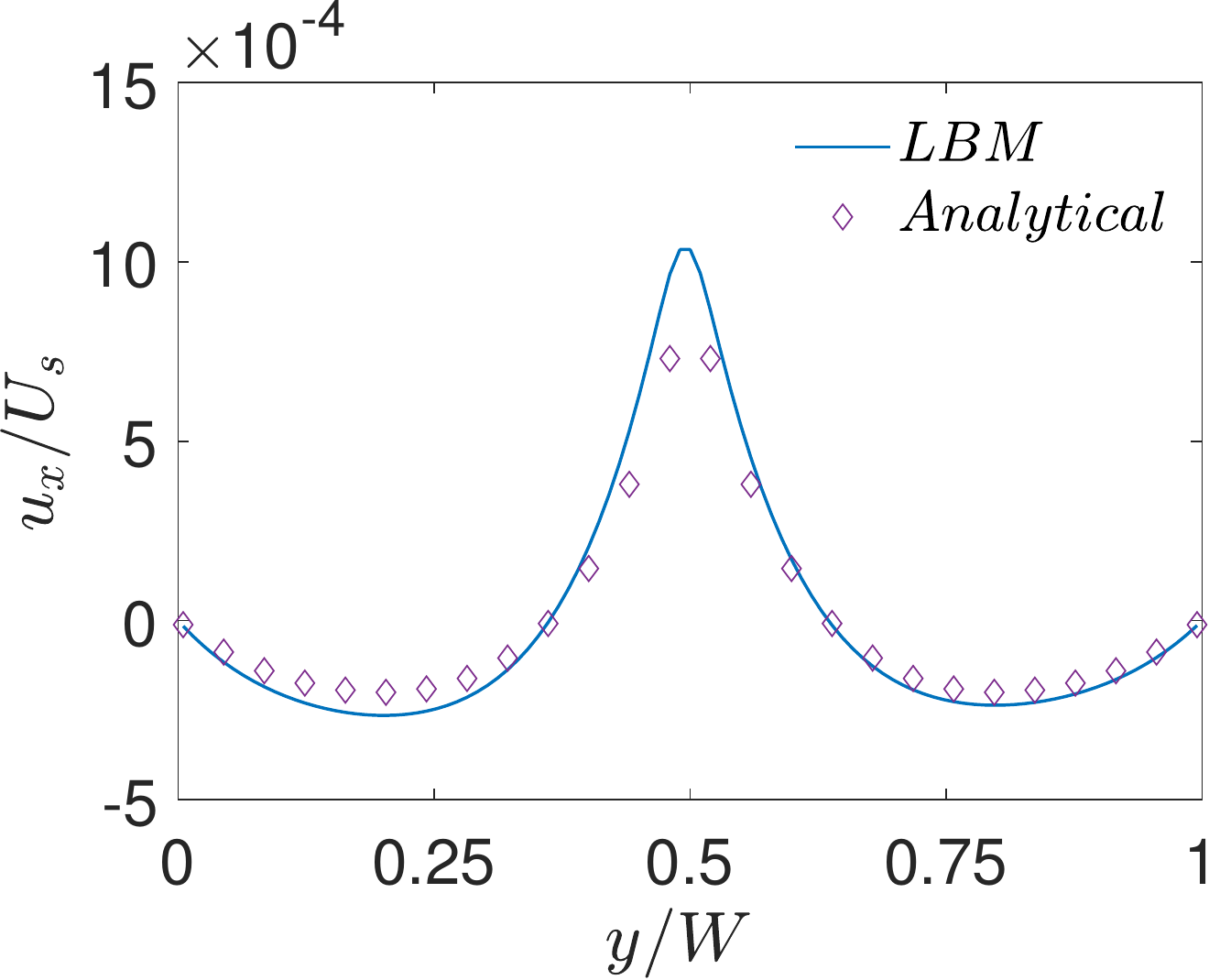}
\end{subfigure}
\begin{subfigure}{0.325\textwidth}
\includegraphics[trim = 0 0 0 0,clip, width = 50mm]{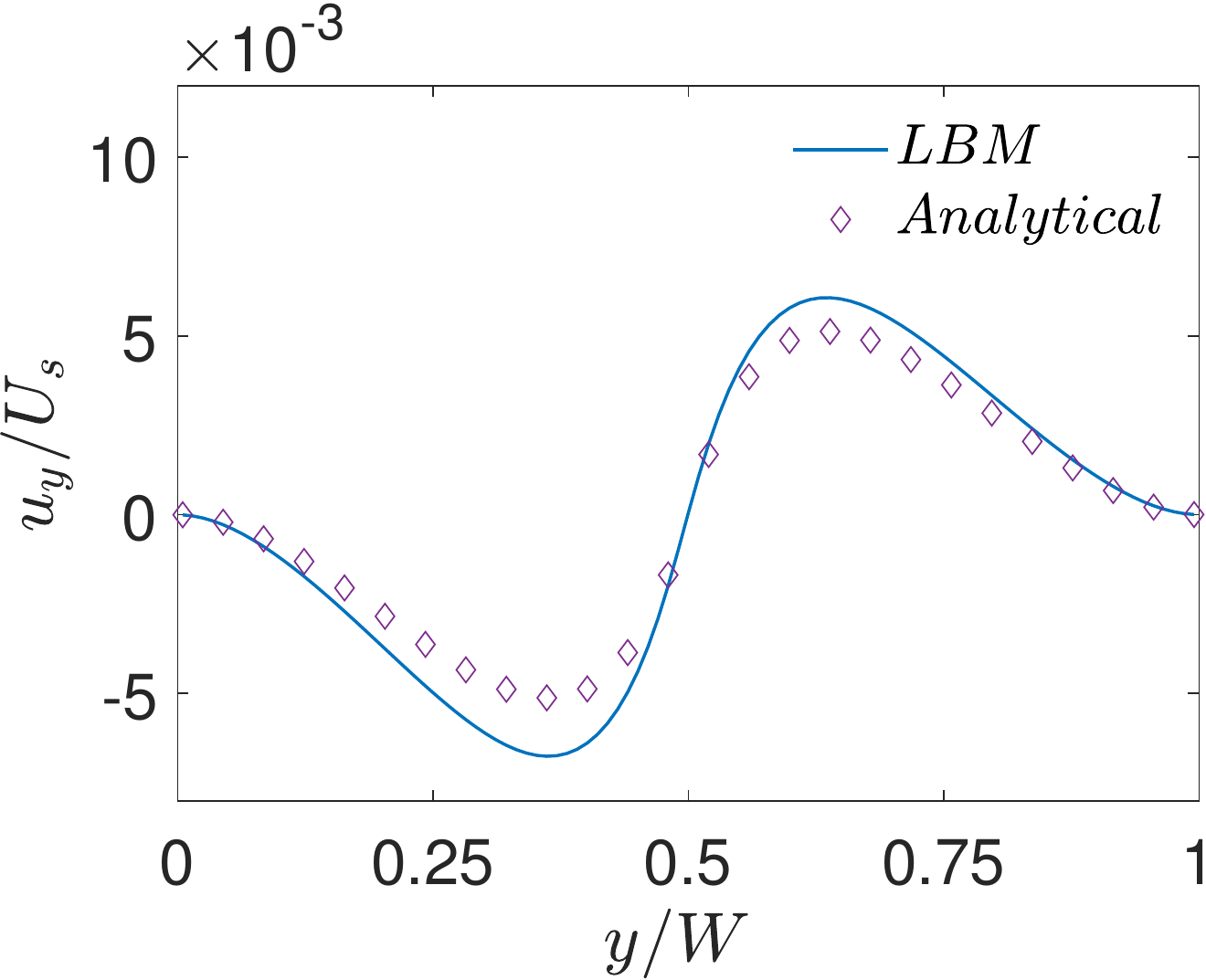}
\end{subfigure}
\caption{Profiles of the temperature and velocity components along the centerline of the domain in the $y$ direction for thermocapillary flow in SRFs for thermal conductivity ratio $\tilde{k} = 5.0$. The purple symbols shown are obtained from the analytical solution and the lines are the LB simulation results. Here, the aspect ratio is $a/b = 1$, viscosity ratio is $\tilde{\mu} = 1$, and the dimensionless surface tension coefficients are $M_1 = 0$ and $M_2 = 1\times 10^{-1}$.}
\label{u_v_T_y_k3}
\end{figure}

Next, we will perform another quantitative study involving the effect of the thermal conductivity ratio $\tilde{k}=k_a/k_b$ on the profiles of temperature and the components of the velocity field in thermocapillary convection in SRFs. In this regard, we fix SRF layers to be of equal thickness, i.e., $a/b=1$, and set $\tilde{\mu} = 1$, $M_1 = 0$ and $M_2 = 1\times 10^{-1}$, and then vary $\tilde{k}$ by considering three representative choices: $\tilde{k}=0.1, 1.0$ and $5.0$. Figures~\ref{u_v_T_x_k13} and \ref{u_v_T_y_k13} show the profiles of the temperature and the components of the velocity field along the centerline of the domain in the $x$ and $y$ directions, respectively, for $\tilde{k} = 0.1$. Similar plots are shown in Figs.~\ref{u_v_T_x_k1} and \ref{u_v_T_y_k1} for $\tilde{k} = 1.0$ and in Figs.~\ref{u_v_T_x_k3} and \ref{u_v_T_y_k3} for $\tilde{k} = 5.0$. First, focusing on the dimensionless temperature profiles $T/T_H$, we notice that $\tilde{k}$ generally does change their overall magnitudes; however, it does change the shape of the temperature profiles in the direction vertical to the interface: while for $\tilde{k} = 1.0$ (see Fig.~\ref{u_v_T_y_k1}) it exhibits a continuous variation, when $\tilde{k}\neq 1$, a discontinuity in the slopes of the temperatures at the interface at $y/W=0.5$ can be observed (see Figs.~\ref{u_v_T_y_k13} and~\ref{u_v_T_y_k3}), which can be interpreted simply based on the continuity of the heat flux and using the Fourier's law. This also explains the observation that for the case when the top fluid layer is significantly more conducting than the bottom fluid layer (i.e., $\tilde{k} = 5.0$), the temperature field changes much more in the former when compared to the latter (see Fig.~\ref{u_v_T_y_k3}). More importantly, the thermal conductivity ratio has more profound influence on the magnitude of the thermocapillary flow fields. While the overall shapes of the components of the velocity fields are generally invariant with $\tilde{k}$, it can be seen that when bottom fluid is thermally more conducting, i.e., when $\tilde{k}<1.0$, the magnitudes of the thermocapillary velocity currents are significantly increased; for example, comparing  Figures~\ref{u_v_T_x_k13} and \ref{u_v_T_y_k13} (for $\tilde{k}=0.1$) with the corresponding Figs.~\ref{u_v_T_x_k3} and \ref{u_v_T_y_k3} (for $\tilde{k} = 5.0$), it can be observed that the Marangoni velocities are significantly larger for the former case when compared the latter. This is a consequence of the fact that when $\tilde{k} < 1.0$, the thermal conductivity of the bottom fluid layer is significantly larger relative to the top fluid layer thereby enhancing heat diffusion to the interface, which in turn sets up significantly larger surface tension gradient induced fluid motion. It is also consistent with the scaling equation for the characteristic thermocapillary velocity $U_s$ given above in Eq.~(\ref{U_s}) based on a stress balance on the interface, which parameterizes it with $\tilde{k}$, among other characteristic parameters. Finally, we also note that the theoretical predictions for the temperature fields as well as the velocity fields in SRFs based on our new analytical solution derived earlier are in good quantitative agreement with the numerical results based on the central moment LB schemes constructed in the previous sections.

\subsection{Effect of characteristic parameters on peak interfacial thermocapillary velocity $U_{max}$ in SRF layers}
\begin{figure}[H]
\vspace{-10pt}
\centering
\includegraphics[trim = 0 0 0 0,clip, width = 100mm]{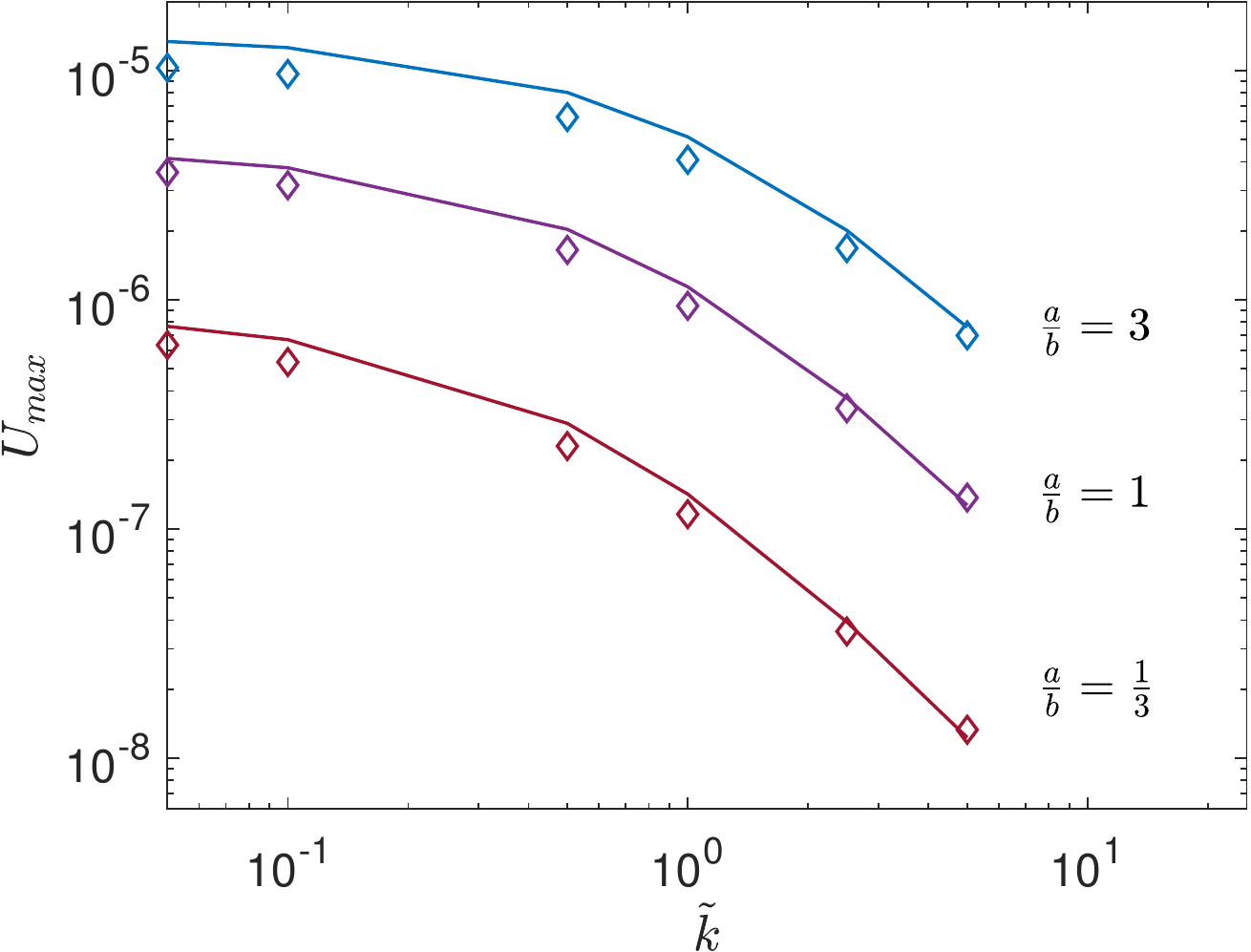}
\caption{Effect of the thermal conductivity ratio $\tilde{k}$ on the maximum thermocapillary velocity at the interface in SRFs for three different values of the aspect ratio $a/b$. The results from the analytical solution are shown as lines and LBM results are shown as symbols. Here, the viscosity ratio is  $\tilde{\mu}=1$, the dimensionless surface tension coefficients are $M_1 = 0$ and $M_2 = 5\times 10^{-2}$, and $\sigma_{0}=1\times 10^{-3}$.}
\label{Ur_Kr}
\end{figure}

Let's now study the effect of various dimensionless variables on the magnitude of the peak velocity generated on the interface $U_{max}$, as a global parameter indicating the strength of the thermocapillary convection in SRFs. First, we investigate the effect of the thermal conductivity ratio $\tilde{k}$ on $U_{max}$. Figure~\ref{Ur_Kr} shows the variation of the dimensionless peak velocity on the interface as a function of the thermal conductivity ratio for three different choices of the aspect ratio, viz., $a/b=1/3$, $1$, and $3$, when $\tilde{\mu}=1$. For a fixed $a/b$, it can be observed that as the thermal conductivity ratio increases, or the top fluid layer is thermally more conducting than the bottom fluid layer, $U_{max}$ is found to decrease monotonically; conversely, notice that the peak thermocapillary convection current can be enhanced by maintaining the thermal conductivity of the top fluid layer constant and increasing the thermal conductivity of the bottom fluid layer, or equivalently by decreasing $\tilde{k}$, which is a consequence larger thermal heat fluxes on the interface from the nonuniformly heated bottom wall. In addition, it is evident from Fig.~\ref{Ur_Kr} that the ratio of fluid thicknesses $a/b$ has a significant effect on $U_{max}$. In general, for a fixed thermal conductivity ratio, when the bottom fluid layer is thinner than the top fluid layer, i.e., $a/b > 1$, their interface lies closer to the heated bottom wall, which in turn enhances thermocapillary convection due to its stronger nonuniform heating, which results in a larger peak Marangoni velocity. For example, when $\tilde{k}=1.0$, by changing $a/b$ from $1/3$ to $3$ increases $U_r$ by more than ten times.

\begin{figure}[H]
\vspace{-10pt}
\centering
\includegraphics[trim = 0 0 0 0,clip, width = 100mm]{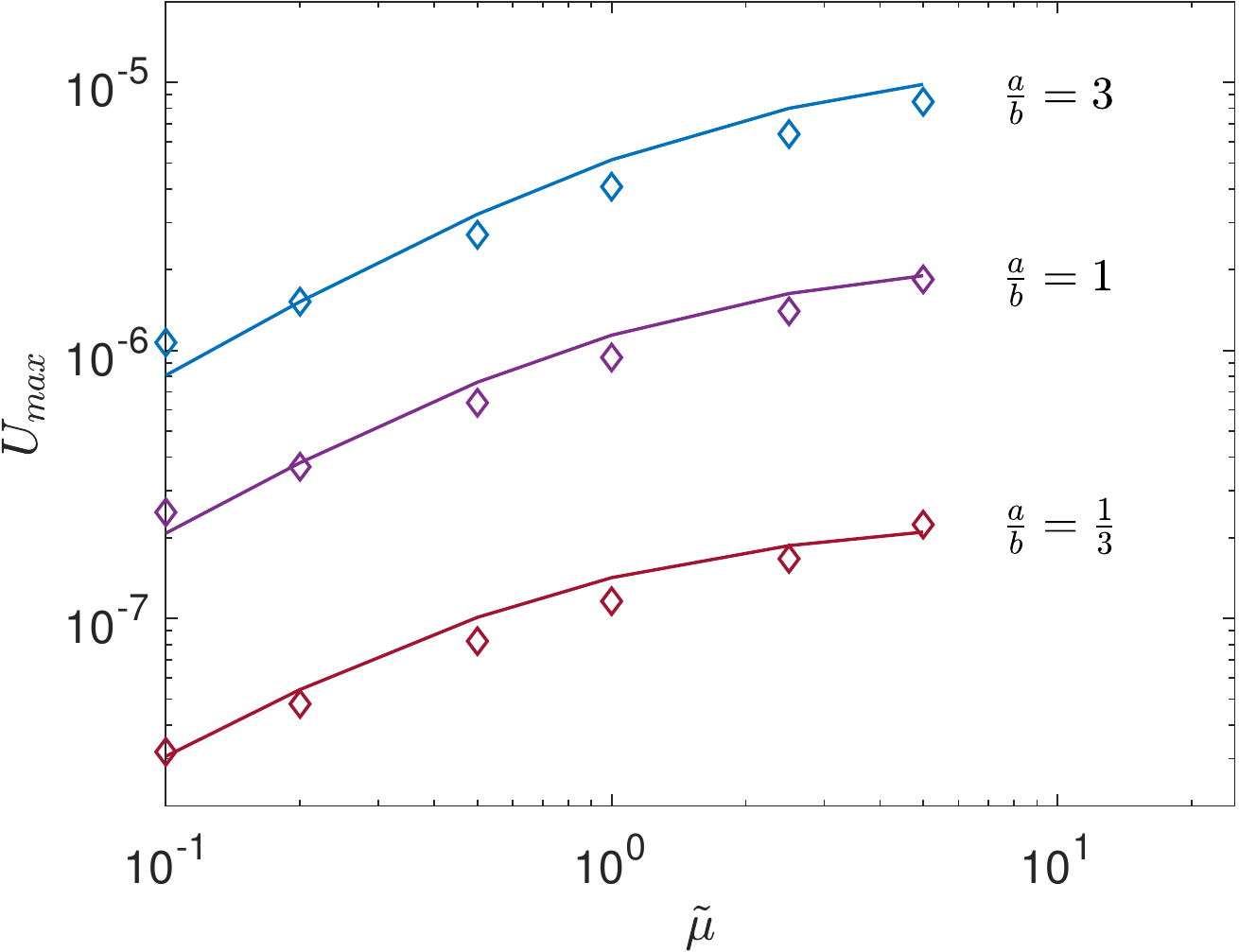}
\caption{Effect of the viscosity ratio $\tilde{\mu}$ on the maximum thermocapillary velocity at the interface in SRFs for three different values of the aspect ratio $a/b$. The results from the analytical solution are shown as lines and LBM results are shown as symbols. Here, the thermal conductivity ratio is $\tilde{k}=1$, the dimensionless surface tension coefficients are $M_1 = 0$ and $M_2 = 5\times 10^{-2}$, and $\sigma_{0}=1\times 10^{-3}$.}
\label{Ur_mur}
\end{figure}

Next, Fig.~\ref{Ur_mur} shows the effect of the dimensionless viscosity ratio $\tilde{\mu}=\mu_a/\mu_b$ on the peak thermocapillary velocity $U_{max}$ for $a/b=1/3$, $1$, and $3$ at a fixed $\tilde{k}=1$. Clearly, the viscosities of the SRFs have profound influence on the strength of thermocapillary convection. In particular, if the bottom fluid layer is less viscous than the top fluid layer, or $\tilde{\mu}>1$, it facilitates the exchange of momentum transfer between layers of the bottom fluid and the interface, which is accompanied by larger peak thermocapillary velocities. Thus, the increasing the viscosity ratio has an opposite effect when compared to the thermal conductivity ratio. On the other hand, for a fixed viscosity ratio, the variations in the thickness ratio $a/b$ has a similar influence as noticed in the previous case.

\begin{figure}[H]
\vspace{-10pt}
\centering
\includegraphics[trim = 0 0 0 0,clip, width = 100mm]{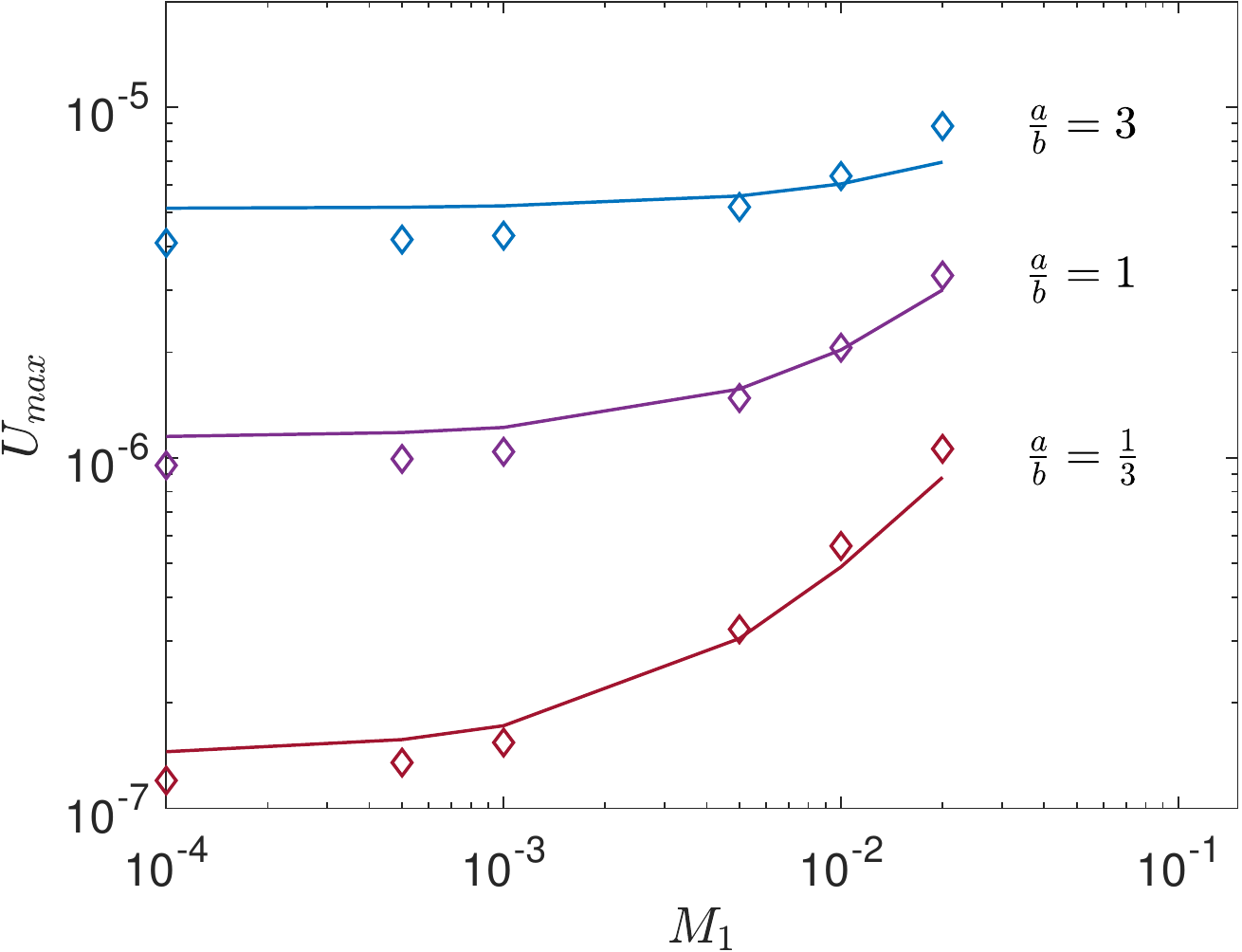}
\caption{Effect of the dimensionless linear coefficient of surface tension $M_1$ on the maximum velocity at the interface in SRFs for three different values of the aspect ratio $a/b$ at $\tilde{k}=1$, $\tilde{\mu}=1$, and $M_2=5 \times 10^{-2}$. The results from the analytical solution are shown as lines and LBM results are shown as symbols.}
\label{Ur_M1}
\end{figure}

\begin{figure}[H]
\vspace{-10pt}
\centering
\includegraphics[trim = 0 0 0 0,clip, width = 100mm]{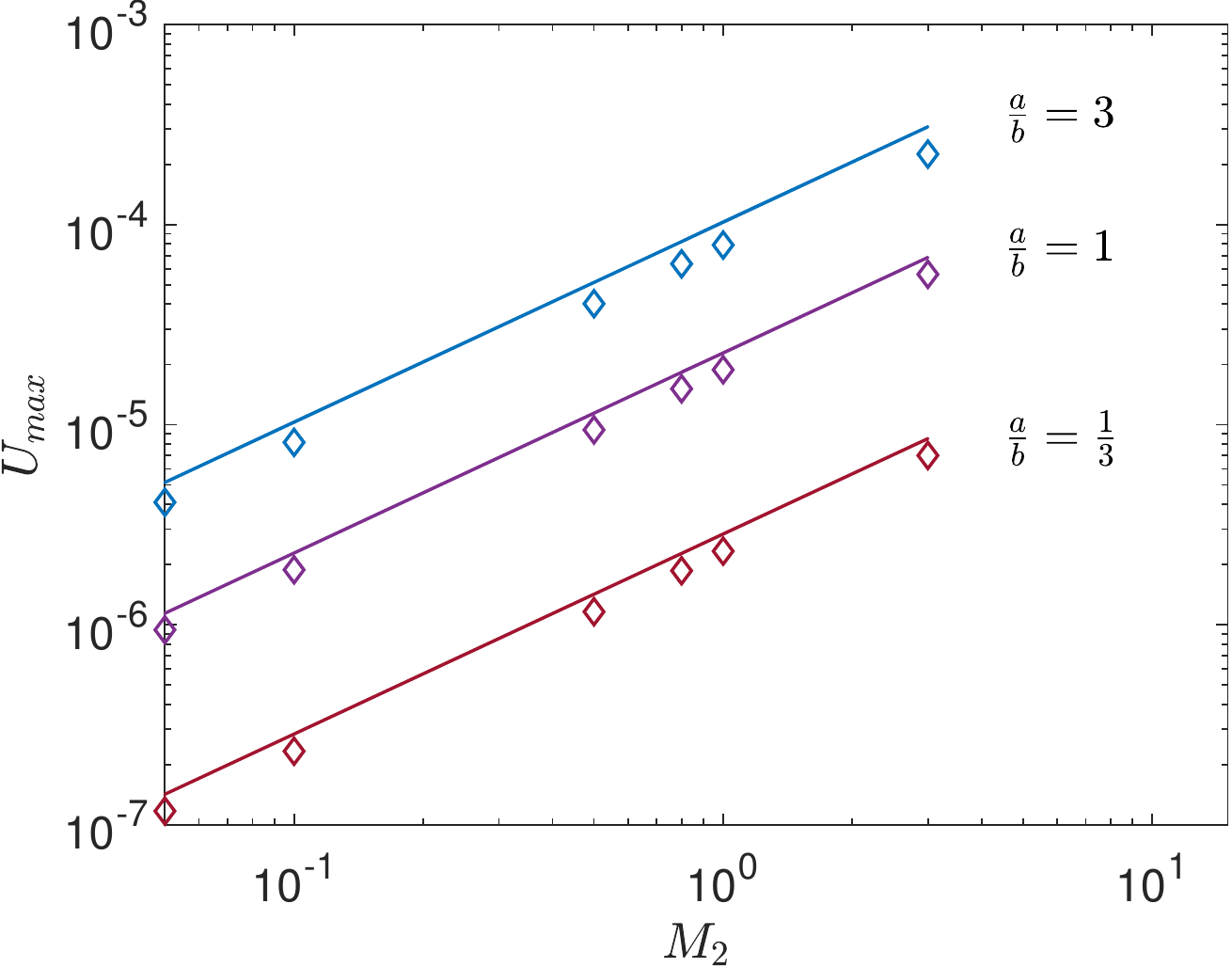}
\caption{Effect of the dimensionless quadratic coefficient of surface tension $M_2$ onthe maximum velocity at the interface in SRFs for three different values of the aspect ratio $a/b$ at $\tilde{k}=1$, $\tilde{\mu}=1$, and $M_1=1 \times 10^{-5}$. The results from the analytical solution are shown as lines and LBM results are shown as symbols.}
\label{Ur_M2}
\end{figure}

Finally, Figs.~\ref{Ur_M1} and \ref{Ur_M2} present the effects of the dimensionless linear and quadratic cofficients, $M_1$ and $M_2$, respectively, on the peak thermocapillary velocity $U_{max}$ for $a/b=1/3$, $1$, and $3$ at fixed $\tilde{k}=1$ and $\tilde{\mu}=1$ (see Eq.~(\ref{three-prime}) for their definitions based on $\sigma_T$ and $\sigma_{TT}$). Evidently, increasing either $M_1$ or $M_2$ increases the strength of the Marangoni velocity, which is a manifestation of the stronger tendency of the SRFs in promoting thermocapillary flow via larger surface tension gradients or Marangoni stresses on the interface. Generally, such effects are more pronounced when the interface is located closer to the heated bottom wall or with increasing $a/b$. Interestingly, it is noted that the effect of variations in the linear surface tension coefficient $M_1$ on $U_{max}$ for fixed $M_2=5\times 10^{-2}$ is greater at $a/b=1/3$ when compared to the other cases; on the other hand, for a fixed $M_1=1\times 10^{-5}$,  $U_{max}$ increases in direct proportion with an increase in the quadratic coefficient $M_2$ with a constant slope (which is consistent with the characteristic velocity dependence on $\sigma_{TT}$ or, equivalently, $M_2$ given in Eq.~(\ref{U_s})) for all choices of the aspect ratio $a/b$.

\subsection{Beyond the analytical solution: Interfacial deformations at higher capillary numbers using lattice Boltzmann simulations}
In the derivation of the new analytical solution for thermocapillary convection in SRF layers, it was assumed that the interface remains flat which is a reasonable assumption at relatively small capillary numbers and applicable for microchannel configurations. It is consistent with those considered in prior work (see e.g.,~\cite{pendse2010analytical}) and, as shown in the previous section, the results obtained from such a theoretical analysis were in good quantitative agreement with the numerical simulations under similar conditions. However, it should be pointed out that the computational approach discussed in Sec.~\ref{Sec.LBschemes} is not restricted by such assumptions and is applicable for more general situations, where the interfaces between the SRFs can deform, which can arise at relatively large capillary numbers. In order to simply illustrate this viewpoint, we have performed some additional simulations involving SRF layers at progressively increasing values of the capillary number $\mbox{Ca}$ while maintaining $M_1= 0 $ and $\sigma_0 = 1 \times 10^{-3}$ with $a/b= 3$ with thermal conductivity ratio $\tilde{k} = 1$ and viscosity ratio $\tilde{\mu}=1$. We chose the interface to be closer to the heated bottom wall by fixing $a/b=3$ so that more pronounced thermocapillary convection are generated, whose magnitudes are controlled by varying the parameter $M_2$ which in turn determines the characteristic velocity used in defining the capillary number.

\begin{figure}[]
\centering
\begin{subfigure}{0.475\textwidth}
\includegraphics[trim = 0 35 0 60,clip, width = 80mm]{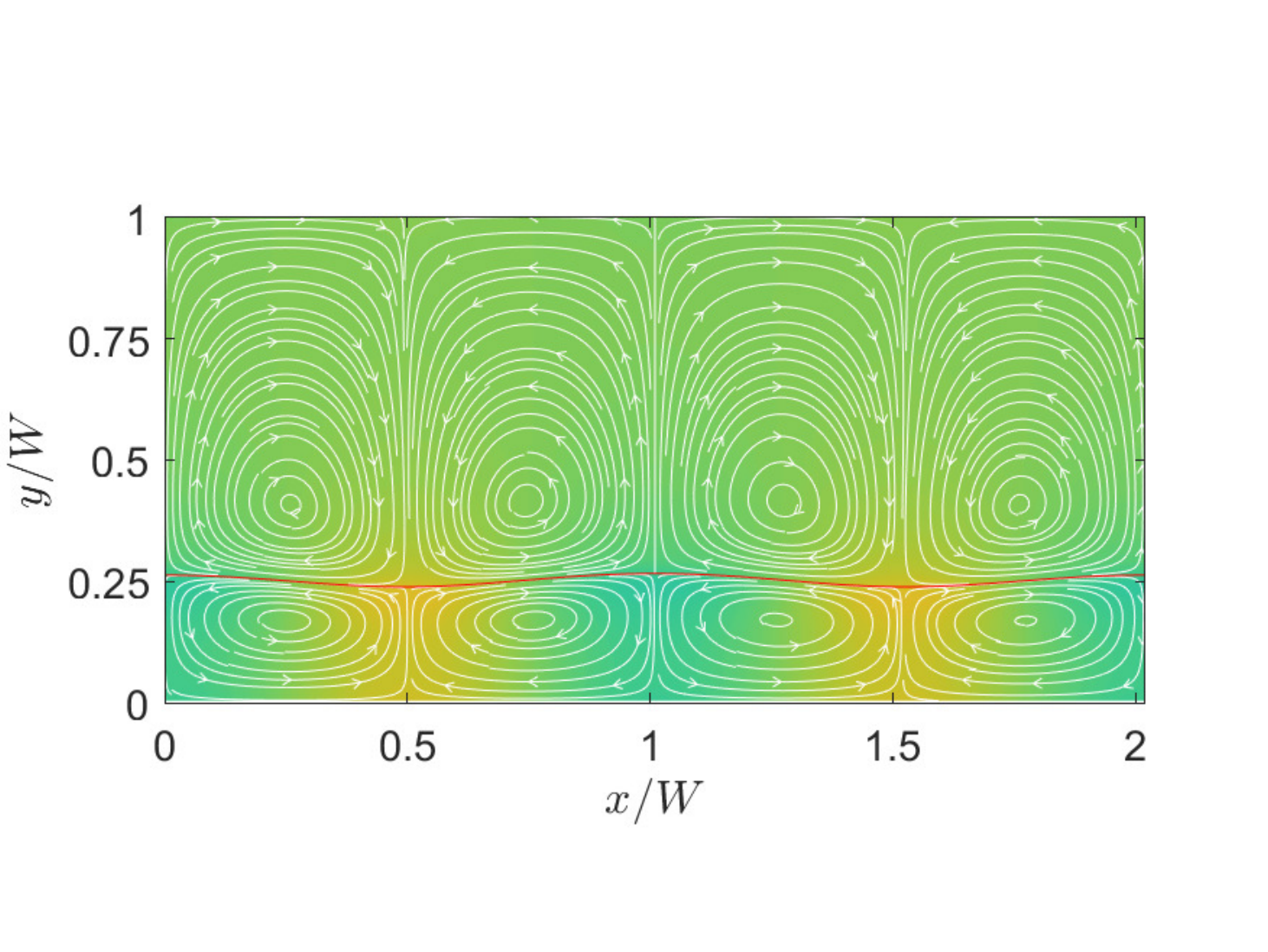}
\caption{$\mbox{Ca} = 0.34$}
\end{subfigure}
\begin{subfigure}{0.475\textwidth}
\includegraphics[trim = 0 0 0 0,clip, width = 72mm]{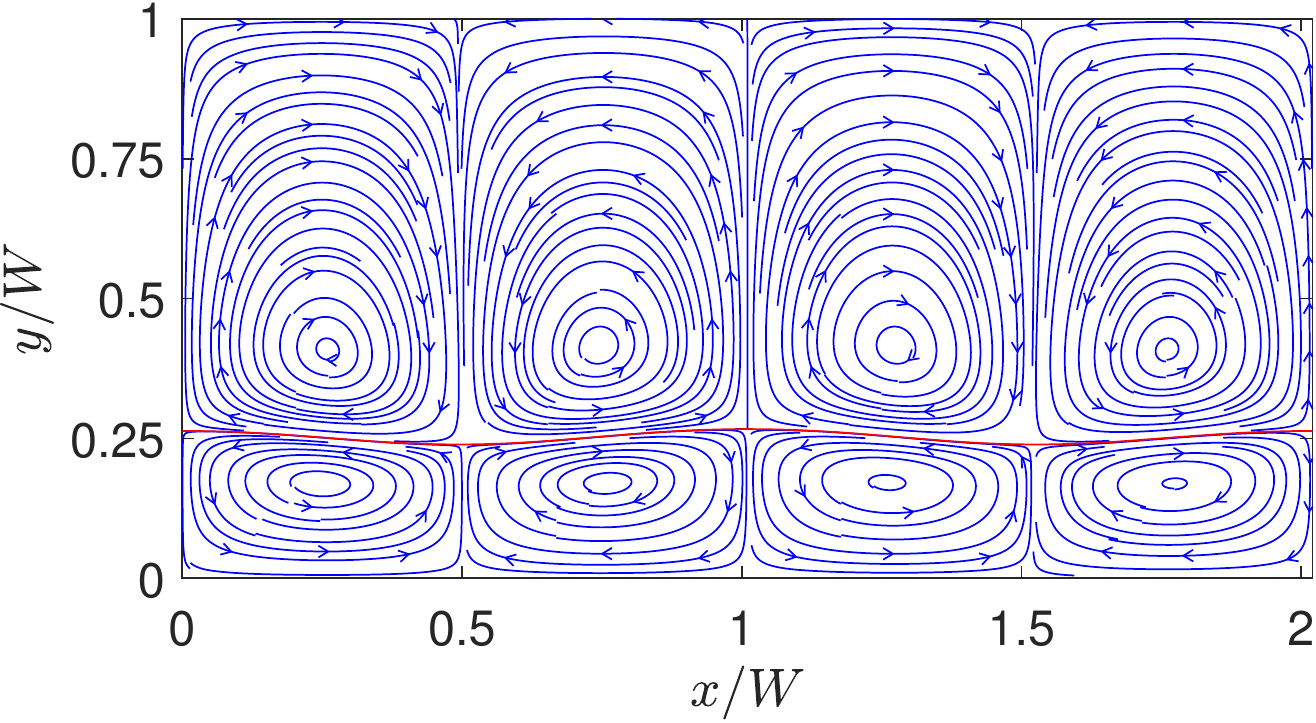}
\caption{$\mbox{Ca} = 0.34$}
\end{subfigure}
\begin{subfigure}{0.475\textwidth}
\includegraphics[trim = 0 35 0 60,clip, width = 80mm]{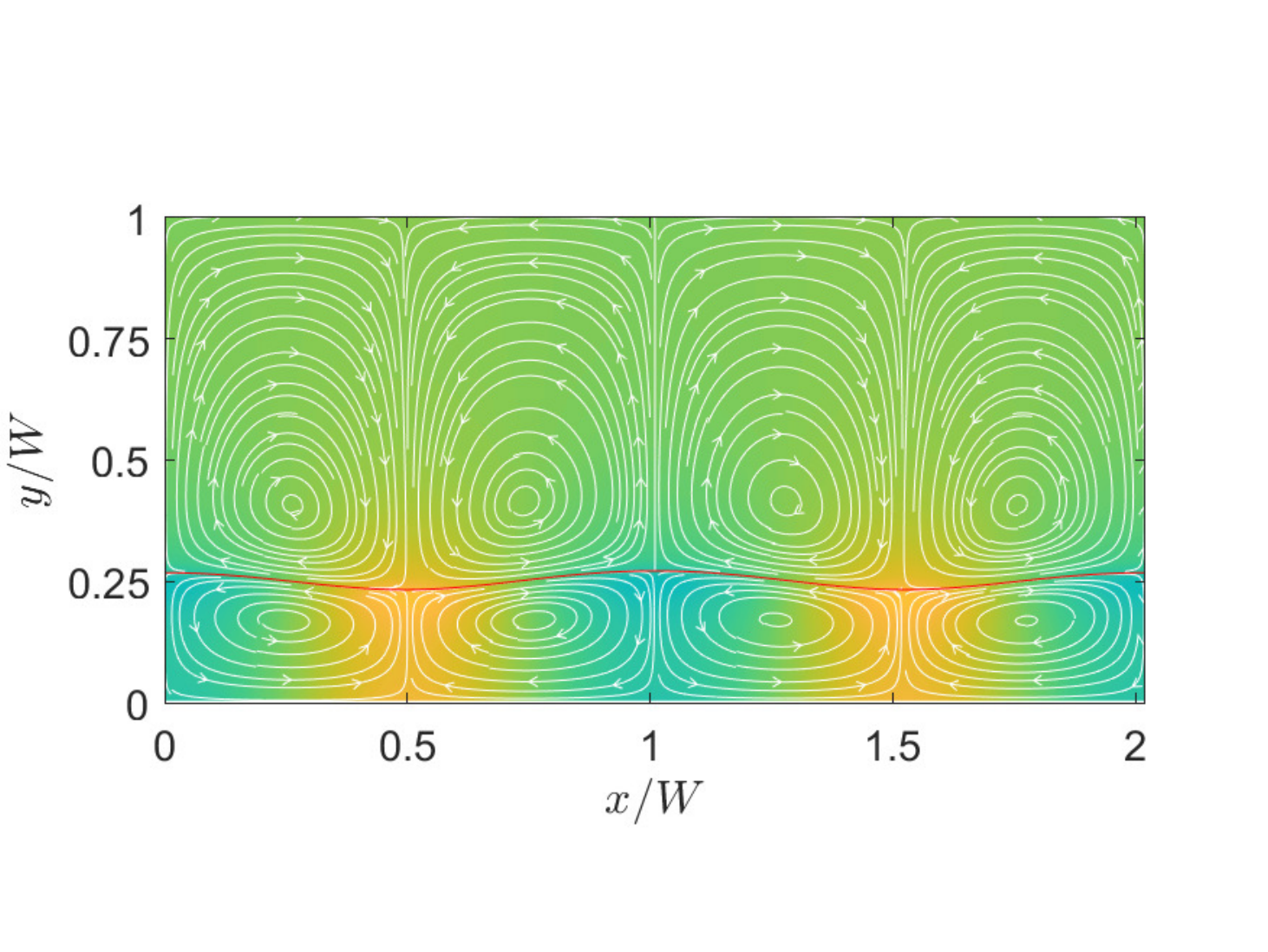}
\caption{$\mbox{Ca} = 0.57$}
\end{subfigure}
\begin{subfigure}{0.475\textwidth}
\includegraphics[trim = 0 0 0 0,clip, width = 72mm]{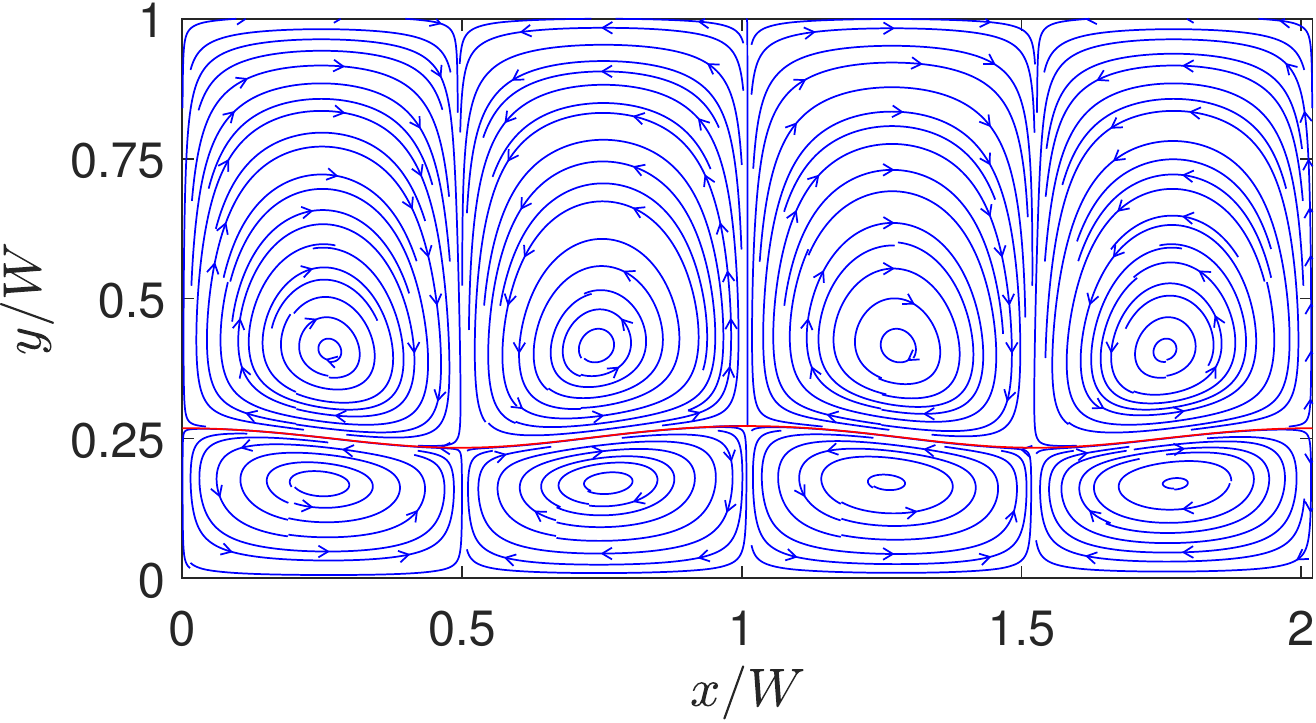}
\caption{$\mbox{Ca} =  0.57$}
\end{subfigure}
\begin{subfigure}{0.475\textwidth}
\includegraphics[trim = 0 35 0 60,clip, width = 80mm]{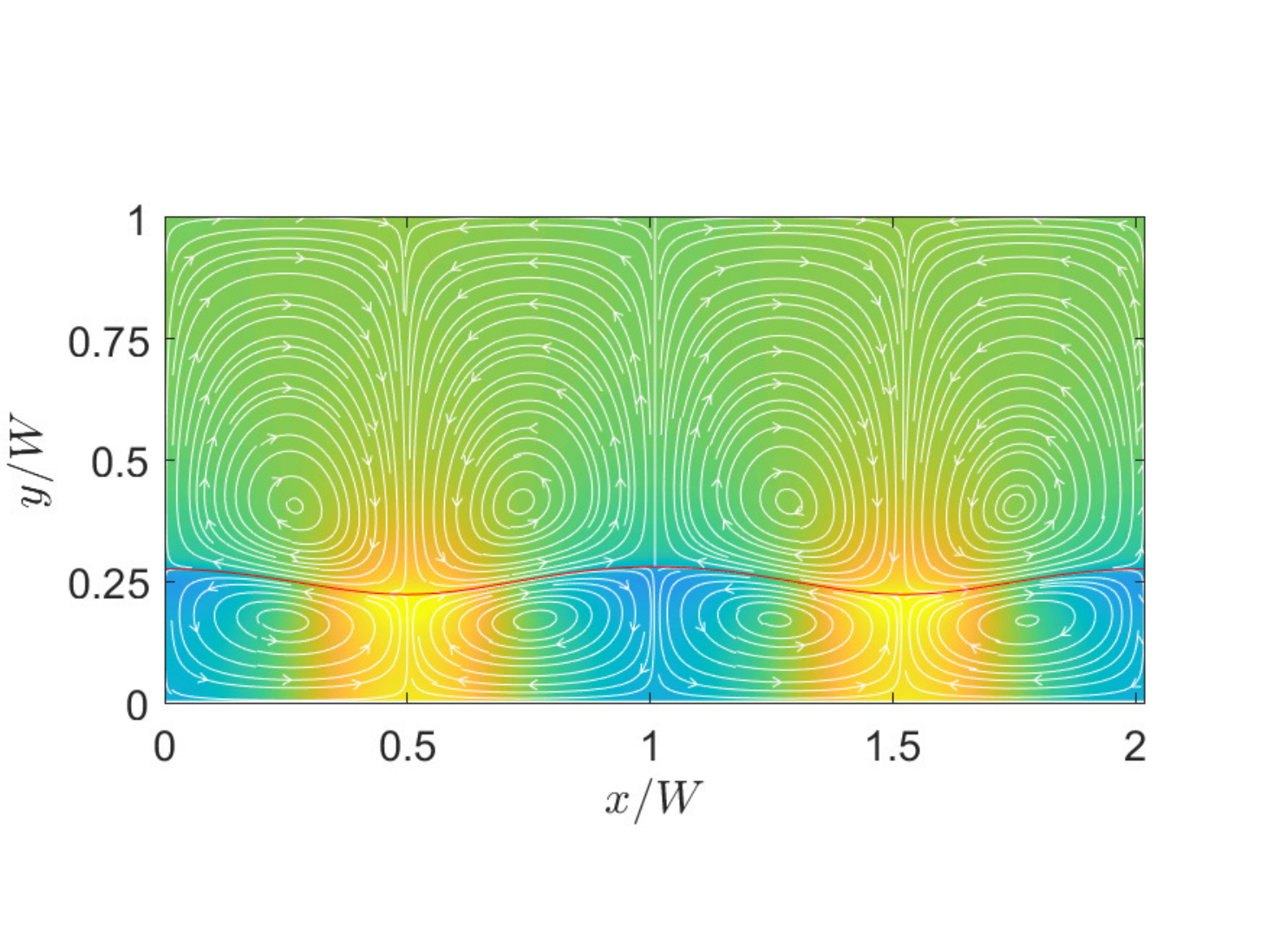}
\caption{$\mbox{Ca} = 1.15$}
\end{subfigure}
\begin{subfigure}{0.475\textwidth}
\includegraphics[trim = 0 0 0 0,clip, width = 72mm]{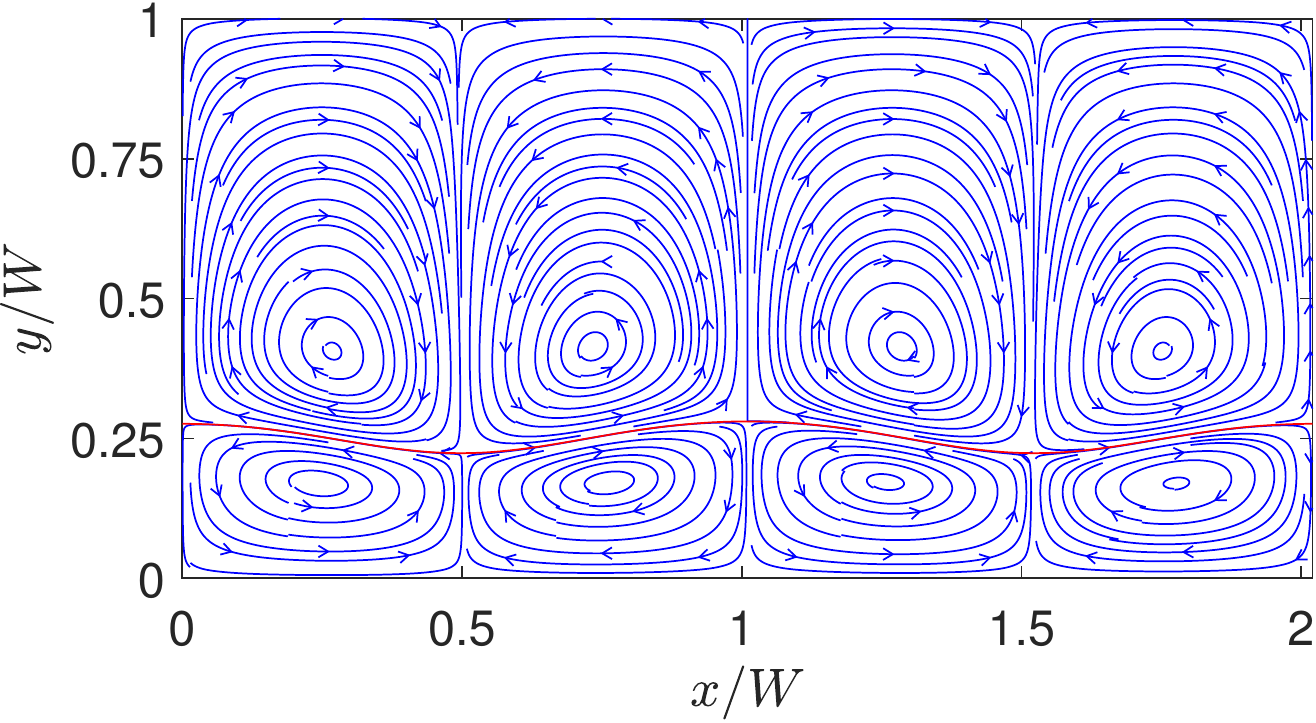}
\caption{$\mbox{Ca} = 1.15$}
\end{subfigure}
\begin{subfigure}{0.475\textwidth}
\includegraphics[trim = 0 35 0 60,clip, width = 80mm]{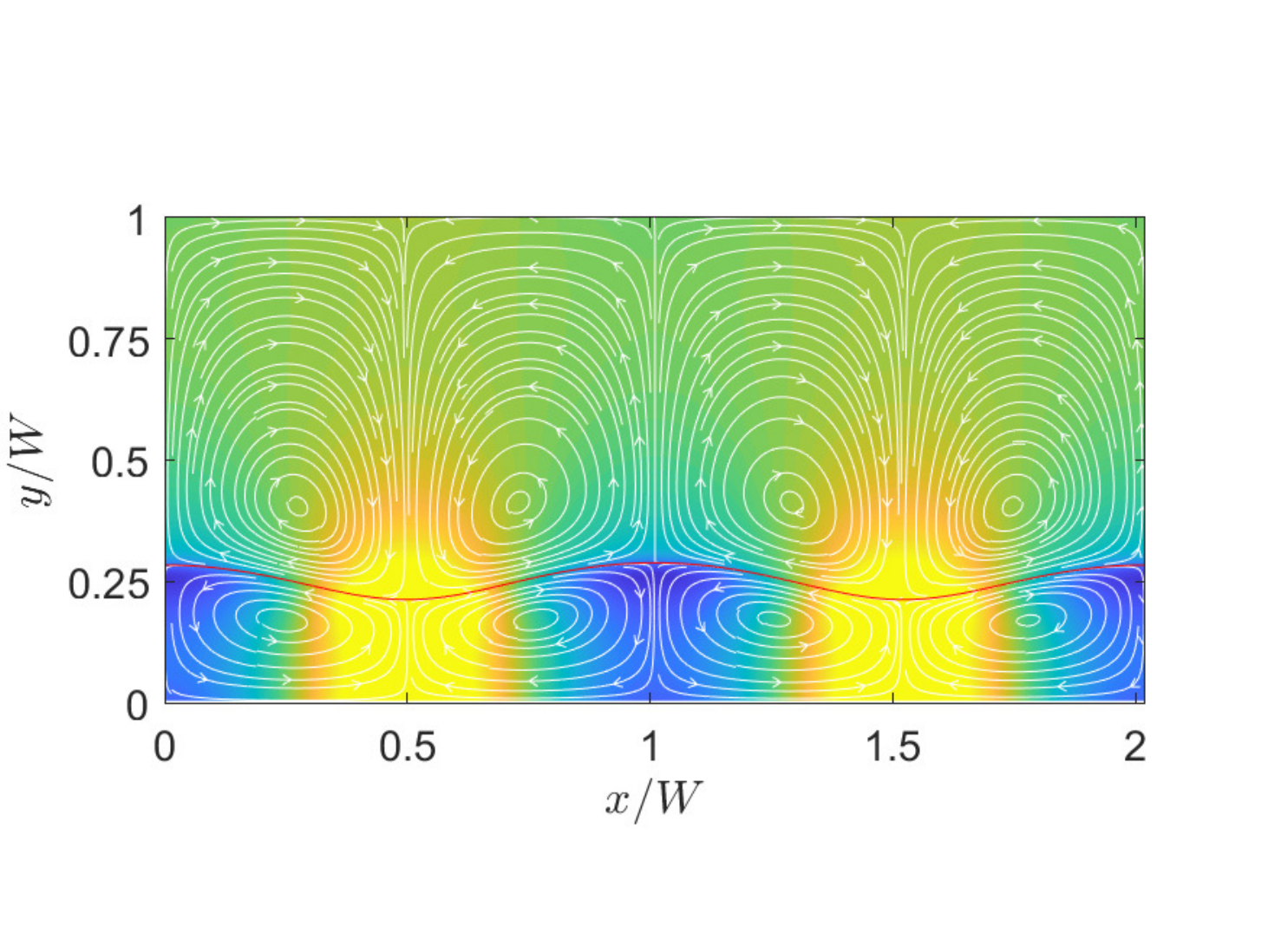}
\caption{$\mbox{Ca} = 2.29$}
\end{subfigure}
\begin{subfigure}{0.475\textwidth}
\includegraphics[trim = 0 0 0 0,clip, width = 72mm]{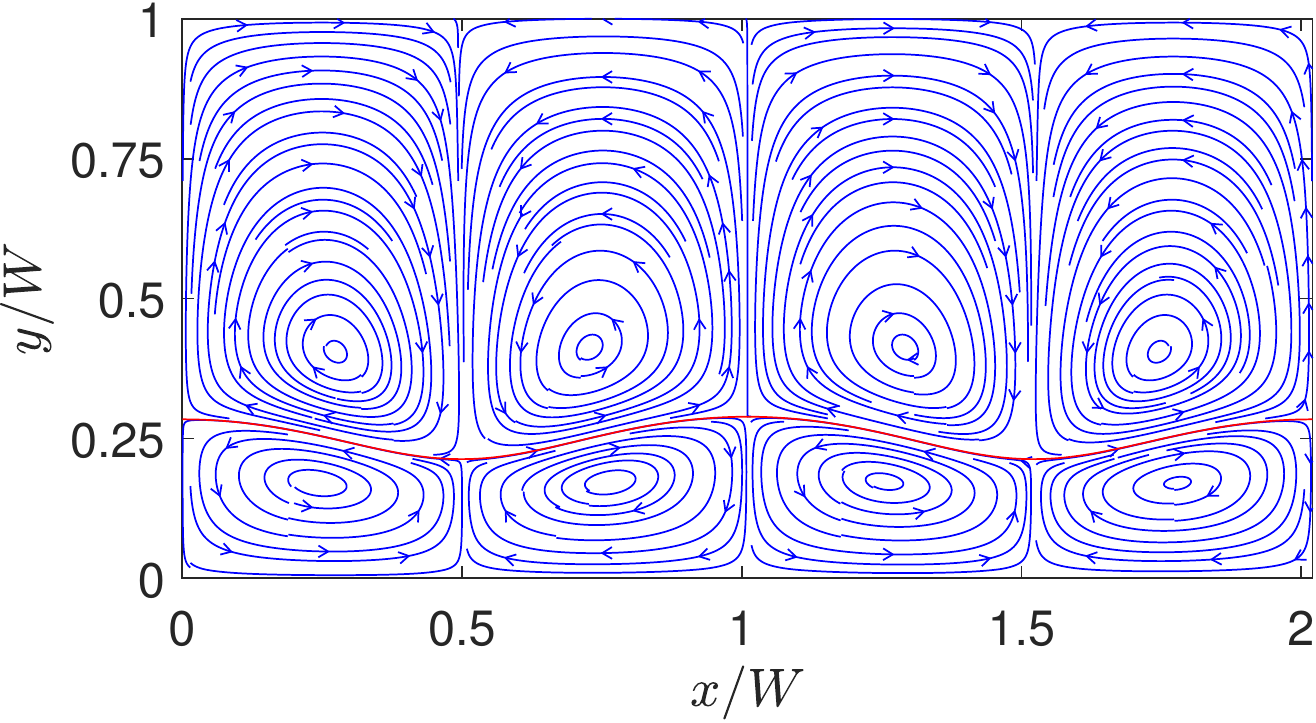}
\caption{$\mbox{Ca} = 2.29$}
\end{subfigure}
\begin{subfigure}{0.475\textwidth}
\includegraphics[trim = 0 0 0 230,clip, width = 80mm]{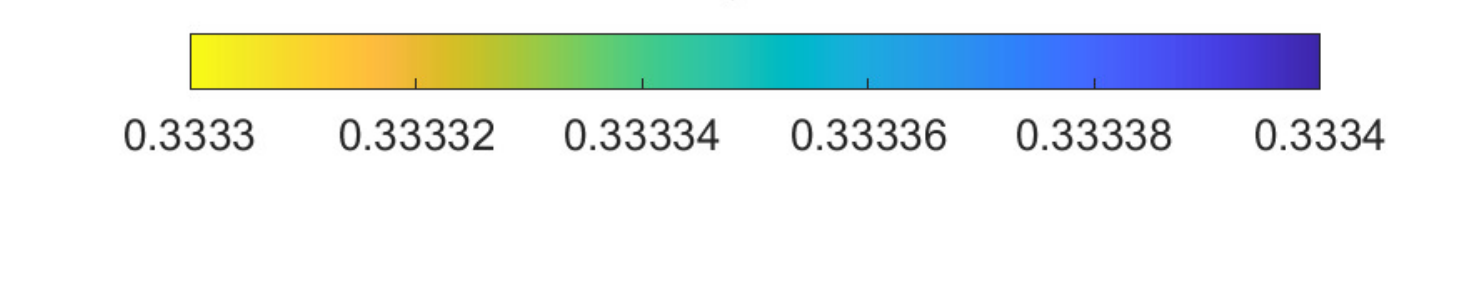}
\end{subfigure}
\caption{Simulations of interfacial deformations using the central moment LB schemes at higher capillary numbers in SRF layers. (a) Pressure contours and (b) streamlines of the thermocapillary flow for aspect ratio $a/b= 3$ with thermal conductivity ratio $\tilde{k} = 1$ and viscosity ratio $\tilde{\mu}=1$ at different capillary numbers $\mbox{Ca}$.}\label{fig:interfacedeformationslargercapillarynumbers}
\end{figure}

Figure~\ref{fig:interfacedeformationslargercapillarynumbers} shows the contours of the pressure field and the streamline patterns in SRF layers computed using the LB schemes at $\mbox{Ca} = 0.34, 0.57, 1.15,$ and $2.29$ via varying $M_2$ as $M_2=3, 5, 10,$ and $20$, respectively. In the results discussed earlier where $\mbox{Ca}<0.1$, the interface was found to be essentially flat and both the analytical solution and the numerical simulations were consistent with each other. By contrast, according to Fig.~\ref{fig:interfacedeformationslargercapillarynumbers}, as $\mbox{Ca}$ is increased to $0.34$, the simulations show that the interfaces undergo relatively small deformations. As $\mbox{Ca}$ is progressively increased further, the interfaces deform more significantly. These result from the differences in the pressure fields between the bottom and top SRF layers which is accompanied by local variations in the curvatures or the normal capillary forces as seen from the pressure contour plots. Clearly, larger the capillary number, the larger is the pressure differences or the greater is the interfacial deformations. Interestingly, despite such interfacial deformations, the thermocapillary flow patterns are seen to be qualitatively similar to that of the flat interface cases considered earlier in that the SRF layers are accompanied with eight counterrotating vortex cells. As such, these demonstrate the capabilities of the central moment LB schemes in computing local variations in the interfacial topologies naturally and their potential of going beyond the possible parametric space of the analytical solution in simulating thermocapillary flow in SRF layers.

\section{Summary and Conclusions} \label{Sec.9}
Surface tension in fluids is a temperature dependent property and is among the main drivers of interfacial transport phenomena. In contrast to the normal fluids (NFs), the self-rewetting fluids (SRFs) exhibit anomalous nonlinear (quadratic) dependence of surface tension on temperature with a minimum and involving a positive gradient. As a result, they are accompanied by certain desirable aspects, such as interfacial fluid motions towards high temperature regions, which can be potentially exploited in various microgravity and terrestrial applications, including microfluidics.

In this paper, we have derived a new analytical solution for thermocapillary convection in superimposed two SRF fluid layers confined within a microchannel that is heated on its bottom side with a sinusoidally varying temperature. Under the creeping flow limit and at small capillary numbers, the derived streamfunction from solving a biharmonic equation consists of a fundamental solution resulting from the linear part of the surface tension and a higher order harmonic solution with a wavenumber that is twice that of the former and arises from the quadratic part of the surface tension variation on the temperature. Moreover, we have also developed a robust numerical technique based on central moment lattice Boltzmann (LB) schemes for interface tracking based on a conservative Allen-Cahn equation, two-fluid motion, and energy transport to simulate thermocapillary convection in SRFs. Such a computational approach was first validated against some other existing benchmark problems, and then applied for the study of thermocapillary convection in SRF layers in a heated microchannel as mentioned above.

It is found that the two SRF layers are accompanied by a set of eight, periodic counterrotating convection cells with the interfacial fluid motion directed towards the high temperature at the center; by contrast, in the two NF layers, only four periodic counterrotating vortices are generated with the fluids moving away from the center along the interface. Such striking differences are well reproduced by both our analytical and computational approaches, and they are found to be in good quantitative agreement. The presence of double the number of convection cells in SRFs when compared that in NFs can be theoretically interpreted as arising from the higher order harmonic solution as noted above. It is shown that the magnitude of the linear coefficient of the surface tension variation with temperature relative to that of the quadratic coefficient of SRFs not only affects the strength of thermocapillary velocities, but also the character of the overall convection patterns. Moreover, a study of the effect of various characteristic parameters such as the thickness ratio of the fluids, thermal conductivity ratio and the viscosity ratio on the magnitude of thermocapillary convection was performed. It is found that the thermocapillary convection currents are more intense when the interface is closer to the heated bottom wall, or if the bottom fluid layer has higher thermal conductivity or lower viscosity when compared to those in the top fluid layer. Furthermore, the larger magnitudes of the surface tension coefficients, which set up greater Marangoni stresses on the interface, are accompanied with higher thermocapillary velocities. By going beyond the analytical solution regime, computations show that at relatively larger capillary numbers the interfaces undergo deformation while maintaining the general flow patterns in SRFs as noted above.

The analytical solution for thermocapillary convection in SRFs derived in this work is useful not only in clarifying the essential transport physics involved, including its ability to predict the doubling of the number of vortex cells in SRFs when compared to that in NFs, but may also serve as a benchmark solution in constructing new numerical techniques for simulating thermocapillary flows in SRFs. The central moment LB schemes are not only quantitatively in agreement with such a solution, but provides an approach to extend it to more general situations involving interfacial deformations. The ability to modulate both the surface-tension driven flow patterns and their magnitudes in SRFs in certain unique manner relative to NFs, such as those shown in this work, could provide new approaches in manipulating interfacial transport phenomena in microfluidic applications, amongst others.

\appendix

\section{Solution of the Energy Equation: Temperature Field}\label{Appendix A}
The solution to the energy equation Eq.~(\ref{energy}) is invariant with the nature of the fluid, i.e., whether it is for a NF or a SRF, and hence the results reported in~\cite{pendse2010analytical} for the temperature field is valid here as well. However, for completeness, we provide all the necessary details involved in the solution procedure in what follows.

The energy equation is not coupled to the momentum equation directly; however, the momentum equation is partially coupled to the energy equation through the equation of state, such that any updates in the temperature field change the surface tension, which then changes the interface Marangoni stress condition and hence the fluid momentum. The energy equation is homogeneous and has periodic boundary conditions in the $x$-direction. Furthermore, the only non-homogeneities are in the upper and lower boundary conditions. To solve this problem, the non-homogeneous boundary conditions can be split across two solutions as we will see next. The energy equation is given as in Eq.(\ref{energy})
\begin{eqnarray} \label{A1}
\frac{\partial^2 T}{\partial x^2} + \frac{\partial^2 T}{\partial y^2} = 0,
\end{eqnarray}
which is subject to the following boundary conditions
\begin{eqnarray}
T^b(x,-b)=T_h+\Delta T \cos(\omega x),\nonumber
\end{eqnarray}
and
\begin{eqnarray}
T^a(x,a)=T_c.\nonumber
\end{eqnarray}
Because of the homogeneity and linearity of the differential equation and that the temperature is periodic in the horizontal direction, the method of separation of variables is used to solve the temperature equation,
\begin{equation} \label{A2}
 T^i(x,y)=P^i(x,y)+Q^i(y), \qquad i = a,b,
\end{equation}
where $P^i(x,y)$ and $Q^i(y)$ are the perturbation and linear temperature fields, respectively.
Substituting Eq.(\ref{A2}) into the energy equation (Eq.(\ref{A1})) gives the following separated equations that we need to solve
\begin{equation}
\frac{\partial^2 P^i}{\partial x^2} + \frac{\partial^2 P^i}{\partial y^2} = 0, \qquad  \text{and} \qquad \frac{\partial^2 Q^i}{\partial y^2} = 0, \qquad i = a,b, \nonumber
\end{equation}
which are subject to the following boundary conditions. \newline
\noindent
i) The temperature is specified at the lower wall:
\begin{eqnarray}
P^b(x,-b)=\Delta T \cos(\omega x), \qquad  \text{and} \qquad Q^b (x,-b) = T_h.\nonumber
\end{eqnarray}
\noindent
ii) The temperature is specified at the upper wall:
\begin{eqnarray}
P^a(x,a)=0, \qquad  \text{and} \qquad Q^a (x,a) = T_c.\nonumber
\end{eqnarray}
\noindent
iii) The temperature is continuous at the interface:
\begin{eqnarray}
P^a(x,0)=P^b(x,0), \qquad  \text{and} \qquad Q^a (x,0) = Q^b (x,0).\nonumber
\end{eqnarray}
\noindent
iv) The heat flux is continuous at the interface:
\begin{eqnarray}
-k_b \frac{\partial P^b}{\partial y}\Biggr\rvert_{y=0} = -k_a \frac{\partial P^a}{\partial y}\Biggr\rvert_{y=0}, \qquad  \text{and} \qquad  -k_b\frac{\partial Q^b}{\partial y}\Biggr\rvert_{y=0} = -k_a \frac{\partial Q^a}{\partial y}\Biggr\rvert_{y=0}.\nonumber
\end{eqnarray}

The solution for the linear temperature field is $Q^i(y)=A^i_1 y + A^i_2$. Applying the above boundary conditions to get the constants of integration which yields in the solution for the lower wall
\begin{equation}
   Q^b(y)=\frac{k_a(T_c-T_h)y+T_ck_ab+T_hk_ba}{(ak_b+bk_a)}.
\end{equation}
Similarly, the solution for the upper wall is :
\begin{equation}
   Q^a(y)=\frac{k_b(T_c-T_h)y+T_ck_ab+T_hk_ba}{(ak_b+bk_a)}.
\end{equation}

Then by the standard separation of variables method, and by looking at the lower boundary condition, the solution for the perturbation in the temperature field $P^i(x,y)$  for the lower fluid is
\begin{equation}\label{A3}
P^b(x,y)=[A_1^b \cosh(\omega y)+A_2^b \sinh(\omega y)]\cos(\omega x).
\end{equation}
Similarly, for the solution for the upper fluid is,
\begin{equation}\label{A4}
P^a(x,y)=[A_1^a \cosh(\omega y)+A_2^a \sinh(\omega y)]\cos(\omega x).
\end{equation}

Now, by applying the above four boundary condition, we get the following constants
\begin{equation}
\begin{split}
&A_1^a=A_1^b=\Delta T \sinh(\tilde a) f(\tilde a,\tilde b,\tilde k), \\
&A_2^a=-\Delta T \cosh(\tilde a) f(\tilde a,\tilde b,\tilde k),\\
& A_2^b=-\Delta T \tilde k \cosh(\tilde a) f(\tilde a,\tilde b,\tilde k), \nonumber \\
\end{split}
\end{equation}
where
\begin{equation} \label{f}
f(\tilde a,\tilde b,\tilde k)=\left[\tilde k \sinh(\tilde b)\cosh(\tilde a)+\sinh(\tilde a)\cosh(\tilde b)\right]^{-1},
\end{equation}
where $\tilde a = a \omega , \quad \textrm{and} \quad \tilde b = b \omega$, and $\tilde{k}=k_a/k_b$. Substitution of the above constants ($A_1^a$, $A_2^a$, $A_1^b$, and $A_2^b$) in Eqs.~(\ref{A3}) and (\ref{A4}) results in the final solution of the perturbation temperature $P^i(x,y)$ in the lower fluid as
\begin{equation}
P^b(x,y)= \Delta T f(\tilde a,\tilde b,\tilde k)[\sinh(\tilde a)\cosh(\omega y)- \tilde k \sinh(\omega y)\cosh(\tilde a)]\cos(\omega x).
\end{equation}
Similarly, for the upper fluid,
\begin{equation}
P^a(x,y)= \Delta T f(\tilde a,\tilde b,\tilde k)\sinh(\tilde a-\omega y)\cos(\omega x).
\end{equation}

\section{Mapping Relations for the Central Moment LB Scheme on a D2Q9 lattice} \label{App B}
Here, we summarize the various mapping relations that are needed prior to and following the collision step, where different central moments are relaxed to their equilibria, in the central moment LB scheme on the D2Q9 lattice.

The transformation matrix $\tensr{P}$ mapping a vector of distribution functions $\mathbf{f}$ to a vector of raw moments $\bm{\kappa^{'}}$ is given by
\begin{equation}\label{eq:tensorP}
\tensr{P} = \begin{bmatrix}
     1  &\quad    1  &\quad    1  &\quad      1  &\quad    1  &\quad     1 &\quad    1  &\quad     1  &\quad     1 \\[10pt]
     0  &\quad    1  &\quad     0  &\quad    \um1  &\quad     0  &\quad    1 &\quad   \um1 &\quad   \um1  &\quad     1 \\[10pt]
     0  &\quad    0  &\quad     1  &\quad     0  &\quad   \um1  &\quad    1  &\quad     1  &\quad    \um1  &\quad  \um1 \\[10pt]
     0  &\quad    1  &\quad     0  &\quad     1  &\quad     0  &\quad    1  &\quad     1  &\quad     1  &\quad     1 \\[10pt]
     0  &\quad    0  &\quad     1  &\quad     0  &\quad     1  &\quad    1  & \quad    1  &\quad     1  &\quad     1 \\[10pt]
     0  &\quad    0  &\quad     0  &\quad     0  &\quad     0  &\quad    1  &\quad    \um1  &\quad     1  &\quad   \um1 \\[10pt]
     0  &\quad    0  &\quad     0  &\quad     0  &\quad     0  &\quad    1  &\quad     1  &\quad  \um1  &\quad  \um1 \\[10pt]
     0  &\quad    0  &\quad     0  &\quad     0  &\quad     0  &\quad    1  &\quad    \um1  &\quad  \um1  &\quad    1 \\[10pt]
     0  &\quad    0  &\quad     0  & \quad    0  &\quad     0  & \quad   1  &\quad     1  &\quad     1  &\quad    1
\end{bmatrix}
\end{equation}
Next, the transformation matrix $\tensr{F}$ mapping a vector of raw moments $\bm{\kappa^{'}}$ to a vector of central moments $\bm{\kappa}$ reads as
\begin{equation}\label{eq:tensorF}
\tensr{F}=
\begin{bmatrix}
      1  &    0  &    0  &     0  &    0  &     0 &    0  &     0  &     0 \\[10pt]

     \um u_x  &   1  &    0  &     0  &    0  &     0 &    0  &     0  &     0 \\[10pt]

      \um u_y  &    0  &   1  &     0  &    0  &     0 &    0  &     0  &     0 \\[10pt]

      u_x^2  &   \um 2u_x  &    0  &     1  &    0  &     0 &    0  &     0  &     0 \\[10pt]

      u_y^2  &    0  &    \um 2u_y  &     0  &    1  &     0 &    0  &     0  &     0 \\[10pt]

     u_x u_y  &   \um u_y  &    \um u_x  &     0  &    0  &     1 &    0  &     0  &     0 \\[10pt]

      \um u_x^2 u_y  &   2u_x u_y  &   u_x^2  &     \um u_y  &    0  &     \um 2u_x &    1  &     0  &     0 \\[10pt]

      \um u_x u_y^2 &    u_y^2  &    2u_x u_y  &     0  &    \um u_x  &    \um 2u_y &    0  &     1  &     0 \\[10pt]

      u_x^2 u_y^2  &    \um u_x u_y^2  &    \um u_x^2 u_y  &    u_y^2  &   u_x^2  &    4u_x u_y &    \um 2 u_y  &     \um 2  u_x  &     1 \\
\end{bmatrix}
\end{equation}
Then, the transformation matrix $\tensr{F}^{-1}$ mapping a vector of (post-collision) central moments $\bm{\tilde{\kappa}}$ to a vector of (post-collision) raw moments $\bm{\tilde{\kappa}}^{'}$ can be written as
\begin{equation}\label{eq:tensorFinverse}
\tensr{F}^{-1}=
\begin{bmatrix}
      1  &    0  &    0  &     0  &    0  &     0 &    0  &     0  &     0 \\[10pt]

      u_x  &   1  &    0  &     0  &    0  &     0 &    0  &     0  &     0 \\[10pt]

      u_y  &    0  &   1  &     0  &    0  &     0 &    0  &     0  &     0 \\[10pt]

      u_x^2  &   2u_x  &    0  &     1  &    0  &     0 &    0  &     0  &     0 \\[10pt]

      u_y^2  &    0  &    2u_y  &     0  &    1  &     0 &    0  &     0  &     0 \\[10pt]

     u_x u_y  &   u_y  &    u_x  &     0  &    0  &     1 &    0  &     0  &     0 \\[10pt]

      u_x^2 u_y  &   2u_x u_y  &   u_x^2  &     u_y  &    0  &     2u_x &    1  &     0  &     0 \\[10pt]

      u_x u_y^2 &    u_y^2  &    2u_x u_y  &     0  &    u_x  &    2u_y &    0  &     1  &     0 \\[10pt]

      u_x^2 u_y^2  &    u_x u_y^2  &    u_x^2 u_y  &    u_y^2  &   u_x^2  &    4u_x u_y &    2 u_y  &     2  u_x  &     1 \\
\end{bmatrix}
\end{equation}
It may be noted that if $\tensr{F}=\tensr{F}(u_x,u_y)$, then $\tensr{F}^{-1}=\tensr{F}(-u_x,-u_y)$ (see~\cite{yahia2021central}).

Finally, we express the transformation matrix $\tensr{P}^{-1}$ mapping a vector of (post-collision) raw moments $\bm{\tilde{\kappa}^{'}}$ to a vector of
(post-collision) distribution functions $\mathbf{\tilde{f}}$ as
\begin{equation}\label{eq:tensorPinverse}
\tensr{P}^{-1} =
\begin{bmatrix}
     1  &\quad    0  &\quad    0  &\quad    \um 1  &\quad   \um 1  &\quad     0 &\quad    0  &\quad     0  &\quad     1 \\[10pt]
     0  &\quad    \frac{1}{2}  &\quad     0  &\quad    \frac{1}{2}  &\quad     0  &\quad    0 &\quad   0 &\quad   \um \frac{1}{2}  &\quad     \um \frac{1}{2} \\[10pt]
     0  &\quad    0  &\quad     \frac{1}{2}  &\quad     0  &\quad   \frac{1}{2}  &\quad    0  &\quad     \um \frac{1}{2}  &    0  &\quad   \um \frac{1}{2} \\[10pt]
     0  &\quad    \um \frac{1}{2}  &\quad     0  &\quad     \frac{1}{2}  &\quad     0  &\quad   0 &\quad     0  &\quad     \frac{1}{2}  &\quad    \um \frac{1}{2} \\[10pt]
     0  &\quad    0  &\quad     \um \frac{1}{2}  &\quad     0  &\quad     \frac{1}{2}  &\quad    0  &\quad     \frac{1}{2} &     0  &\quad     \um \frac{1}{2} \\[10pt]
     0  &\quad    0  &\quad     0  &\quad     0  &\quad     0  &\quad    \frac{1}{4}  &\quad    \frac{1}{4}  &\quad     \frac{1}{4}  &\quad   \frac{1}{4} \\[10pt]
     0  &\quad    0  &\quad     0  &\quad     0  &\quad     0  &\quad    \um \frac{1}{4}  &\quad     \frac{1}{4}  &\quad  \um \frac{1}{4}  &\quad \frac{1}{4} \\[10pt]
     0  &\quad    0  &\quad     0  &\quad     0  &\quad     0  &\quad    \frac{1}{4}  &\quad    \um \frac{1}{4} &\quad  \um \frac{1}{4}  &\quad    \frac{1}{4} \\[10pt]
     0  &\quad    0  &\quad     0  &\quad     0  &\quad     0  &\quad    \um \frac{1}{4}  &\quad     \um \frac{1}{4}  &\quad     \frac{1}{4}  &\quad    \frac{1}{4}
\end{bmatrix}
\end{equation}

\section{Characteristic Thermocapillary Velocity Scale on the Interface in Self-rewetting Fluids}\label{AppendixC}
The characteristic velocity scale can be derived by considering the balance of shear stress with the Marangoni stress due to the surface tension gradient on the interface. The shear stress scales as
\begin{eqnarray*}
\tau^b_{\mu} \sim \frac{\mu^b U_s}{b},
\end{eqnarray*}
where $b$ is the thickness of the lower fluid and $U_s$ is the unknown characteristic velocity scale to be determined in what follows. Furthermore, the surface tension gradient scales as
\begin{eqnarray*}
\frac{d \sigma}{d x} \sim \frac{d \sigma}{d T} \frac{\Delta T}{l},
\end{eqnarray*}
where $l$ is the length of the microchannel, and for SRFs with a quadratic dependence of surface tension on temperature (see Eq.~(\ref{ST_SRF}))
\begin{eqnarray*}
\frac{d \sigma}{d T} = \sigma_T + 2 \sigma_{TT}\left( T - T_{ref} \right).
\end{eqnarray*}
Thus, the velocity scale can be deduced by from the first two equations above by setting $\tau^b_{\mu} \sim d \sigma/d x$ and then substituting the last equation for $d \sigma/d T$ as
\begin{eqnarray}\label{eq:velocityscale}
U_s \sim \frac{\Delta T}{\mu^b} \left(\frac{b}{l}\right) \frac{d \sigma}{d T} = \frac{\Delta T}{\mu^b}\left(\frac{b}{l}\right)  \left[  \sigma_T + 2 \sigma_{TT}\left( T - T_{ref} \right) \right]
\end{eqnarray}
Here, we need a scale for the temperature on the interface, which we take it to be temperature field at $x=0$ and $y=0$, where it reaches a maximum. Now, using the temperature along the interface given as (see Appendix~\ref{Appendix A})
\begin{eqnarray*}
T(x,y=0) = C_1 + C_2 \cos\left(\omega x\right),
\end{eqnarray*}
and evaluating it at $x=0$, we get
\begin{eqnarray}
T(x=0,y=0) =  C_1 + C_2,
\end{eqnarray}
where
\begin{eqnarray}
C_1 = \frac{T_h \left(\frac{a}{b}\right) +T_c \tilde{k}}{\left(\frac{a}{b}\right) + \tilde{k}}, \qquad C_2 = \frac{\Delta T \sinh(\tilde{a})}{\tilde{k} \cosh (\tilde{a}) \sinh (\tilde{b}) + \cosh (\tilde{b}) \sinh (\tilde{a})},
\end{eqnarray}
Here, $\tilde{k} = k_a/k_b$, $\tilde{a} = a\omega$, and  $\tilde{b} = b\omega$. Substituting the above estimate for the temperature scale on the interface in Eq.~(\ref{eq:velocityscale}), we finally obtain the characteristic thermocapillary velocity scale in SRFs as
\begin{eqnarray}
U_s \sim  \frac{\Delta T}{\mu^b}\left(\frac{b}{l}\right)  \left[ \sigma_T + 2 \sigma_{TT}\left(C_1+C_2 - T_{ref} \right) \right].
\end{eqnarray}

\clearpage
\bibliographystyle{unsrt}


\end{document}